\documentclass[a4paper,12pt]{article}
\usepackage{epsfig}

\usepackage{citesort}
 \normalsize

\newcommand{\SD}{\scriptscriptstyle SD}
\newcommand{\IB}{\scriptscriptstyle IB}
\newcommand{\INT}{\scriptscriptstyle INT}
\newcommand{\bea}{\begin{eqnarray}}
\newcommand{\eea}{\end{eqnarray}}

\begin{document}

\def\lsim{\raise0.3ex\hbox{$\;<$\kern-0.75em\raise-1.1ex\hbox{$\sim\;$}}} 

\def\gsim{\raise0.3ex\hbox{$\;>$\kern-0.75em\raise-1.1ex\hbox{$\sim\;$}}}

\def\Frac#1#2{\frac{\displaystyle{#1}}{\displaystyle{#2}}}
\def\no{\nonumber\\}

\def\dobox#1#2{\centerline{\epsfxsize=#1\epsfig{file=#2, width=10cm,
height=5.5cm, angle=0}}}

\def\dofigure#1#2{\centerline{\epsfxsize=#1\epsfig{file=#2, width=13cm, 
height=5cm, angle=0}}}
%
\def\dofig#1#2{\centerline{\epsfxsize=#1\epsfig{file=#2, width=15cm, 
height=5cm, angle=0}}}
\def\dofigA#1#2{\centerline{\epsfxsize=#1\epsfig{file=#2, width=15cm, 
height=6cm, angle=0}\vspace{-1cm}}}
\def\dofigs#1#2#3{\centerline{\epsfxsize=#1\epsfig{file=#2, width=6cm, 
height=7.5cm, angle=-90}\hspace{0cm}
\hfil\epsfxsize=#1\epsfig{file=#3,  width=6cm, height=7.5cm, angle=-90}}}
\def\dotwofigs#1#2#3{\centerline{\epsfxsize=#1\epsfig{file=#2, width=8cm, 
height=6cm, angle=0}\hspace{0cm}
\hfil\epsfxsize=#1\epsfig{file=#3,  width=8cm, height=8cm, angle=0}}}
\def\dofourfigs#1#2#3#4#5{\centerline{
\epsfxsize=#1\epsfig{file=#2, width=6cm,height=7.5cm, angle=-90}
\hspace{0cm}
\hfil
\epsfxsize=#1\epsfig{file=#3,  width=6cm, height=7.5cm, angle=-90}}

\vspace{0.5cm}
\centerline{
\epsfxsize=#1\epsfig{file=#4, width=6cm,height=7.5cm, angle=-90}
\hspace{0cm}
\hfil
\epsfxsize=#1\epsfig{file=#5,  width=6cm, height=7.5cm, angle=-90}}
}

\def\dosixfigs#1#2#3#4#5#6#7{\centerline{
\epsfxsize=#1\epsfig{file=#2, width=6cm,height=7cm, angle=-90}
\hspace{-1cm}
\hfil
\epsfxsize=#1\epsfig{file=#3,  width=6cm, height=7cm, angle=-90}}

\centerline{
\epsfxsize=#1\epsfig{file=#4, width=6cm,height=7cm, angle=-90}
\hspace{-1cm}
\hfil
\epsfxsize=#1\epsfig{file=#5,  width=6cm, height=7cm, angle=-90}}

\centerline{
\epsfxsize=#1\epsfig{file=#6, width=6cm,height=7cm, angle=-90}
\hspace{-1cm}
\hfil
\epsfxsize=#1\epsfig{file=#7,  width=6cm, height=7cm, angle=-90}}
}

\def\no{\nonumber\\}
\def\slash#1{\ooalign{\hfil/\hfil\crcr$#1$}}
\def\ep{\eta^{\prime}}

%
\begin{titlepage}
\vspace*{-2cm}
\begin{flushright}
\end{flushright}

{\Large
\begin{center}
{\bf Light Mesons and Muon Radiative Decays and Photon Polarization Asymmetry}
\end{center} 
}
\vspace{.5cm}

\begin{center}
{Emidio Gabrielli$^{a}$ and Luca Trentadue$^{b}$}
\\[5mm]
{$^{a}$\textit{Helsinki Institute of Physics,
P.O.B. 64, 00014 University of  Helsinki, Finland }}\\[0pt]
{$^{b}$\textit{Dipartimento di Fisica, Universit\'a di Parma, \\ and \\ INFN, Gruppo Collegato di Parma, Viale delle Scienze 7, 43100 Parma, Italy}}\\
[10pt]
\vspace{1.5cm}

\begin{abstract}
We systematically compute and discuss meson and muon polarized radiative 
decays. Doubly differential distributions in terms of momenta and helicities 
of the final lepton and photon are explicitly computed.
The undergoing dynamics giving rise to lepton and photon 
polarizations is examined and analyzed in the soft and hard region of 
momenta. The  particular configurations made by right-handed leptons with 
accompanying photons are investigated and interpreted as a manifestation 
of the axial anomaly. The photon polarization asymmetry is evaluated. 
Finiteness of polarized amplitudes against infrared and collinear singularities
is shown to take place with mechanisms distinguishing between right handed and 
left handed final leptons. 
We propose a possible test using photon polarization to clarify a recently 
observed discrepancy in radiative meson decays.
\end{abstract}

\end{center}

\end{titlepage}

\noindent

\vspace{0.5in}

\section{Introduction}
Radiative decays of light mesons and leptons have been 
widely studied both experimentally and theoretically. 
They represent an excellent source of information on the 
experimental side as well as a benchmark for  theoretical speculations. 
Extensive comparisons have been carried on in the past between experiments 
and theoretical predictions for meson radiative decays 
(see for example Ref.\cite{bdl}). 
A while ago, radiative polarized leptonic decays of mesons \cite{cwsk,tv}
and muons \cite{fgkm,ss} have also been considered. 
Recently special attention has been given to the role played 
by the final lepton mass $m_l$ in the threshold region of the 
decay and to the $m_l\rightarrow 0$ limit concerning the helicity 
amplitudes for mesons \cite{tv} and leptons Refs.\cite{fgkm,ss}. 
The $O(\alpha)$ radiative corrections generate an helicity flip of 
the final lepton even in the zero mass limit
\cite{ln} provided the lepton mass is kept from the beginning into account. 
Following the interpretation due to Dolgov and Zakharov \cite{dz} of the 
axial anomaly the final states with opposite helicity can be interpreted 
\cite{s,fs,tv}, as a manifestation of the axial anomaly giving rise to a 
peculiar mass-singularity cancellation for the right-handed 
polarized final lepton amplitudes. 

We consider in this work the case of polarized radiative decays of the 
pion and kaon meson and of the muon more extensively by taking into 
account polarizations of final lepton and photon degrees of freedom. 
Contrary to the previous case \cite{cwsk}, in meson decays we consider
the polarization states of both lepton and photon final states.

This approach, containing a complete description of the final
momenta and helicities,  may give further and more detailed  information on 
the final state with respect to the inclusively polarized and
unpolarized cases.  
Furthermore, the agreement with the more inclusive results 
previously obtained in the literature can be easily recovered by 
summing over the emitted final states polarizations.
It is worth noticing that
this approach allows to describe more closely the interplay 
between several peculiar features of the dynamics involved. As, 
for instance, to pinpoint the role played by angular momentum conservation 
and its connection with hard and soft photon momenta, and
to consider the role played by the parity conservation in weak decays. 
All these aspects related to angular momentum dynamics
may be effectively described in terms of the photon polarization asymmetry.

Here we emphasize that the knowledge of the helicity amplitudes 
of the final leptons and photons, in addition to an explicit 
test  of the angular momentum conservation, shows the relative 
rates of the partial helicity amplitudes. 
Indeed, in the total rate, different helicity amplitudes, 
depending on the range of momenta, enter  with varying weights. 
Therefore, this behavior gives the opportunity to isolate peculiar 
polarized configurations in order to maximize or minimize 
them according to favorable intervals of momenta. 
As far as phenomenological applications are concerned, 
this may be, as will be discussed later, 
an effective way to compare theory and experiment on a new basis. 
The case of the photon polarization asymmetry, proposed in this work, 
allows, in this respect, a new approach to inspect interaction dynamics 
via a finite and universal quantity which is also directly associated to parity 
violation. Moreover, the photon polarization asymmetry
is very sensitive, in radiative meson decays, to the hadronic structure, 
allowing for a more precise determination of the electromagnetic form factors
with respect to the one obtained so far.

Some of the results achieved in this paper can be shortly listed: 
we explicitly calculate amplitudes and final distributions 
in terms of lepton and photon momenta at fixed final lepton 
and photon helicities.
Double differential expressions in terms of lepton and photon momenta are 
also provided together with the partial helicity amplitudes 
for the meson and muon cases respectively.
Moreover, we analyze how the cancellation pattern of mass singularities works
on polarized processes.
In the inclusive quantities this is a sensible test of 
the consistency of the results. Once inclusive distributions are 
obtained by integrating over final momenta, we observe the 
cancellation of all mass singularities both infrared and collinear. 
In particular, a peculiar pattern of mass singularity cancellation 
is shown to take place, which differs for the left-handed helicity final 
lepton states with respect to the right-handed ones. 
The same behavior can be observed for the meson as well as for the muon case.

Finally, we discuss a possible interpretation in terms of tensorial 
coupling of the results obtained recently at the PIBETA 
experiment \cite{pibeta} for the radiative pion decay in electron channel.
It is argued that polarized radiative processes may 
constitute a sensible test to resolve the controversial 
issue of tensorial couplings in radiative pion decay, 
allowing also for a sensitive test in the corresponding kaon decays as well.

The paper is organized as follows: In Section 2 we consider the case 
of the meson polarized radiative decay. We discuss the contributing 
amplitudes and the underlying theoretical tools. 
We also define the gauge invariant set of matrix 
elements together with the definition of the Lorentz invariant quantities. 
Allowed and forbidden helicities configurations 
are here analyzed as well. In Section 3 the polarized radiative muon 
decay is discussed. In Section 4 we define distributions of branching 
ratios in the photon and electron energies and the photon polarization 
asymmetry. Numerical results for the distributions of branching ratios 
and polarization asymmetries are provided in subsections 4.1, 4.2, 
and 4.3 for the cases of pion, kaon, and muon decays respectively. 
Results for the polarized electron energy spectra are presented in Section  5.
In Section 6 the dependence of the photon polarization asymmetry,
induced by tensorial couplings, is discussed in the radiative 
pion and kaon decays.
The peculiar pattern of mass singularities cancellation in 
polarized radiative decays is described in Section 7, while
conclusions are presented in Section 8. 
General results and the corresponding formulae for the polarized 
radiative meson and muon decays are collected in Appendix A and B respectively.

\section{The polarized radiative meson decay }
\begin{figure}[tpb]
\begin{center}
{\centerline{\epsfxsize=3.1in\epsfig{file=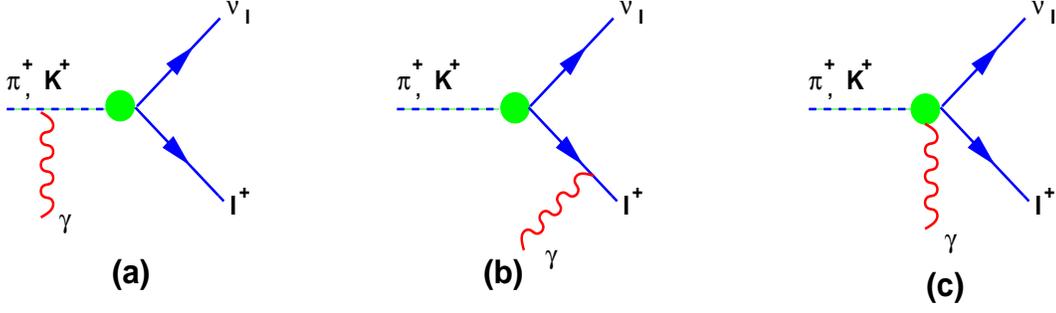, width=14cm, 
height=4cm, angle=0}}}
\end{center}
\caption{\small  Feynman diagrams for $(\pi^+, K^+) \to \nu_{l} l^+ \gamma$
decay, where $l=e,\mu$}
\label{Pion}
\end{figure}

We start this section with the calculation of the polarized amplitude 
for the process 
\bea
M^+(p)\to \nu_l(p_{\nu})~ l^+(p_l,\lambda_l)~ \gamma(k,\lambda_{\gamma})\, ,
\eea
where $M^+=\pi^+~ (K^+)$ and $l=e~ (\mu)$ stand 
for pion (kaon) and electron (muon) respectively,
with $\nu_{l=e,\mu}$ the corresponding neutrinos.
The four momenta $p, p_{\nu}, p_l$ correspond 
to meson $M$, neutrino, and charged lepton, while
$\lambda_l, \lambda_{\gamma}$ indicate the charged lepton and photon helicity, 
respectively. The neutrino is
assumed massless and therefore is a pure left-handed polarized state.
The Feynman diagrams at tree-level 
for this process are shown in Fig.\ref{Pion}, where the green bubble 
just indicates the Fermi interaction.
The first two diagrams
Figs.1a-b correspond to the so-called 
inner bremsstrahlung (IB) diagrams, 
where the photon is emitted from external lines and the meson behaves
as a point-like scalar particle. 
The third diagram Fig.1c is the so-called 
structure-dependent (SD) diagram, where the photon is emitted from 
an intermediate hadronic state and the matrix element will depend on
the vectorial (V) and axial (A) meson form factors.
The total amplitude for this process can be split in two gauge invariant
contributions
\bea
{\rm M}^{(\lambda_l,\lambda_{\gamma})}=
{\rm M}_{IB}^{(\lambda_l,\lambda_{\gamma})}+
{\rm M}_{SD}^{(\lambda_l,\lambda_{\gamma})}\, ,
\eea
where ${\rm M}_{IB}$ and ${\rm M}_{SD}$ 
correspond to the IB and SD part of the amplitude.
The IB amplitude is given by \cite{EG,Ksemilept} 
\bea
{\rm M}_{IB}^{(\lambda_l,\lambda_{\gamma})}
=\frac{i e G_F}{\sqrt{2}}m_l f_M V_{uq}~
\epsilon^{\star}_{\mu}(k,\lambda_{\gamma})
\left[\bar{u}(p_{\nu})\left(\frac{p^{\mu}}{\left(p\cdot k\right)}
-\frac{/\hspace{-0.25cm} 
k \gamma^{\mu}+2 p^{\mu}_l}{2\left(p_l\cdot k\right)}\right)
\left(1+\gamma_5\right)v(p_l,\lambda_l)\right]\, ,
\label{MIB}
\eea
where $/\hspace{-0.25cm}k=\gamma^{\alpha} k_{\alpha}$,
$\epsilon_{\mu}(k,\lambda_{\gamma})$ stands for the 
photon polarization vector of momentum $k$ and helicity $\lambda_{\gamma}$, 
while $\bar{u}(p_{\nu})$ and $v(p_l,\lambda_l)$ are the 
bispinors of final neutrino and charged lepton respectively.
Explanations of other symbols appearing above are in order.
The $G_F$ is the Fermi constant, $m_l$ is the charged lepton mass, 
$f_M$ is the meson decay constant, where $f_\pi\simeq 131$ MeV 
and $f_K\simeq 161$ MeV, and
$V_{uq}$ is the Cabibbo-Kobayashi-Maskawa matrix element corresponding to
$u\to q=d$ and $u\to q=s$ quark transitions for pion and kaon decays 
respectively.

The SD part of the amplitude contains 
vectorial $(V)$  and axial $(A)$ form factors, that
clearly depend on the kind of initial meson, but not on the 
lepton final states. Indeed,
they are connected to the matrix elements of the 
electromagnetic hadron current $V_{\rm em}^{\mu}$ and the 
axial and vectorial weak currents $A^{\mu}$ and $V^{\mu}$ respectively, as
\bea
({\rm V,A})^{\mu\nu}(p,k)\equiv 
\int d^4x e^{i kx} \langle 0|~ T V^{\mu}_{\rm em}(x)
(V(0),A(0))^{\nu} ~|M^+(p)\rangle\, .
\eea
Using Lorentz covariance and electromagnetic gauge invariance, it follows that:
\bea
{\rm V}^{\mu\nu}(p,k)&=&i \frac{V}{m_M} \epsilon^{\mu\nu\alpha\beta} 
k_{\alpha} p_{\beta}
\nonumber\\
{\rm A}^{\mu\nu}(p,k)&=&(p\cdot k) \frac{A}{m_M}\left(\eta^{\mu\nu} -
\frac{p^{\mu} k^{\nu}}{(p\cdot k)}\right)
-f_M\left(\eta^{\mu\nu}
+\frac{p^{\mu}\left(p^{\nu}-k^{\nu}\right)}{(p\cdot k)}\right)\, ,
\eea
where $\eta^{\mu \nu}={\rm diag}(1,-1,-1,-1)$ is the Minkowski metric, and
$\epsilon^{\mu\nu\alpha\beta}$ is the totally antisymmetric tensor
\footnote{In our notation, the $\epsilon^{\mu\nu\alpha\beta}$ is defined as
$\epsilon^{0 1 2 3}=1$ and $\epsilon_{0 1 2 3}=-1$, 
when generic four-vectors $v_{\mu}$ 
are $v_{\mu}=(v_0,\vec{v})$ and $v^{\mu}=(p_0,-\vec{v})$.}.
Finally, the SD part of the amplitude is given by \cite{EG,Ksemilept} 
\bea
{\rm M}_{SD}^{(\lambda_l,\lambda_{\gamma})}
&=&-\frac{i e G_F}{\sqrt{2}} V_{uq}~
\epsilon^{\star}_{\mu}(k,\lambda_{\gamma})
\left\{\left(p\cdot k\right) \frac{A}{m_M}
\left(-\eta^{\mu\nu}+\frac{p^{\mu} k^{\nu}}{\left(p\cdot k\right)}\right)
+i \epsilon^{\mu\nu\alpha \beta}\frac{V}{m_M} k_{\alpha} p_{\beta}\right\}
\nonumber\\
&\times&
\Big[
\bar{u}(p_{\nu})\gamma_{\nu}\left(1-\gamma_5\right)v(p_l,\lambda_l)
\Big]\, ,
\label{MSD}
\eea
where  $m_M$ stands for the meson mass, while
$V$ and $A$ are the meson vectorial and axial form factors respectively.
Notice that both the terms $M_{IB}$ and $M_{SD}$ are separately
gauge invariant, as can be easily checked by making the substitution
$\epsilon^{\star}_{\mu}(k,\lambda_{\gamma})\to 
\epsilon^{\star}_{\mu}(k,\lambda_{\gamma}) + k_{\mu}$ in 
Eqs.(\ref{MIB}) and (\ref{MSD}).

Now we provide the corresponding expressions for the polarized amplitude
in the center of mass (c.m.) frame of the fermion pair 
(neutrino and charged lepton), 
namely $\vec{p}_{l}+\vec{p}_{\nu}=0$.
We choose a frame where the 3-momenta of neutrino and photon have
the following components in polar coordinates
\bea
\vec{p}_{\nu}=E_{\nu}~\left(\sin{\theta} \cos{\varphi}, 
\sin{\theta} \sin{\varphi}, \cos{\theta}\right),~~~
\vec{p}_{l}=-\vec{p}_{\nu},~~~~
\vec{k}=E_{\gamma}~(0,0,1)\, ,
\eea
where $E_{\nu}$ and $E_{\gamma}$ 
are the neutrino and photon energies
respectively and $\theta,\varphi$ are the usual polar angles.
For the photon polarization vectors we choose helicity eigenstates
($\epsilon(k,\lambda)$), which in this frame are  given by 
\bea
\epsilon_{\mu}(k,\lambda_{\gamma})=\frac{1}{\sqrt{2}}
\left(0,1,i\lambda_{\gamma},0\right)
\eea
whose helicity eigenvalues correspond 
to $\lambda_{\gamma}=-1$ left-handed (L) and 
$\lambda_{\gamma}=1$ right-handed (R) circular polarizations.
Photon polarization vectors satisfy the transversality 
condition $k^{\mu} \epsilon_{\mu}(k,\lambda_{\gamma})=0$.
Regarding the polarization vectors of fermions,
it is convenient to use the solution of the Dirac equation
for the particle ($u$) and antiparticle ($v$) bispinors in 
the momentum space \cite{Landau}. In the standard basis \footnote{
In the standard basis representation, 
$\gamma_0={\rm Diag}({\bf 1},-{\bf 1})$, and 
$\vec{\gamma}=\left ( \begin{array}{c} 
~0~~~\vec{\sigma}
\\ 
-\vec{\sigma}~~0
\end{array} 
\right)
$, and $\gamma_5=\left ( \begin{array}{c} 
0~~{\bf 1}
\\ 
{\bf 1}~~0
\end{array} 
\right)$, where ${\bf 1}={\rm Diag}(1,1)$ and $\vec{\sigma}$ are as usual the 
Pauli matrices.
} we have:
\bea
u(p,\lambda)=\left ( \begin{array}{c} 
\sqrt{E+m}\,\,  \omega_{\lambda}(\vec{n})
\\ 
\sqrt{E-m}\,\,  
\left(\vec{\sigma}\cdot \vec{n}\right) \omega_{\lambda}(\vec{n})
\end{array} 
\right)
~~~~~
v(p,-\lambda)=\left ( \begin{array}{c} 
\sqrt{E-m}\,\, 
\left(\vec{\sigma}\cdot \vec{n}\right) \omega_{\lambda}(\vec{n})
\\ 
\sqrt{E+m}\,\,  \omega_{\lambda}(\vec{n}) 
\end{array} 
\right)\,  ,
\label{spi}
\eea
where the 2-component spinors $\omega_{\lambda}(\vec{n})$ 
(with helicity $\lambda=\pm 1$) are the eigenstates of the helicity operator
$\left(\vec{\sigma}\cdot \vec{n}\right) 
\omega_{\lambda}(\vec{n})
=\lambda\, \omega_{\lambda}(\vec{n})$, and
$\; \sigma_i$ are the Pauli matrices. Here, 
$\vec{n}\equiv \vec{p}/|\vec{p}|$, where $\vec{p}$ is the 3-momentum and
$E=\sqrt{|\vec{p}|^2+m^2}$ is the corresponding energy. If
$\vec{p}=|\vec{p}|\left(\sin{\theta}\cos{\varphi},
\sin{\theta}\sin{\varphi},\cos{\theta}\right)$, then
in polar coordinates, $\omega_{\lambda}(\vec{n})$ can be expressed as
\bea
\omega_{+1}(\vec{n})=\left ( \begin{array}{c} 
e^{-i\frac{\varphi}{2}}\, \cos{\frac{\theta}{2}}
\\ 
e^{i\frac{\varphi}{2}}\, \sin{\frac{\theta}{2}}
\end{array} 
\right)\, ,
~~~~~
\omega_{-1}(\vec{n})=\left ( \begin{array}{c} 
-e^{-i\frac{\varphi}{2}}\, \sin{\frac{\theta}{2}}
\\ 
e^{i\frac{\varphi}{2}}\, \cos{\frac{\theta}{2}}
\end{array} 
\right)\, .
\eea

At this point it is convenient to introduce the following
Lorentz invariant quantities
\bea
x\equiv \frac{2 p\cdot k}{m_M^2},~~~~
y\equiv \frac{2 p\cdot p_l}{m_M^2},~~~~
z\equiv \frac{2 p_l\cdot k}{m_M^2}=y-1+x-r_l
\label{variables}
\eea
where in the meson rest frame, $x$ and $y$ are just 
proportional to the photon and charged lepton energies respectively
and $r_l=m_l^2/m_M^2$.
Finally, after a straightforward algebra, the $IB$ and $SD$ contributions to 
the polarized amplitude in the fermion pair c.m. frame are given by
\bea
{\rm M}_{IB}^{(\lambda_l,\lambda_{\gamma})}&=&
e G_F m_{l} f_MV_{uq}~\frac{2}{z}
\left\{\delta_{\lambda_l,-1}\left(
\delta_{\lambda_{\gamma},-1}~\hat{E}_{\gamma}
+\hat{E}_{\nu}\right)R_{+}\sin{\theta}
\right. \nonumber\\
&+&\left.\delta_{\lambda_l,+1}~\delta_{\lambda_{\gamma},-1}~
\hat{E}_{\gamma}
R_{-}\left(1-\cos{\theta}\right)
\right\} e^{i\lambda_{\gamma} \varphi}
\nonumber\\
\nonumber\\
{\rm M}_{SD^{\pm}}^{(\lambda_l,\lambda_{\gamma})}&=&
e G_F m_M^2 V_{uq}~
\frac{\left(V\pm A\right)}{2}~\delta_{\lambda_{\gamma},\pm 1}~ x
\Big\{\mp\delta_{\lambda_l,-1}~R_{-}\sin{\theta}
\nonumber\\
&\pm&
\delta_{\lambda_l,+1}R_{+}\left(\cos{\theta}\pm 1\right)
\Big\}~ e^{i\lambda_{\gamma} \varphi}\, ,
\label{Mpol}
\eea
where the structure dependent part is given by 
${\rm M}_{SD}^{(\lambda_l,\lambda_{\gamma})}=
{\rm M}_{SD^+}^{(\lambda_l,\lambda_{\gamma})}+
{\rm M}_{SD^-}^{(\lambda_l,\lambda_{\gamma})}$
and the symbol
$R_{\pm}\equiv \sqrt{\hat{E}_{\nu}}\left(\sqrt{\hat{E}_l+\sqrt{r_l}}
\pm \sqrt{\hat{E}_l-\sqrt{r_l}}\right)$, with
$\hat{E}_{i}\equiv E_i/m_M$ and $E_l$ is 
the energy of the final charged lepton.
In this frame, the energies normalized to the meson mass are given by
\bea
\hat{E}_{\gamma}&=&\frac{x}{2\sqrt{1-x}},~~~~
\hat{E}_{\nu}=\frac{1-x-r_l}{2\sqrt{1-x}},~~~
\hat{E}_{l}=\frac{1-x+r_l}{2\sqrt{1-x}},
\nonumber\\
\cos{\theta}&=&\frac{(x-2)(1-x+r_l)+2y(1-x)}{x(1-r_l-x)}
\label{energies}
\eea
and
\bea
R_{+}&=&\sqrt{1-r_l-x},~~~R_{-}=\sqrt{r_l\frac{1-r_l-x}{1-x}}\, .
\eea
Notice that, as expected from general arguments,
the azimuthal angle $\varphi$ factorizes in the overall phase
of the amplitude. At this point it is important to stress that the
SD terms in the amplitude, proportional to 
$V+A$ and $V-A$, correspond to pure right-handed and left-handed 
photon polarizations respectively, while the IB one is a mixture of both.
In particular, the terms proportional to pure left-handed photon polarizations 
in the $M_{IB}$, come only from the tensorial structure in Eq.(\ref{MIB}), 
namely from terms proportional to 
$[\bar{u}_{\nu}~ \sigma_{\mu \nu}(1+\gamma_5)~ v_l ]$, 
while scalar contributions of type 
$[\bar{u}_{\nu}~(1+\gamma_5)~ v_l]$
do not select any specific photon polarization. We will return
on this point in the following when anomalous tensorial 
coupling in radiative pion, and kaon decays will be discussed.

By using Eqs.(\ref{Mpol}) and (\ref{energies}), it is now straightforward to 
evaluate the square modulus of the amplitude. 
Below we will provide its
expression summed over the charged lepton polarizations, as a function
of the photon helicities. After integrating over the phase space, we obtain 
for the photon polarized decay rate $\Gamma^{\lambda_{\gamma}}$, 
the following result:
\bea
\frac{d^2 \Gamma^{(\lambda_{\gamma})}}{dx~ d\lambda}=
\frac{m_{\rm M}}{256 \pi^3} \sum_{\lambda_l=\pm 1} 
|{\rm M}^{(\lambda_l,\lambda_{\gamma})}|^2=
\rho^{(\lambda_{\gamma})}(x,\lambda)\, .
\eea
Here  $m_{\rm M}$ stands for the generic meson mass
$m_{{\rm M}=\pi,K}$, and $\lambda\equiv z/x$.
The Dalitz plot densities 
$\rho^{\lambda_{\gamma}}(x,\lambda)$ for the polarized decay 
are Lorentz invariant functions, and are given by
\bea
\rho^{(-1)}(x,\lambda)&=&
A_{\IB}~ f_{\IB}^{\rm L}(x,\lambda)
+A_{\SD}~ \frac{1}{2} (V-A)^2 f_{\SD}^{\rm L}(x,\lambda)
+A_{\INT} (V-A) f_{\INT}^{\rm L}(x,\lambda)~~~ 
\eea
\bea
\rho^{(+1)}(x,\lambda)&=&
A_{\IB}~ f_{\IB}^{\rm R}(x,\lambda)
+A_{\SD}~  \frac{1}{2}(V+A)^2 f_{\SD}^{\rm R}(x,\lambda)
+A_{\INT} (V+A) f_{\INT}^{\rm R}(x,\lambda)\, , ~~~  
\eea
where
\bea
A_{\IB}&=&2\,r_l \left(\frac{f_M}{m_M}\right)^2 A_{\SD},~~~ 
A_{\INT}=2\, r_l \, \frac{f_M}{m_M}\, A_{\SD} 
\nonumber\\
A_{\SD}&=& \frac{\alpha}{32 \pi^2}\,G_F^2\,m_M^5\,|V_{uq}|^2\, .
\label{ASD}
\eea
In the following, for later convenience, we will introduce the labels R and L
corresponding to photon helicities $\lambda_{\gamma}=1$ and 
$\lambda_{\gamma}=-1$ respectively.
The functions $f^{\rm L,R}_{\IB}(x,\lambda),~f_{\SD}^{L,R}(x,\lambda)$, and
$f_{\INT}^{L,R}(x,\lambda)$ are given by
\bea
f^{\rm L}_{\IB}(x,\lambda)&=&\frac{1-\lambda}{x\lambda}
\left(1+r_l\left(x-1\right)-\frac{r_l}{\lambda}\left(1+x-r_l\right)
\right)
\nonumber\\
f^{\rm R}_{\IB}(x,\lambda)&=&\frac{1-\lambda}{x\lambda}
\left(x-1+\frac{r_l}{\lambda}\right)\left(x-1+r_l\right)
\nonumber\\
f_{\SD}^{\rm R}(x,\lambda)&=&x^2\lambda\left(\left(1-x\right)
\left(x\lambda+r_l\right)-r_l\right)
\nonumber\\
f_{\SD}^{\rm L}(x,\lambda)&=&x^2\left(1-\lambda\right)\left(\left(x-1\right)
\left(r_l+x\left(\lambda-1\right)\right)+r_l\right)
\nonumber\\
f_{\INT}^{\rm R}(x,\lambda)&=&\frac{1-\lambda}{\lambda}\left(\left(x-1\right)
\left(x\lambda+r_l\right)+r_l\right)
\nonumber\\
f_{\INT}^{\rm L}(x,\lambda)&=&
\frac{1-\lambda}{\lambda}\left(x^2+\left(1-x\right)
\left(x\lambda+r_l\right)-r_l\right)\, .
\label{density}
\eea
The function
$f^{\rm L}_{\IB}(x,\lambda)+f^{\rm R}_{\IB}(x,\lambda)$
coincides with the corresponding
IB function $f_{\IB}(x,\lambda)$
for the unpolarized case provided in \cite{Pobl1,EG,Ksemilept},  
as well as
$f_{\SD}(x,\lambda)=f_{\SD}^L(x,\lambda)+ f_{\SD}^R(x,\lambda)$ 
and $f_{\INT}(x,\lambda)= f_{\INT}^L(x,\lambda)+f_{\INT}^R(x,\lambda)$.
More general results for the complete polarized radiative decay rate, 
including also the charged lepton helicity in the pion rest frame, 
are provided in appendix A.

In order to obtain the differential branching ratio (${\rm BR}$) 
it is convenient to rewrite the term $A_{SD}$ in Eq.(\ref{ASD}) as 
\bea
A_{\SD}=\frac{\alpha}{4\pi} \frac{1}{r_l\left(1-r_l\right)^2}
\left(\frac{m_M}{f_M}\right)^2 \Gamma_0
\eea
where $\Gamma_0=\Gamma(M\to l \nu_l)$ is Born contribution to 
the total width of non radiative decay $M\to l \nu_l$, in particular
\bea
\Gamma_0(M^+(p)\to \nu_l+l^+)=\frac{G_F^2\,f_{M}^2\,m_M}{8\pi}|V_{uq}]^2
r_l\,(1-r_l)^2\, .
\eea
Then
\bea
\frac{d^2 {\rm BR}}{{dx~ d\lambda}} ={\rm BR}(M\to l \nu_l)~
\frac{1}{\Gamma_0}\frac{d^2 \Gamma}{dx~ d\lambda}\, ,
\label{BRM}
\eea
where ${\rm BR}(M\to l \nu_l)$ is the total branching ratio 
of the corresponding non radiative decay.
Finally, the total branching ratio ${\rm BR}$ is obtained by integrating 
Eq.(\ref{BRM}) in the full kinematical range as follows
\bea
{\rm BR} =\int dx \int d\lambda~ \frac{d^2 {\rm BR}}
{{dx~d\lambda}}
\eea
\bea
0\le x \le 1-r_l,~~~~~ \frac{r_l}{1-x}\le \lambda \le 1\, .
\eea
In case in which kinematical cuts ($x^{\rm min}$, and $\lambda^{\rm min}$)
should be applied, the minima of integrations should be replaced as
\bea
x^{\rm min}\le x \le 1-r_l,~~~~~ {\rm max}\left\{\lambda^{\rm min}, 
~\frac{r_l}{1-x}\right\}\le \lambda \le 1\, .
\eea

\begin{figure}[tpb]
\begin{center}
\dofig{3.1in}{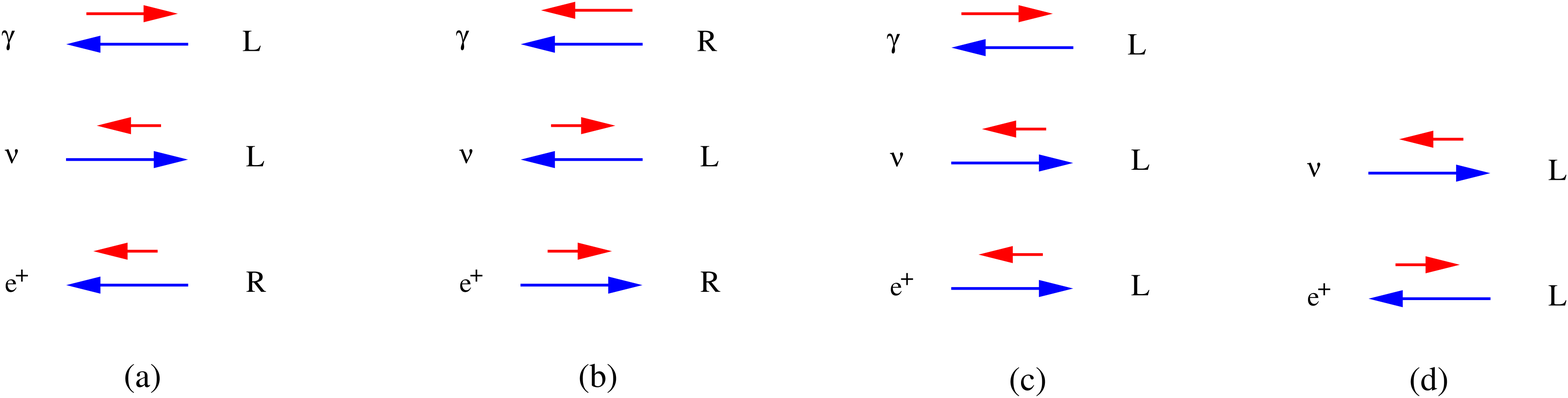}
\end{center}
\caption{\small Allowed helicity (in red) configurations of $\gamma$, $\nu$ and
$e^+$  for $\pi^+\to e^+\nu_e\gamma$ decay in $\pi^+$ rest frame, 
figures (a),~  (b),~  (c), when all momenta
(in blue) are aligned on the same axis. 
Direction of photon momentum is fixed by convention. Figure (d) 
corresponds to the non radiative decay  $\pi^+\to e^+\nu_e$.
Analogous spin configurations hold for the corresponding $K^+$ decays as well.
}
\label{spin}
\end{figure}

Let us now consider the positron decay mode, where, as a good approximation,
the lepton mass can be neglected in comparison to the pion one.
A remarkable aspect of the results 
in Eq.(\ref{density}), is that in the
limit in which  $\lambda\to 0$ and $x\to 1$, which corresponds
to the emission of low energy positron and
hard photons at relative small angles in the meson rest frame,
the contribution proportional
to $f_{\IB}^{\rm L}(x,\lambda)$ and to $f_{\INT}^{\rm L}(x,\lambda)$ 
distributions dominates in the decay rate. In other words,
hard photons will be mainly produced with left-handed polarizations.
This behavior, as will be shown in more details in section 4, 
is just a consequence of the V-A nature of weak interactions 
and of the angular momentum conservation. This can be easily understood as follows.
In the $\pi^+$ rest frame, neglecting the lepton mass, 
we have 
$x=2E_{\gamma}/m_M$, and $\lambda=E_{e}/m_M(1-\cos{\theta_{\gamma e}})$, 
where $\theta_{\gamma e}$ is the angle between positron 
and photon momenta.
Let us consider the 
kinematical region in which $\lambda \to 0$, which corresponds to
$E_e\to 0$ and/or $\theta_{\gamma e}\to 0$.
Due to the conservation of total momentum, and to the fact that 
$|\vec{p}_e|/m_M\ll 1$ and $\theta_{\gamma e}\to 0$, 
the neutrino must be emitted in this region
almost backward with respect to the photon direction, 
therefore final momenta are almost
aligned on the same axis. This configuration is 
shown in Fig.\ref{spin}a, where all momenta are chosen to be 
aligned on the same axis. 
One then could easily check the conservation of the spin $(S_X)$   
along the direction of the photon momentum which in Fig.\ref{spin} is set by convention 
on the negative $X$-axis.
Since neutrino is a purely left-handed state,
its spin projection along $X$-axis would be 
$S_X(\nu_l)=+1/2$. As a consequence 
of the angular momentum conservation ($S_X(\pi)=0$ for pion), the photon must 
also be left-handed polarized giving $S_X(\gamma)=-1$. In this case the
positron, whose momentum is parallel to the one of the photon,
must be right-handed in order to satisfy the total sum
$S_X(\gamma) + S_X(\nu_e)+ S_X(e^+)=0$. 
Notice that, in this particular kinematical limit $\lambda\to 0$, 
photons with right-handed polarization
would be suppressed, since the total sum of spins along $X$-axis would give
in that case $S_X(\gamma)+S_X(\nu_l)=3/2$ thus spoiling angular momentum conservation. 
It is worth noticing that also in 
the case of $\pi^+ \to \mu^+ \nu \gamma$ decay mode, where the muon mass
cannot be neglected in comparison to the one of the  pion,
the left-handed photon amplitude still dominates for 
high energy photons.
This fact can be explained as follows. When the photon energy approaches its
maximum value, 
being neutrino massless,  in order to conserve total momentum,
the production of the $\mu^+$ at rest it is favored. In this case the momentum of the neutrino
should be opposite to one of the photon. As explained above, for this kinematical
configuration, the photon is therefore favored to be produced as left-handed 
in order to conserve total angular momentum.
Same considerations apply to the corresponding kaon decays as well.

Another interesting case is the one in which the photon energy tends to zero, 
namely $x\to 0$. In this singular kinematical region, 
one should expect soft photons to behave as scalar 
particles, carrying no spin. Then in this limit 
both the left-handed or right-handed distributions should tend to the same
value, as indeed can be verified by performing the limit $x\to 0$ on the
density distributions in Eq.(\ref{density}). 
In the following we will show how this property could be relevant in order 
to define an observable which is free from infrared 
($E_{\gamma}\to 0$) singularity, 
namely the photon polarization asymmetry.
%
\section{The polarized radiative muon decay}
\begin{figure}[tpb]
\begin{center}
{\centerline{\epsfxsize=3.1in\epsfig{file=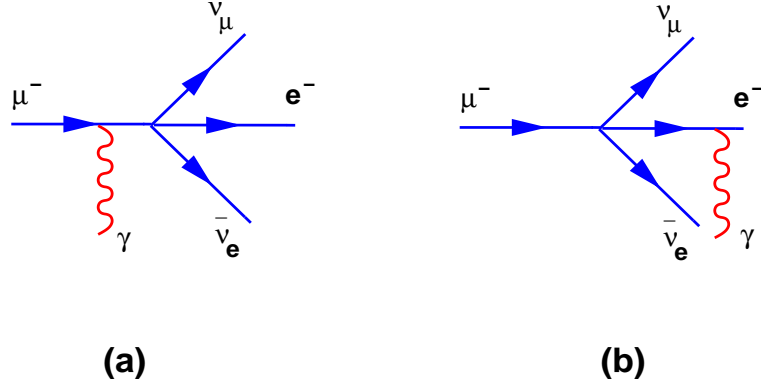, width=10cm, 
height=5cm, angle=0}}}
\end{center}
\caption{\small Feynman diagrams for $\mu^-\to \nu_{\mu} \bar{\nu}_{e} 
e^- \gamma$ decay.}
\label{Muon}
\end{figure}

Here we analyze the radiative muon decay 
\bea
\mu^-(p)\to \nu_{\mu}(q_1)\, \bar{\nu}_e(q_2)\, e^-(p_e)\, \gamma(k)
\eea
in which both photon and electron final states are polarized.
The corresponding Feynman diagrams for this process are shown in 
Fig.\ref{Muon}, where
$p,\, p_e,\, q_{1,2},\, k$ are the corresponding momenta.
This decay is obtained from the non radiative one
$\mu^-\to \nu_{\mu}\, \bar{\nu}_e\, e^-$, by simply attaching the photon to 
the muon and electron external lines. Due to the V-A nature 
of weak interactions and a simple Fierz rearrangement,
the square modulus of the polarized amplitude can be factorized as follows
\cite{fgkm,ss,fks}
\bea
|M^{(\lambda_{\gamma},\lambda_e)}|^2=
\frac{G_F^2}{2}\,\Big[M_{\alpha}^{(\lambda_{\gamma},\lambda_e)\dag}
\,M_{\beta}^{(\lambda_{\gamma},\lambda_e)}\Big]\, 
\Big[N^{\alpha \dag}\,N^{\beta}\Big]\, ,
\label{M2muon}
\eea
where $M^{\alpha}(\lambda_e,\lambda_{\gamma})$ 
corresponds to the ${\cal O}(\alpha)$
amplitude in which photon is either radiated off the electron or 
off the muon, and $N^{\alpha}$ to the neutrino amplitude respectively
\bea
M^{\alpha}(\lambda_{\gamma},\lambda_e)&=&e\,\bar{u}_e(p_e,\lambda_e)\left(
\gamma^{\delta}\frac{\slash p_e+ \slash k+m_e}{(p_e+k)^2-m_e^2}
\gamma^{\alpha}_L
+\gamma^{\alpha}_L
\frac{\slash p - \slash k+m_{\mu}}{(p_{\mu}-k)^2-m_{\mu}^2}
\gamma^{\delta}
\right)u_{\mu}(p)\,
\epsilon_{\delta}^{\star}(k,\lambda_{\gamma})
\nonumber\\
N^{\alpha}&=&\bar{u}_{\nu}(q_1)\,\gamma^{\alpha}_L v_{\nu}(q_2)\, ,
\eea
where $u_{\mu}(p)$, $u_e(p_e,\lambda_e)$,
$u_{\nu}(q_{1,2})$ correspond to the muon, electron, and neutrinos
four-spinors in momentum space respectively, with 
$\lambda_{\gamma,e}$ and $\lambda_{\gamma}$ the corresponding helicities, and
$\gamma_{L/R}^{\alpha}\equiv (1\mp \gamma_5)\, \gamma^{\alpha}$.
Due to the factorization property of the amplitude in Eq.(\ref{M2muon}),
one can easily calculate the sum over spins and 
the integral in phase space of neutrinos.
At this purpose it is convenient to introduce 
the following tensor $N_{\alpha\beta}$
\bea
N^{\alpha\beta}\equiv \int\, \frac{d^3\,q_1}{2\,E_1}
\frac{d^3\,q_2}{2\,E_2}\, \delta^4(p-p_e-k-q_1-q_2)\,
\sum_{\rm spins} N^{\alpha\dag}\, N^{\beta}\,.
\label{Ntensor1}
\eea 
By making use of Lorentz covariance, one easily gets \cite{fks}
\bea
N^{\alpha\beta}=\frac{4\pi}{3}\left((p-p_e-k)^{\alpha}
(p-p_e-k)^{\beta}-g^{\alpha\beta}(p-p_e-k)^2\right)\,.
\label{Ntensor2}
\eea
where $E_{1,2}$ are the neutrinos energies.
In order to describe the kinematic of muon radiative decay, 
we introduce the following independent variables
\bea
x=\frac{2p\cdot k}{m_{\mu}^2},~~ 
y=\frac{2p\cdot p_e}{m_{\mu}^2},~~ 
z=\frac{2k\cdot p_e}{m_{\mu}^2}\, ,
\eea
where $m_{\mu}$ is the muon mass.
In terms of these variables, 
the differential decay width, normalized to its tree-level non-radiative
decay $\Gamma_0$, is given by
\bea
\frac{1}{\Gamma_0}
\frac{d \Gamma^{(\lambda_{\gamma}\,,\,\lambda_e)}}{dx\,dy\,dz}=
-\frac{\alpha}{2\pi}\, 
\Big\{\frac{M_{\alpha}^{(\lambda_{\gamma},\lambda_e)\dag}
M_{\beta}^{(\lambda_{\gamma},\lambda_e)}
\,N^{\alpha\beta}}
{4\,m_{\mu}^2}\Big\}\, .
\eea
Now we provide the expressions for the differential decay width
in the rest frame of the muon, at fixed helicities of electron 
$(\lambda_e)$ and 
photon $(\lambda_{\gamma})$, where as in previous section
$L$ and $R$ symbols correspond to 
$\lambda_{e,\gamma}=-1$ and $\lambda_{e,\gamma}=1$ respectively. 
In particular, in
the muon rest frame one has
\bea
x=\frac{2E_{\gamma}}{m_{\mu}},~~~~
y=\frac{2E_e}{m_{\mu}},~~~~ 
z=\frac{x}{2}\left(y-{\rm A_e}\,\cos{\theta}\right)\, ,
\eea
where $E_e$, $E_{\gamma}$ are the energies of electron and photon, 
$\cos{\theta}$ the angle between their 3-momenta, and 
${\rm A_e}\equiv \sqrt{y^2-4r}$,
with $r\equiv m_e^2/m_{\mu}^2$. 
The allowed 
kinematical ranges for the above variables are
\bea
0\,\le\, x\, \le\, 2\, \left(\frac{1+r-y}{2-y+{\rm A_e}\cos{\theta}}\right)\, 
,~~~~~~~~~~~
2\,\sqrt{r}\,\le\, y\, \le\, 1+r\, ,
\label{kinmuon0}
\eea
while $-1\le \cos{\theta}\le 1$.
Notice that the upper limit  of $x$ depends on $\cos{\theta}$.
It is therefore not possible to perform 
the integration on $\cos{\theta}$  first.
After the $x$-integration the dependence on $\cos{\theta}$
it is also quite complicated.
In the present analysis we are mainly interested 
in analyzing the structure of the leading logarithmic 
terms absorbing the regularized infrared
and collinear singularities as well as the one of the 
finite terms for the right-handed electron rate in the $m_e\to 0$ limit.
For this purpose it is convenient to choose a particular region of the 
phase space where the
analytical calculations are further simplified provided that, on the same
time, all the leading logarithmic  terms are preserved. 
As shown in Ref.\cite{ss}, a suitable choice consists
in taking the upper limit of $x$ evaluated at $\cos{\theta}=1$, 
corresponding to its minimum value, as follows
\bea
0\,\le\, x\, \le\, 2\, \left(\frac{1+r-y}{2-y+{\rm A_e}}\right)\, 
,~~~~~~~~~~~
2\,\sqrt{r}\,\le\, y\, \le\, 1+r\, ,
\label{kinmuon1}
\eea
or equivalently
\bea
2\,\sqrt{r}\,\le\, y\, 
\le\, \frac{r+(1-x)^2}{1-x}\, ,~~~~~~~~~~~0\,\le\, x\, \le\, 1-\sqrt{r}\, .
\label{kinmuon2}
\eea
In this way, the integrals on $x$ and $\cos{\theta}$ can be exchanged, 
giving a consistent simplification of the analytical integrations.

For comparison, we will also provide 
the analytical expressions for the $y$-distributions
and the total rates obtained by integrating over the full phase-space, 
but in the approximation of neglecting terms of ${\cal O}(r)$. 
The corresponding analytical results at any order in $r$, 
will be presented elsewhere \cite{gt}.
Regarding the corresponding numerical
results, as shown in section 4, these are obtained by integrating the exact expression of the matrix density (given in Appendix B)
at any order in $r$ and over the full phase space.

In the muon rest frame,
the differential decay width is  given by
\bea
\frac{1}{\Gamma_0}
\frac{d \Gamma^{(\lambda_{\gamma}\,,\,\lambda_e)}}{dx\,dy\,d\cos{\theta}}=
\frac{\alpha}{8\pi}\,\frac{1}{x\,z^2}\,
\Big[{\rm A}_e\left(g_0+\lambda_{\gamma}\, \bar{g}_0\right)+\lambda_e\,
(g_1+\lambda_{\gamma}\, \bar{g}_1)\Big]\, ,
\label{BRmuon}
\eea
where the exact expressions at any order in $r$
of the functions $g_{0,1}$ and $\bar{g}_{0,1}$, which 
depend on $x,y,z$, are provided in Appendix B.

Notice that the $r$ independent terms in the functions $g_{0,1}$ 
and  $\bar{g}_{0,1}$ are proportional to $z$, partly compensating the $1/z^2$
in front of the right hand side of Eq.(\ref{BRmuon}). This is not true, however, for
the $r$ dependent terms which leave the distribution to 
be proportional to $1/z^2$.
These terms generate a singular behavior in the $r\to 0$ limit for the
distribution rate of the right-handed polarized electron, as it is in 
the analogous case of pion decay.
Indeed, if the $m_e\to 0$ limit is taken after 
integrating over $\cos{\theta}$,
due to the property that $\int d\cos{\theta}\, \frac{1}{z^2}\, \, 
\propto 1/r $, terms proportional to $r/z^2$ lead to a non-vanishing contribution in the integrated width. 
By taking into account the electron mass effects, 
the (polarized) integrated rate distributions 
($\Gamma^{(\lambda_{\gamma}\,,\,\lambda_e)}_{\rm res}$) on the restricted range
in Eq.(\ref{kinmuon1}) are given by
\bea
\frac{1}{\Gamma_0}
\frac{d \Gamma^{(\lambda_{\gamma}\,,\,\lambda_e)}_{\rm res}}{dx\,dy}=
\frac{\alpha}{24\pi}\,\frac{1}{{\rm A_e}\, x}\,
\Big[G_0+\lambda_{\gamma}\, \bar{G}_0+\lambda_e\,
(G_1+\lambda_{\gamma}\, \bar{G}_1)\Big]\, ,
\eea
where the expressions for the functions $G_{0,1}$ and $\bar{G}_{0,1}$
depending on $x,y$ and $r$ variables, are reported in Appendix B.

The same phenomenon appearing in the meson decay for the
right-handed electron \cite{tv}, it is also manifest 
here as a discontinuity in the electron mass. In particular, 
for right-handed electrons, the integrated rate in $\cos{\theta}$ 
does not vanish in the $m_e\to 0$ limit, as one should expect from the massless theory. 
This discontinuity, firstly noticed in Ref.\cite{ln}, can be associated to the axial anomaly \cite{s,fs,tv} according to the interpretation of the axial anomaly given by Dolgov and Zakharov \cite{dz}.
This anomalous behavior can be easily seen from the
$r\to 0$ limit of the functions $G_0+G_1$ and $\bar{G}_0+\bar{G}_1$
reported in Appendix B.
For this purpose, we will provide below the expressions for the
polarized differential decay width in the $m_e\to 0$ limit, in particular: 
\bea
\lim_{r\to 0}\,
\frac{1}{\Gamma_0}
\frac{d \Gamma^{(R,L)}_{\rm res}}{dx\,dy}&=&
\frac{\alpha}{3\pi}\,
\frac{y^2}{x}\Big\{
 -3\,\left( x -3\right) \,
        \left( {x^2} -2 \right)  + 
       \left( 12 + x\,\left( 9 + 
             x\,\left( 2\,x -5\right)  \right)  \right) \,y 
\nonumber\\
&+&\left( 2\,x + 2\,y -3\right) \,(3\,\log(r)-6\,\log (y)) 
\Big\}
\nonumber\\
\lim_{r\to 0}\,
\frac{1}{\Gamma_0}
\frac{d \Gamma^{(R,R)}_{\rm res}}{dx\,dy}&=&0
\nonumber\\
\lim_{r\to 0}\,
\frac{1}{\Gamma_0}
\frac{d \Gamma^{(L,L)}_{\rm res}}{dx\,dy}&=&
\frac{\alpha}{2\pi}\,
\frac{1}{x}\Big\{4\,{x^3}\,
        \Big( 1 + \log (r) + y \Big)  + 
       2\,x\,\left( y -1\right) \,y\,
        \Big( 12 + 6\,\log (r) + y \Big)  
\nonumber\\
&+& 
       2\,\left( 2 + \log (r) \right) \,{y^2}\,
        \left(2\,y -3\right)  + 
       {x^2}\,\Big(
          6\,\log(r)\,\left( 2\,y -1\right)  
\nonumber\\
&+& 
          y\,\left( 16 + 5\,y \right)  -6\Big)  
- 4\,{{\left( x + y \right) }^2}\,
        \left( 2\,x + 2\,y -3\right) \,\log (y) \Big\}
\nonumber\\
\lim_{r\to 0}\,
\frac{1}{\Gamma_0}
\frac{d \Gamma^{(L,R)}_{\rm res}}{dx\,dy}&=&
\frac{\alpha}{\pi}\,
\,x\,\left( 3 - 2\,x - 2\,y \right)\, ,
\label{mudistr}
\eea
where the $\log(r)$ terms are retained in order to regularize the
collinear divergences.

The lepton is intrinsically left-handed, due to the nature 
of the coupling and parity violation. However,
a final right-handed electron can also be produced 
with a sizeable rate in the limit $r\to 0$ \cite{ss,fs}.
As we can see from Eqs.(\ref{mudistr}), in the 
$r\to 0$ limit the photon is purely left-handed polarized when 
final electron is right-handed, as expected from angular momentum conservation.

QED obeys parity conservation and, therefore, 
the photon polarization is a "measure" of parity violation.
In the limit $x\to 0$ however the photon should behave as 
if its wavelength does not any longer resolve the process itself, leaving to a spin decoupling phenomenon as already observed in the radiative meson decay.
Therefore, in this case, the soft photon 
does not take part to the angular momentum conservation of the whole process.
Then, as in the meson case, 
in the $x \to 0 $ limit the 
partial amplitudes corresponding to the distributions $RL$ and $LL$, 
first and third above, tend to the same limit as well as the $RR$ and $LR$
ones. This property again shows the soft photon decoupling
from any spin-related process.

Now we provide the analytical results for the differential distribution in the
electron energy, in the approximation of neglecting terms
of ${\cal O}(r)$, obtained after integrating over $x$ 
the distributions in Eq.(\ref{mudistr}). 
Since the total integral in $x$ contains 
the well known infrared divergence when $x\to 0$, 
due to the emission of soft photons, 
we should provide integrated results by fixing a cut in the photon energy, 
namely $x_0$, corresponding to the
experimental energy resolution of photon detector.
Then, in the $r\to 0$  limit, the kinematical range of 
$x$ is $x_0 < x < 1-y$. After integrating over the $x$ range, and
by retaining only the leading terms in $x_0$ and $r$, the result is
\bea
\frac{1}{\Gamma_0}
\frac{d \Gamma^{(R,L)}_{\rm res}}{dy}&=&\frac{\alpha}{\pi}\,y^2
\, \Big\{\Big[\log (x_0)-\log (1 - y)\Big]
\,\left(3-2\,y \right)
       \left( 2 + \log(r) - 2\,\log (y) \right)
\nonumber\\
&+&\frac{1}{18}\,(1-y)\Big(
57+36\,\log(r)+28\, y+y^2+4\,y^3-72\,\log(y)\Big)
\Big\}
\nonumber\\
\nonumber\\
\frac{1}{\Gamma_0}
\frac{d \Gamma^{(R,R)}_{\rm res}}{dy}&=&0
\nonumber\\
\nonumber\\
\frac{1}{\Gamma_0}
\frac{d \Gamma^{(L,L)}_{\rm res}}{dy}&=&\frac{\alpha}{\pi}\,
\, \Big\{\Big[\log (x_0)-\log (1 - y)\Big]
{y^2}\,\left(3-2\,y \right)
       \left( 2 + \log(r) - 2\,\log (y) \right)
\nonumber\\
& -& \frac{1}{12}
    \left( y -1\right)^2\,
        \Big( 10 + 96\,y + 5\,{y^2} + 
          2\,\log(r)\,\left( 5 + 22\,y \right)  
\nonumber\\
&-& 
          4\,\left( 5 + 22\,y \right) \,\log (y) \Big)\Big\}
\nonumber\\
\nonumber\\
\frac{1}{\Gamma_0}
\frac{d \Gamma^{(L,R)}_{\rm res}}{dy}&=&
\frac{\alpha}{6\pi}\,
\,(1-y)^2\,(5-2y)\, .
\label{integrated}
\eea
Notice that the coefficient of the term proportional to $\log(x_0)$ in
Eqs.(\ref{integrated}), should cancel the same term appearing in the
one-loop corrections to the non-radiative Born decay, as shown in section 7.
Here we would like to stress that the coefficients of the 
terms proportional to  $\log(x_0)$, appearing only in the 
expressions of $(R,L)$ and $(L,L)$ 
in Eqs.(\ref{integrated}),
are the same for both Left- and Right-handed photon contributions.
This, again, shows the property that the photon spin must decouple
in the infrared limit.

In the zero lepton mass limit, the kinematical range 
of $y$ are now $0<y< 1-x_0$ and it is easy to check that the 
electron energy distribution vanishes at the end points.
As a cross check of our results we integrate over $y$ the 
non-vanishing expressions above and obtain
\bea
\frac{\Gamma^{(R,L)}_{\rm res}}{\Gamma_0}
&=&\frac{\alpha}{\pi}\,
\Big\{\, \log(r)\left(\frac{1}{2}\log(x_0)+\frac{23}{24}\right)+
\frac{17}{12}\,\log(x_0)-\frac{\pi^2}{6}+\frac{10399}{2520}\Big\}\, ,~~
\nonumber
\\\frac{\Gamma^{(L,L)}_{\rm res}}{{\Gamma_0}}
&=&\frac{\alpha}{\pi}\,
\Big\{\, \log(r)\left(\frac{1}{2}\log(x_0)+\frac{5}{24}\right)+
\frac{17}{12}\,\log(x_0)-\frac{\pi^2}{6}+\frac{17}{18}\Big\}\, ,~~
\nonumber
\\\frac{\Gamma^{(L,R)}_{\rm res}}{{\Gamma_0}}
&=&\frac{\alpha}{4\pi}\, .
\label{GammaPOLred}
\eea
Finally, the total width for the radiative muon decay,
integrated over the restricted phase space in Eq.(\ref{kinmuon1}), 
summed over all polarizations is
\bea
\frac{\Gamma_{\rm res}}{\Gamma_0}&=&\frac{\alpha}{\pi}\,
\Big\{\, \log(r)\left(\log(x_0)+\frac{7}{6}\right)+
\frac{17}{6}\,\log(x_0)-\frac{\pi^2}{3}+\frac{13409}{2520}\Big\}\, ,~~
\label{totalmu}
\eea
where terms of order ${\cal O}(r)$ and  ${\cal O}(x_0)$ were 
neglected
\footnote{
Here we would like to stress that this expression agrees with 
the corresponding one reported
in Ref.\cite{ss}, but differs from the
old results in Refs.\cite{ep,ks1}. We remind here that 
the total width calculated in
Refs.\cite{ep,ks1} is obtained by integrating
over the full phase space. 
Therefore, the coefficients of the logarithmic terms coincide 
with the corresponding ones in Refs.\cite{ep,ks1}, 
as expected since both infrared and collinear
singularities are included in the phase space region of
Eq.(\ref{kinmuon1}),
Therefore, the total width in Refs.\cite{ep,ks1}  will differ 
with respect to Eq.(\ref{totalmu}) 
by finite non logarithmic terms in the $x_0\to 0$ and $r\to 0$ limits.
}. 

For comparison, we report below the $y$-distributions
integrated over the full phase space.
As in Eqs.(\ref{integrated}), we use the approximation of 
neglecting terms of order ${\cal O}(r)$ and ${\cal O}(x_0)$. 
In particular, for the polarized differential rates in the
electron energy, we have
\bea
\frac{1}{\Gamma_0}
\frac{d \Gamma^{(\lambda_{\gamma},\lambda_{e})}}{dy}&=&\frac{1}{\Gamma_0}
\frac{d \Gamma^{(\lambda_{\gamma},\lambda_{e})}_{\rm res}}{dy}
+\frac{\alpha}{\pi}\, \Delta^{(\lambda_{\gamma},\lambda_{e})}\, ,
\label{exact}
\eea
where the additional terms $\Delta^{(\lambda_{\gamma},\lambda_{e})}$, arising
from the extra phase-space integration, are given by 
\bea
\Delta^{(R,L)}&=&
\frac{1}{18}\Big\{ -30\,y\, + 3\,y^2\,
         \left( 7 + 3\,{\rm L}_3(y) \right)  - y^3
        \left( 37 + 6\,{\rm L}_3(y) \right) + 
        27\,{y^4} - 3\,{y^5} + 4\,{y^6}\Big\}
\nonumber
\\
&+& \frac{1}{3}\,\log (1 - y)\,\left\{ -5 + 6\,y + 
        3\,{y^2}\,\left( -3 + 2\,y \right) \,\log (y)  -y^3\right\}\, ,
\nonumber \\
\Delta^{(R,R)}&=& 0\, ,
\nonumber \\
\Delta^{(L,L)}&=&
\frac{1}{12}\Big\{32\, y - \left( 95 - 6\, L_3 \right)
           \,y^2 + \left( 46 - 4\,L_3(y) \right) \,
         {y^3} + 5\,{y^4} \Big\}
\nonumber
\\
&+& 
     \frac{1}{3}\Big\{\log (1 - y)\,\left( 5 - 24\,y + 30\,{y^2} - 
        11\,{y^3} - 3\,{y^2}\,\left( 3 - 2\,y \right) \,
         \log (y) \right) \Big\}\, ,
\nonumber \\
\Delta^{(L,R)}&=& 0\, 
\label{F}
\eea
with ${\rm L}_3(y)\equiv \pi^2 -6\, {\rm Li}_2(1-y)$.
As shown in Eqs.(\ref{F}), only the LL and RL distributions
get an extra contribution which is non-vanishing in the $r\to 0$ limit, 
while for
the corresponding  RL and RR ones this is of order ${\cal O}(r)$.
This is due to the fact that in the radiative muon decay 
the right-handed electron is mainly produced at $\theta \simeq 0$. Hence, 
regarding the right-handed-electron production,
the maximum of the $x$ range integration can be well 
approximated by $x^{\rm max}(\cos{\theta}) \simeq
x^{\rm max}(\cos{\theta}=1)$. This
approximation, adopted in Ref.\cite{ss}, corresponds to consider the
restricted phase space in Eq.(\ref{kinmuon1}).

Finally, by integrating the distributions in Eqs.(\ref{F})
over the full range $0\le y \le 1$, we obtain
\bea
\frac{\Gamma^{(R,L)}}{\Gamma_0}
&=&\frac{\alpha}{\pi}\,
\Big\{\, \log(r)\left(\frac{1}{2}\log(x_0)+\frac{23}{24}\right)+
\frac{17}{12}\,\log(x_0)-\frac{\pi^2}{12}+\frac{997}{288}\Big\}\, ,~~
\nonumber
\\\frac{\Gamma^{(L,L)}}{{\Gamma_0}}
&=&\frac{\alpha}{\pi}\,
\Big\{\, \log(r)\left(\frac{1}{2}\log(x_0)+\frac{5}{24}\right)+
\frac{17}{12}\,\log(x_0)-\frac{\pi^2}{12}+\frac{133}{288}\Big\}\, ,~~
\nonumber
\\\frac{\Gamma^{(L,R)}}{{\Gamma_0}}
&=&\frac{\alpha}{4\pi}\, .
\label{GammaMU}
\eea
Then, the total rate summed over all polarizations is given by
\bea
\frac{\Gamma}{\Gamma_0}&=&
\frac{\alpha}{\pi}\,
\Big\{\, \log(r)\left(\log(x_0)+\frac{7}{6}\right)+
\frac{17}{6}\,\log(x_0)-\frac{\pi^2}{6}+\frac{601}{144}\Big\}\, .
\label{GammaMUTOT}
\eea
which is in agreement with the previous result obtained 
in \cite{bfs,ks1,ep}.
We would like to stress here that
the structure of the leading logarithmic terms 
is also preserved in the polarized rates when the restricted 
phase-space-integration is considered, as can be seen by 
comparing Eqs.(\ref{GammaPOLred}) and (\ref{GammaMU}).

%
\section{Distributions and polarization asymmetries}
In this section we present the numerical results for the
distributions of branching ratios in the
photon and electron energies.
In both cases we will sum over the 
fermion polarizations, leaving fixed only the photon polarizations.
For this purpose, it is very useful to introduce also an observable which 
provides a direct measurement of the amount of parity violation in the 
weak decays, namely the distribution of photon polarization asymmetry
${\rm A}_{\gamma}$, defined as follows
\bea
\frac{\rm d A_{\gamma}}{\rm d\xi }~\equiv~
\frac{ {\rm d_\xi ( BR_L)}-{\rm d_\xi (BR_R)}}{
{\rm d_\xi (BR_L)}+{\rm d_\xi (BR_R)}}
\label{asym}
\eea
where ${\rm d_\xi (BR_{L,R})}\equiv \frac{\rm d\, BR_{L,R}}{\rm d\xi}$ 
stands for the differential 
branching ratio (BR) in 
$\xi=\{x,y\}$, where $x=2E_{\gamma}/M$ and $y=2E_l/M$ in the rest frame
of the decaying particle of mass $M$.
Labels $L$ and $R$ indicate left- and right-handed photon 
polarizations respectively.
Here we would like to stress that $\frac{dA_{\gamma}}{d\xi}$
is a finite quantity, free from infrared divergences. 
Indeed, when the photon energy goes to zero, 
the distribution $\frac{dA_{\gamma}}{dx}$ tends to zero, since
\bea
\lim_{x\to 0}\, \left\{
{\rm \rho_L}(x,y)-
{\rm \rho_R}(x,y)\right\}&\to& {\cal O}(x)\,~~~{\rm and}~~~
\label{soft}\lim_{x\to 0}\, \left\{
{\rm \rho_L}(x,y)+
{\rm \rho_R}(x,y)\right\}\to \log(x)\, ,
\eea
where $\rho_{L,R}(x,y)$ indicates a generic Dalitz plot distribution for
the polarized decay with left- (L) or right-handed (R) photons.
Therefore, the total integral of Eq.(\ref{asym}) 
is a finite and universal quantity. It
does not depend on the photon energy resolution of the
detector, and provides a direct measure of parity 
violation.
Moreover, being $A_{\gamma}$ quite sensitive to the hadronic
structure of radiative meson decays, it is also
an useful tool for accurate measurements of $V$ and $A$ form factors.

In the following sections 4.1, 4.2, and 4.3, we will show our results
separately for the case of pion, kaon, and muon decays respectively.
Let us start with pion decay.
\subsection{Radiative $\pi^+$ decays}
Here we report the numerical results obtained for
the distributions of BRs and asymmetries for the case of 
radiative pion decays $\pi^+\to e^+ \nu_e \, \gamma$ and
$\pi^+\to \mu^+ \nu_{\mu} \, \gamma$.
As shown in section 2, the corresponding amplitudes 
contain only two free parameters which enter in the hadronic structure 
dependent terms (SD), that is $V$ and $A$ form factors.
However, being $V$ and $A$ non-perturbative hadronic quantities, 
they cannot be evaluated in QCD perturbation theory. An alternative 
approach like the one of  effective field theories  as, 
for instance, Chiral Perturbation Theory (ChPT), or the lattice QCD,
should be employed. 
On the other hand, $V$ and $A$ could be directly measured by experiments
\cite{pion_exp1,pion_exp2,istra,pibeta}.

These form factors are not constant over the allowed phase space.
Nevertheless, in radiative pion decays, the momentum dependence
in $V(W^2)$ and $A(W^2)$, parametrized by $W^2\equiv (1-x)\, m_{\pi}^2$,
is expected to be very small, not exceeding a few per cent of the 
allowed phase space.
This expectation is also supported by ChPT, 
since at the leading order in ChPT $V$ and $A$ are constant.
Recent calculations at next-to-leading order in ChPT \cite{bij_tal}, 
which included terms up to 
${\cal O}(p^6)$, where $p$ indicates a generic momentum involved 
in the decay, show a mild dependence on momenta, 
confirming the above expectations.
In our analysis, we will assume 
$V$ and $A$ to be constant in the full kinematical range\footnote{
Taking into account the effect of momentum dependence in the form factors goes
beyond the purpose of the present work, since
we are mainly interested in analyzing the dependence of photon polarization
asymmetries by the photon and electron energies. For more accurate
predictions of BRs and asymmetries, these effects should be included, 
especially in the kaon decay where they are expected to be sizeable.}.
Now we summarize the present status of form factors determination in pion 
decay. 

The vectorial form factor $V$ can be extracted in a model independent
way from $\pi^0\to \gamma \gamma$.  By using the conservation of 
vectorial current (CVC) hypothesis, one can relate the vectorial form factor
to the lifetime of the neutral pion \cite{vaks_ioffe}
\bea
|V|=\frac{1}{\alpha}\sqrt{\frac{2 \Gamma(\pi^0 \to \gamma \gamma)}
{\pi\, m_{\pi^0}}}=0.0259\pm 0.0005\, ,
\eea
where $ \Gamma(\pi^0 \to \gamma \gamma)$ is the total width of 
$\pi^0 \to \gamma \gamma$ decay and $V$ is assumed constant.
On the other hand, the axial $A$ form factor can be measured via the ratio
$\gamma=V/A$.
In previous experiments \cite{pion_exp1}, using the stopped pion technique,
the radiative pion decay has been measured in a 
limited phase space region where $V+A$ contributions dominate, leaving
to an ambiguity on the sign of $\gamma$.
In more recent experiments \cite{pion_exp2,istra}, 
the investigated larger portion of the phase space
allowed to determine the sign of $\gamma$ as well, which has also been
confirmed by the $\pi^+\to e^+\nu e^+e^-$ measurement \cite{egli}.

The most recent measurements of radiative pion decay, using the stopped
pion technique,
has been performed by the PIBETA collaboration with a good accuracy
\cite{pibeta}.
There, the CVC hypothesis has been used for the determination of $\gamma$.
The preliminary results of PIBETA experiment indicate 
a deficit of events in the observed $\pi\to e\nu\gamma$ decay \cite{pibeta}, 
suggesting for a new tensorial four-fermion interaction beyond the V-A theory. 
An analogous effect was first observed in a previous experiment at
ISTRA facility in early 90s \cite{istra}, 
in which pion decays where studied in flight. 
We will return on this point in section 6, where
the potential role of new tensorial couplings, suggested 
in order to accommodate experimental data, will be discussed.

The results contained in this section have been obtained by using 
for $\gamma$  the central 
value of the best CVC fit reported by the PIBETA experiment, namely
\bea
\gamma=0.443\pm 0.015, ~~~~~~~~{\rm with}~~~~~~~V\equiv 0.0259
\eea
which is also consistent with predictions in ChPT.

\begin{figure}[tpb]
\begin{center}
\dofourfigs{3.1in}{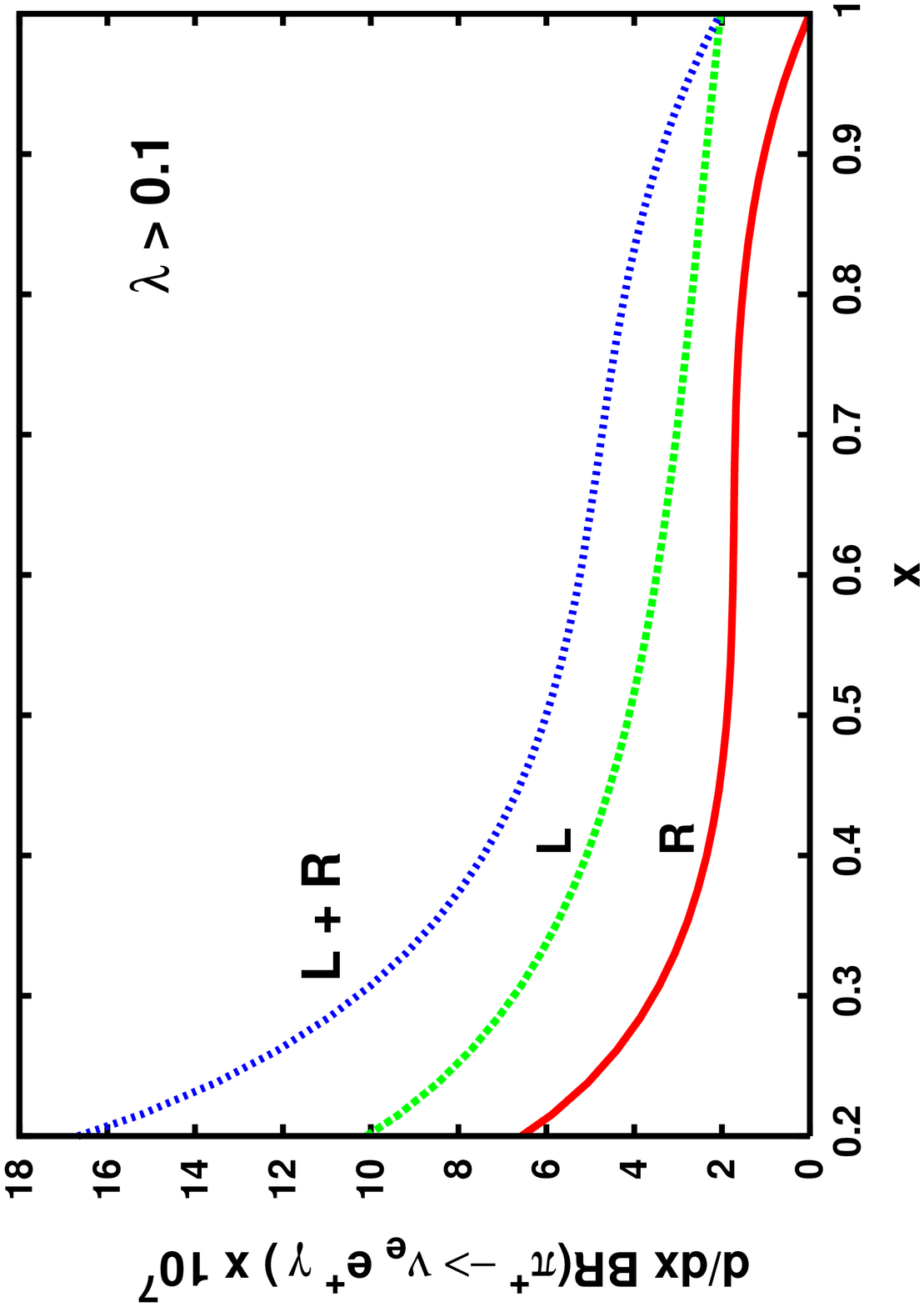}{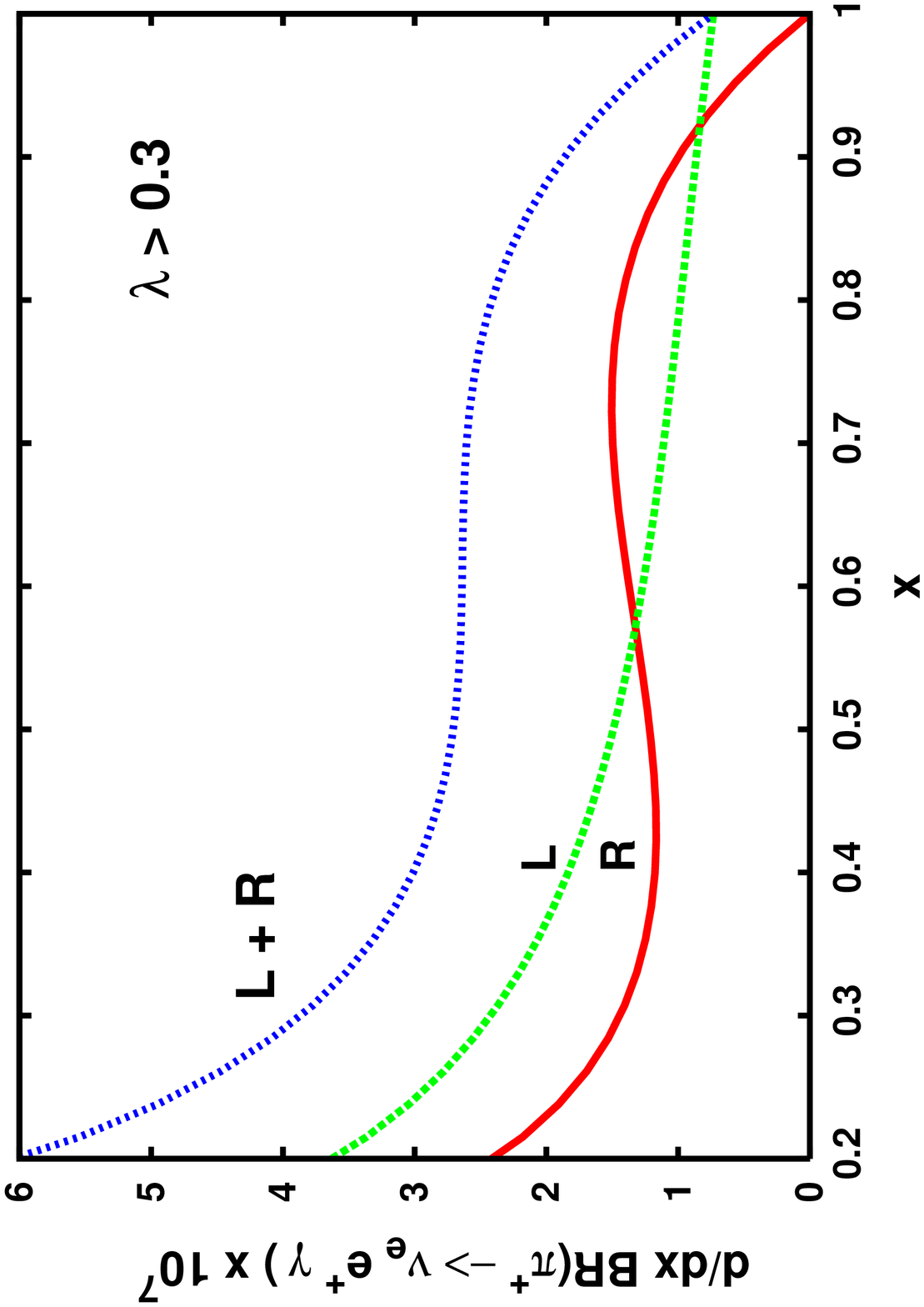}{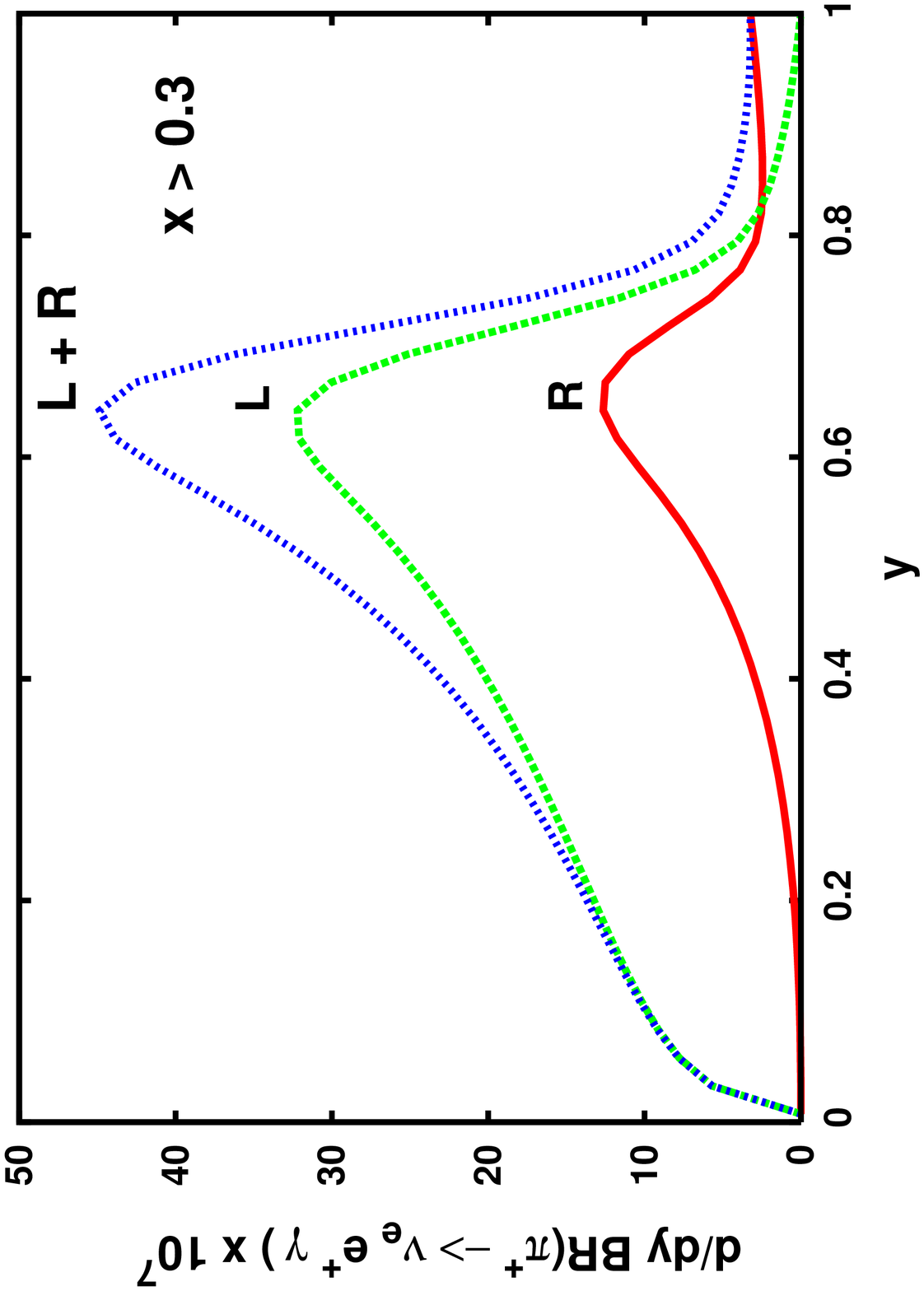}{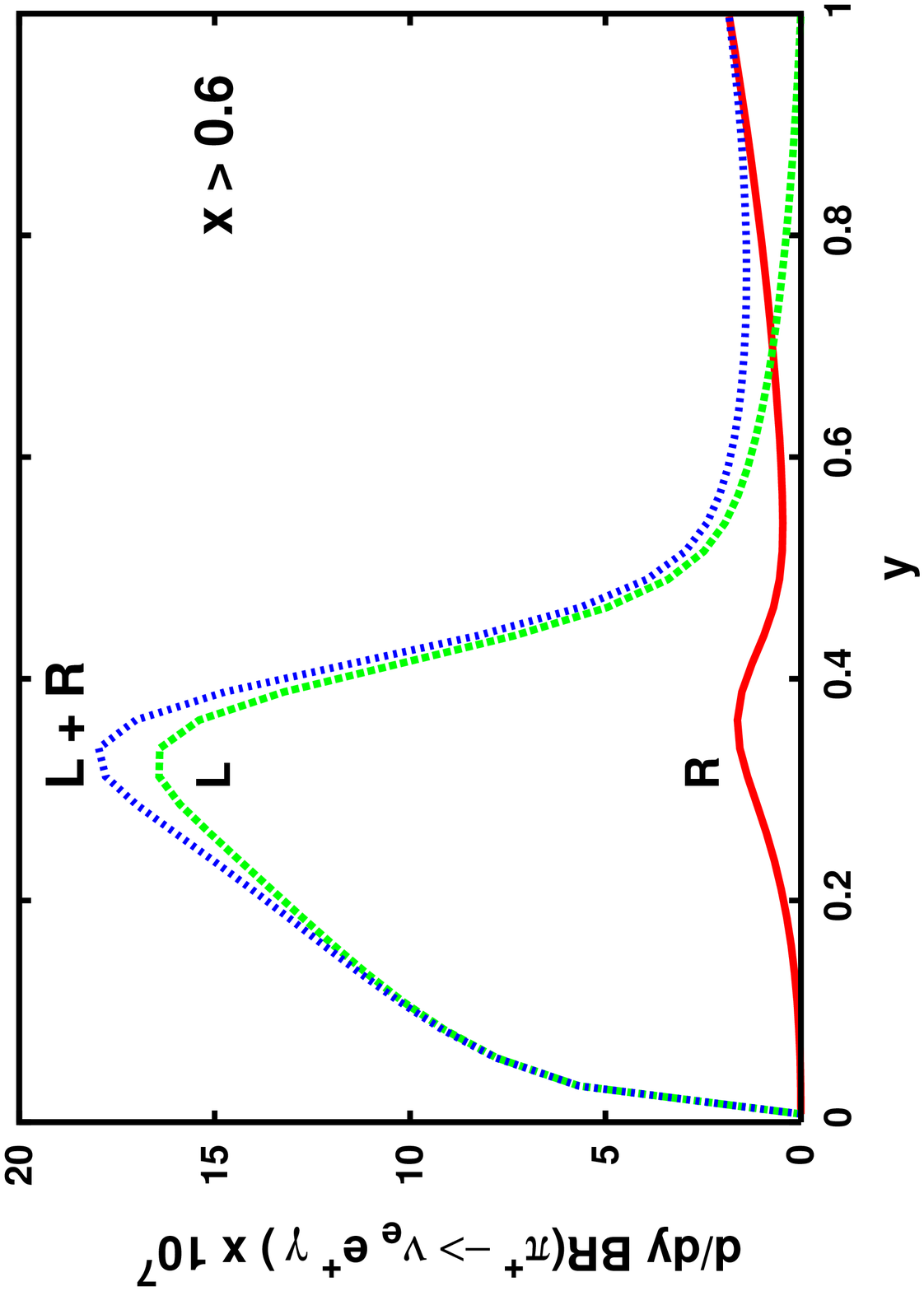}
\end{center}
\caption{\small The photon energy spectrum
$\frac{\rm d BR_{\gamma}}{\rm dx}$ versus $x$ 
(top plots) and electron energy spectrum 
$\frac{\rm d BR_{\gamma}}{\rm dy}$ versus $y$ (bottom plots), 
for pion decay $\pi^+ \to \nu_{e} e^+ \gamma$.
The labels $L$ and $R$ attached to 
the curves indicate pure left-handed and right-handed photon polarizations 
contributions respectively, while $L+R$ correspond to the sum.
Kinematical cuts $\lambda > 0.1$ (top-left), $\lambda > 0.3$ (top-right) 
and $x> 0.3$ (bottom-left), $x> 0.6$ (bottom-right) are applied respectively.}
\label{dBRx_Pe}
\end{figure}

In Fig.\ref{dBRx_Pe} we show the distributions of BR for pion
decay in electron channel. In particular, $d{\rm BR}/dx$ and $d{\rm BR}/dy$ 
are reported in the ``top'' and ``bottom'' plots respectively\footnote{
In the following, in all the plots involving the distributions of BRs, 
green and red curves correspond to left- (L) and
right-handed (R) photon polarizations respectively, as also indicated 
in each curve. Unpolarized decays are 
drawn as blue curves, with L+R label associated to them.}.
In the plots we have integrated the phase space
over $\lambda$ and $x$ respectively. Results are obtained
by imposing kinematial cuts on $\lambda$ or $x$, as indicated in the plots. 
Looking at the $d{\rm BR}/dx$ distributions in Fig.\ref{dBRx_Pe},
the general behavior emerging from these results is the following.
When cuts on 
$\lambda$ are relaxed, the left-handed photon polarizations dominate
in all range of values of $x>0.2$, corresponding to photon energies
$E_{\gamma}> 14$ MeV. 
On the other hand, the contribution of right-handed photons  can be 
increased by imposing larger cuts on $\lambda$, 
as can be seen by comparing left-top and right-top plots in Fig.\ref{dBRx_Pe}.
For example, by requiring that $\lambda >0.3$,
right-handed polarizations could dominate in the region of hard 
photons $0.6<x<0.9$.
These results can be explained
by using angular momentum conservation.
When cuts on $\lambda$ are relaxed, the main contribution to the integral
in $d\lambda$ comes from the region of low $\lambda$, where the IB effects
dominate with respect to SD terms. Low values of $\lambda$ should correspond 
to small angles between photons
and $e^+$, but could also correspond to low positron energies. 
In the former case,
neutrinos are likely to be produced with opposite direction
with respect to the photon momentum, 
in order to compensate for the missing momentum in the pion rest frame.
Since neutrinos are always left-handed polarized, photons must be 
left-handed as well, as required by angular momentum conservation. 
The spin configuration for  this case is shown in  Fig.\ref{spin}a.
On the other hand, small values of $\lambda$ could also correspond, 
in the latter case, to the spin configuration shown in  Fig.\ref{spin}c,
where positron and photon are backward.
There, the photon should be mainly emitted 
from the $\pi^+$ line, leaving  positron and photon both left-handed
polarized. As we will show in section 5, 
after integrating over $x$ with cuts $x>0.3$, 
the dominant effect will be given by this last configuration.

On the contrary, when cuts on $\lambda$ are very large, 
IB effects are reduced and SD terms become sizeable.
In this case, the positron is mainly produced right-handed, due to the fact
that SD terms are not chiral suppressed, 
neutrino momentum is favored to be directed forward with 
respect to the photon one,
leading to a right-handed photon as shown in Fig.\ref{spin}b.
However, as we can see from the top plots in Fig.\ref{dBRx_Pe}, 
there is a region of large $x$ where 
left-handed photon contributions are also sizeable, in particular 
for $0.9<x<1$.
This peculiar behavior in the end point region of photon energy 
can be explained as follows. 
When photon energy approaches its maximum,
positrons start to be produced almost at rest, if $\lambda$ is small.
Then, in order to conserve the total momentum, neutrino should be 
mainly emitted backward with respect to the photon direction,
see Figs.\ref{spin}a and \ref{spin}c,
leading to left-handed photon polarizations as required by angular
momentum conservation. However, we should stress that, depending on the
cuts on $\lambda$, the right-handed photon 
polarization could dominate even near the end-point region of 
photon energy.\footnote{Even if it is not shown in the plot,
at the end point $x=1-r_e$ the L curves (as well as R and L+R ones ) 
corresponding to the $dBR/dx$ distributions in Fig.\ref{dBRx_Pe} vanish.
However, the L curves go to zero more slowly than the corresponding R ones. }

In the bottom plots of  Fig.\ref{dBRx_Pe}, we report the BRs distributions on
positron energy versus $y$, for two representative 
choices of cuts, namely $x>0.3$ (left-plot) and $x>0.6$ (right-plot).
As we can see from these results, the left-handed photon polarizations 
dominate in the region $y<0.5$, while the gap between
left-handed and right-handed contributions increases
by using stronger cuts on the photon energies.
This behavior can be roughly understood as follows. At fixed 
positron energy, the larger the photon energy is the more the neutrinos 
are produced parallel and backward to the photon 
direction, in order to conserve total momentum. Therefore, as explained 
above, conservation of angular momentum favors
in this case the left-handed photon polarizations. However, at the end point 
of positron energy, when cuts on $x> 0.3$ and $x> 0.6$ are imposed,
the scenario could be reversed. As a consequence of the strong cuts on $x$, 
the IB contribution can be made very small, and near the region of $y=y_{max}$,
the SD terms 
should dominate favoring the production of a right-handed positron.
Clearly, when photon and positron are both very energetic
they tend to be emitted with opposite direction in order to 
conserve total momentum, 
approaching, in the case of a right-handed positron, to the spin 
configuration in Fig.\ref{spin}b.
Therefore, due to angular momentum
conservation, photons are mainly right-handed in the region 
$y> 0.8$ and $x>0.3$. Here we would like to stress that 
the relative gap between left- and right-handed contributions of hard photons,
near the region $y > 0.8$,
should be very sensitive to the values of hadronic form factors.

\begin{figure}[tpb]
\begin{center}
\dofourfigs{3.1in}{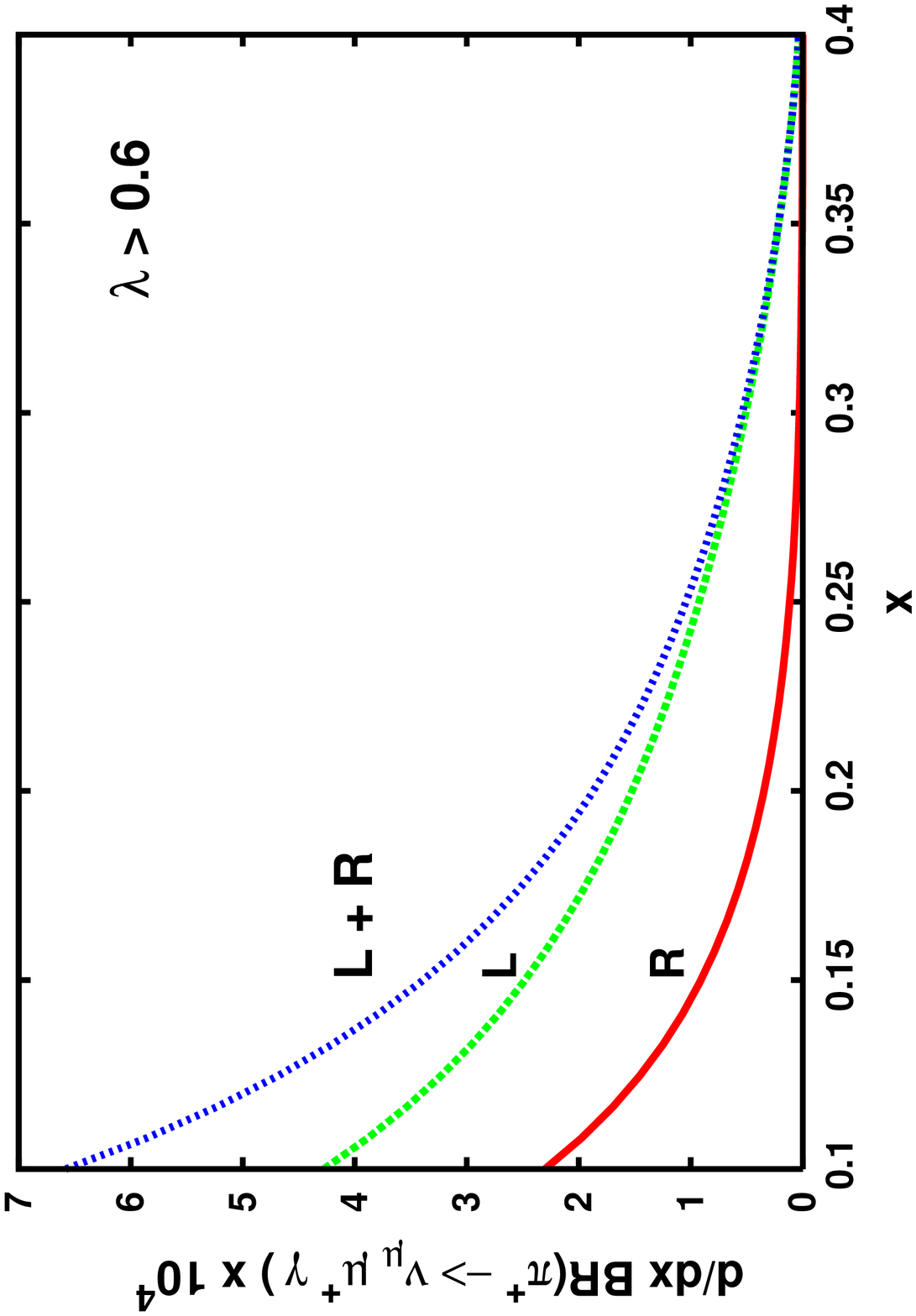}{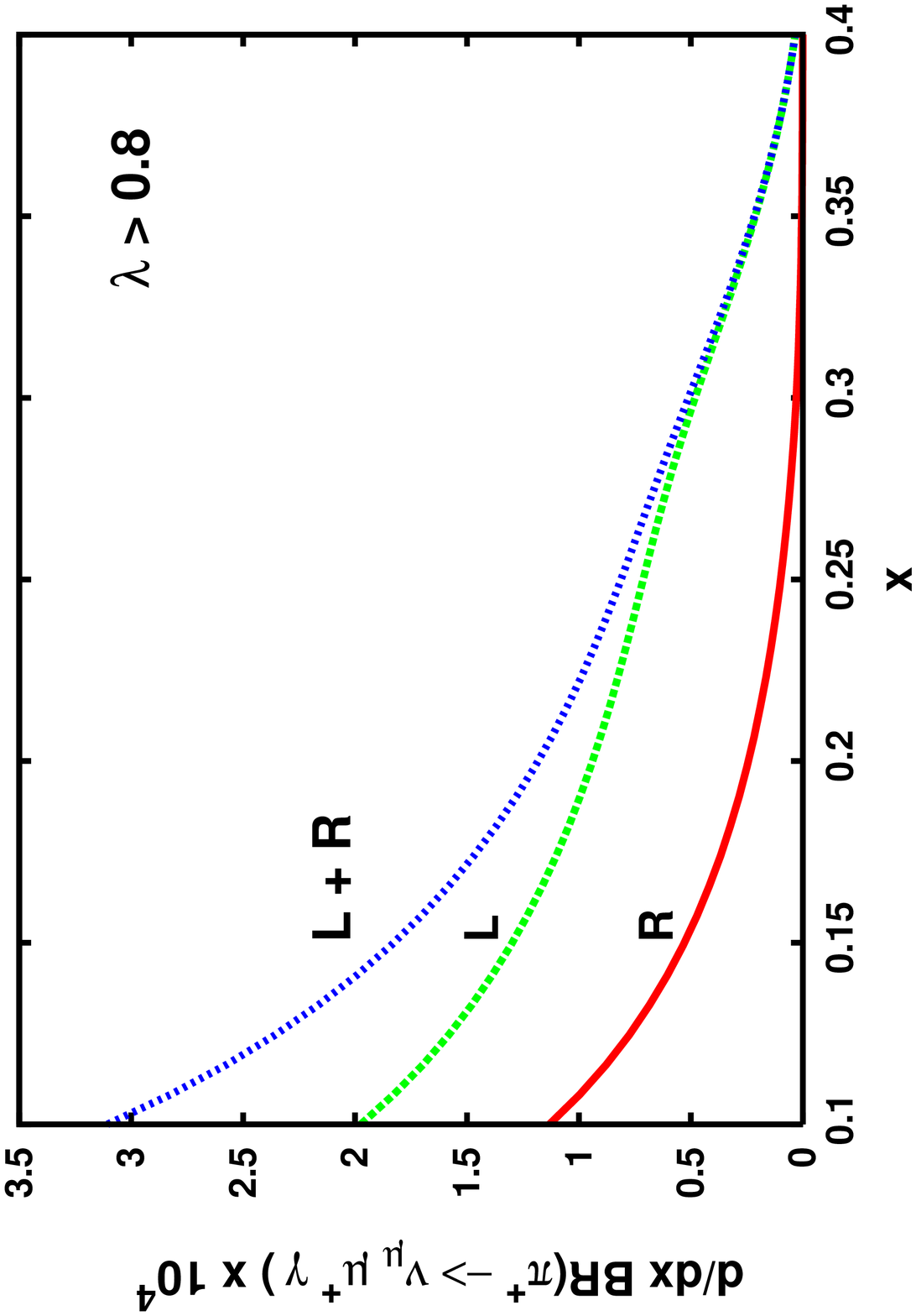}{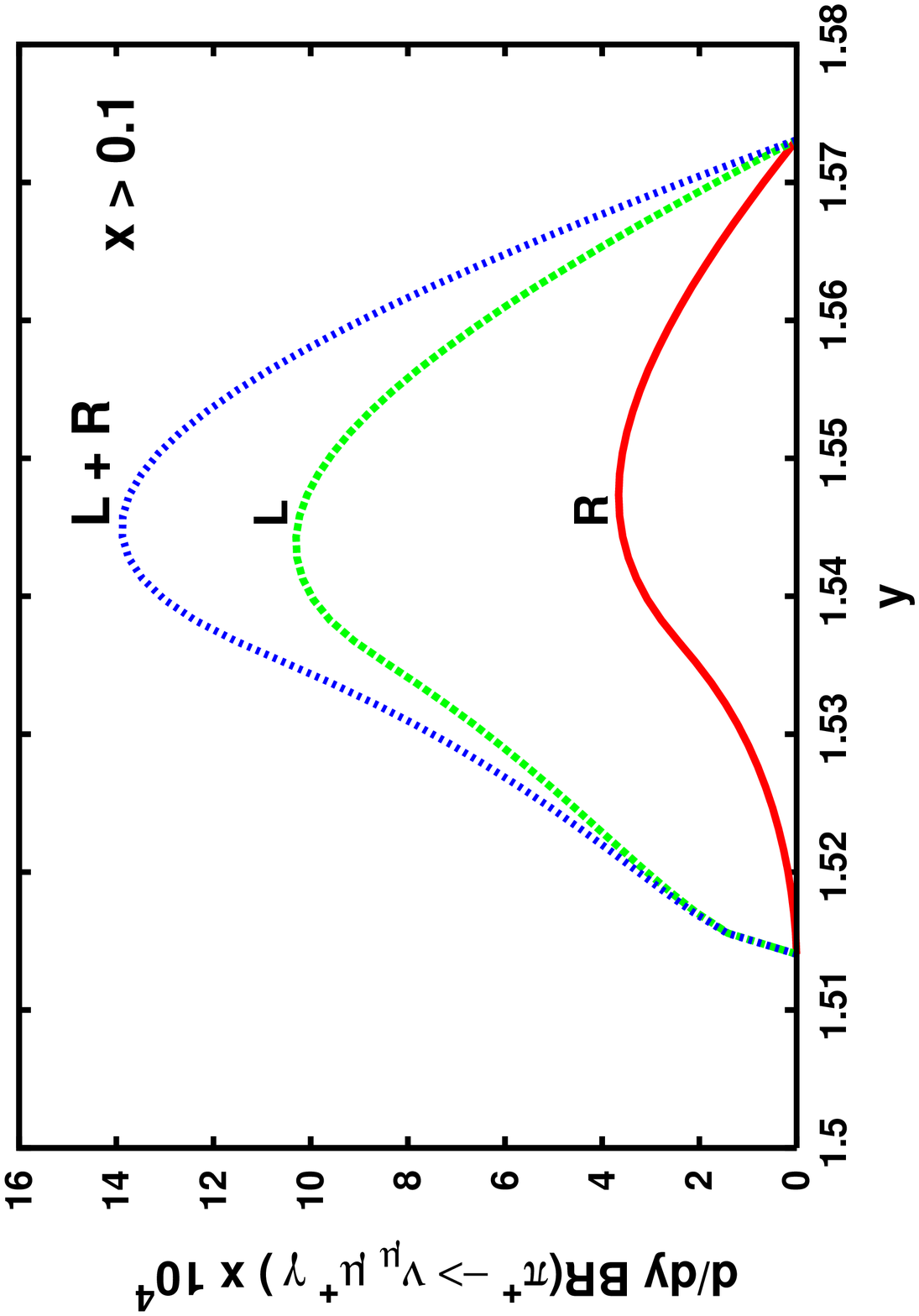}{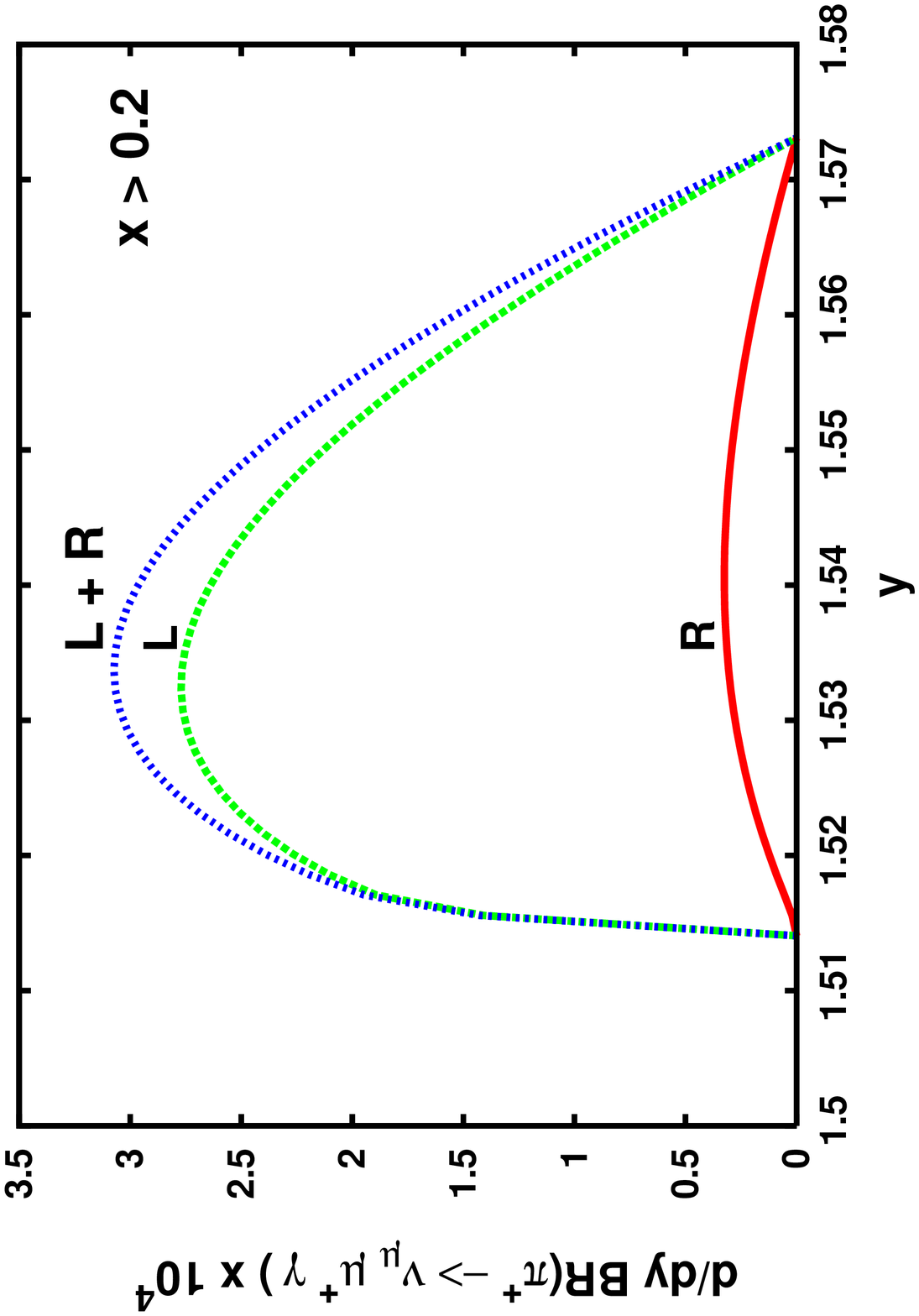}
\end{center}
\caption{\small As in Fig.~\ref{dBRx_Pe}, but for 
pion decay $\pi^+ \to \nu_{\mu} \mu^+ \gamma$, 
and with kinematical cuts $\lambda > 0.6$ (top-left), 
$\lambda > 0.8$ (top-right) and $x> 0.1$ (bottom-left), $x> 0.2$ 
(bottom-right).}
\label{dBRx_Pmu}
\end{figure}
In Fig.\ref{dBRx_Pmu} we show the corresponding results for the pion decay 
in the muon channel. In this case we can see that the left-handed 
photon polarizations always dominate over the entire phase space, 
while right-handed ones are quite suppressed.
Notice that, being the muon mass very close to the pion one, 
the IB contribution is not chiral suppressed as in the electron channel 
and it is larger than the SD one almost over all the allowed phase space.
This implies that $\mu^+$ 
is mainly produced with left-handed polarization.
Moreover, due to the fact that here the minimum 
allowed value of $\lambda$ (for $x>2$) is $\lambda_{\min}\simeq 0.7$, 
the $\mu^+$ and photons are mainly produced at large angles.
Then, if left-handed $\mu^+$ tends to be produced backward 
with respect to photon momentum, 
this last one must be also left-handed in order to 
conserve total angular momentum, as shown in Fig.\ref{spin}c.

\begin{figure}[tpb]
\begin{center}
\dofourfigs{3.1in}{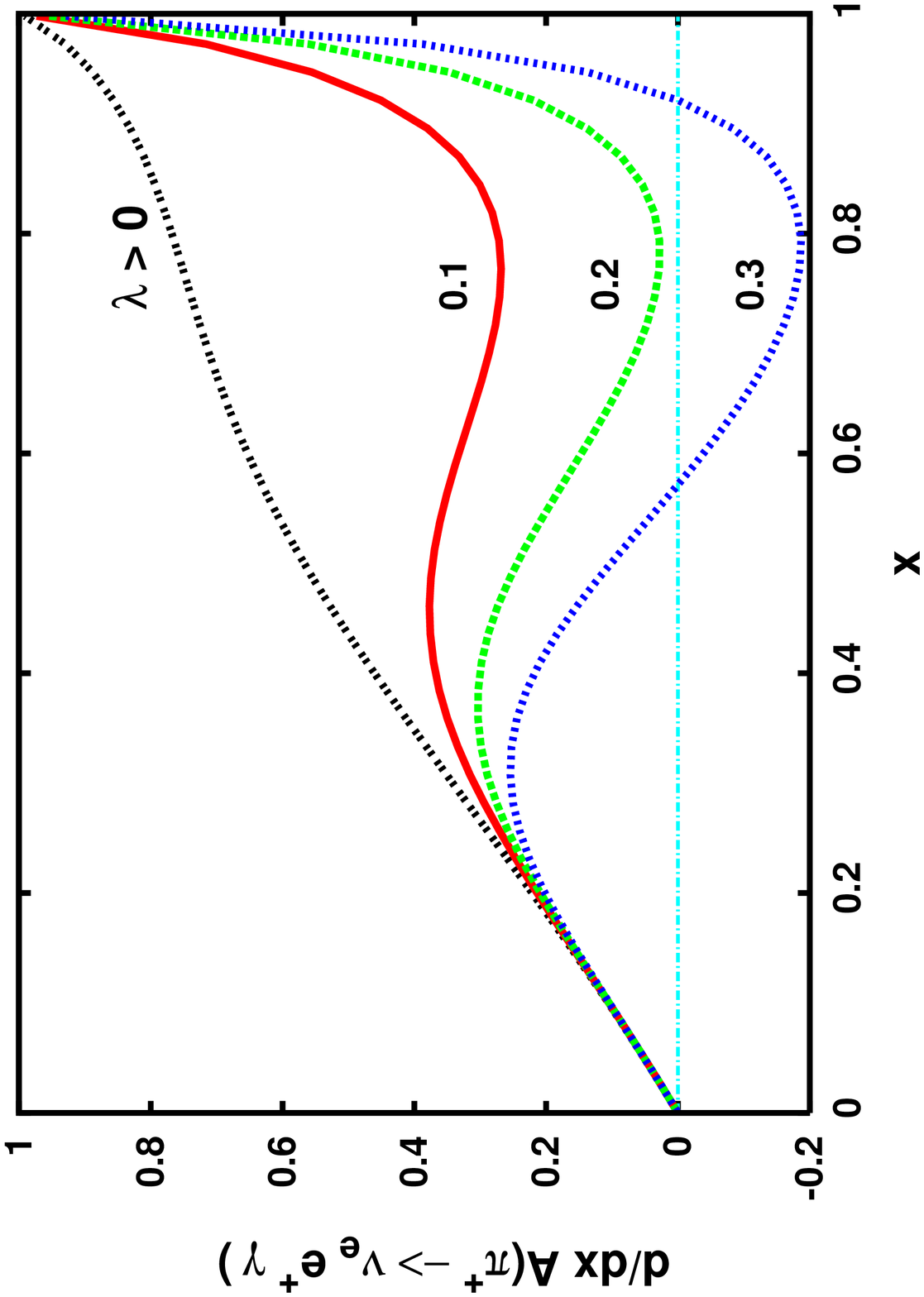}{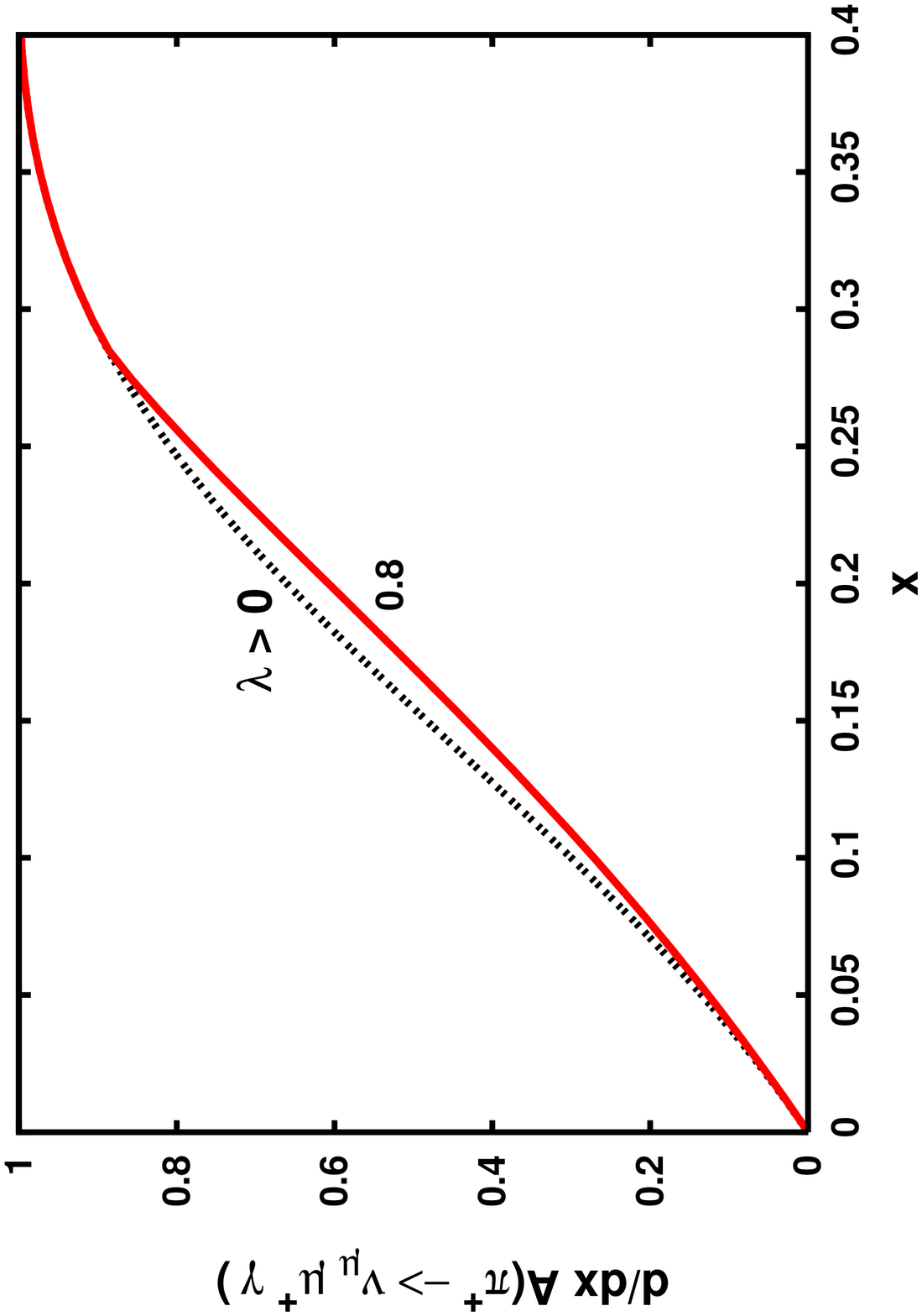}{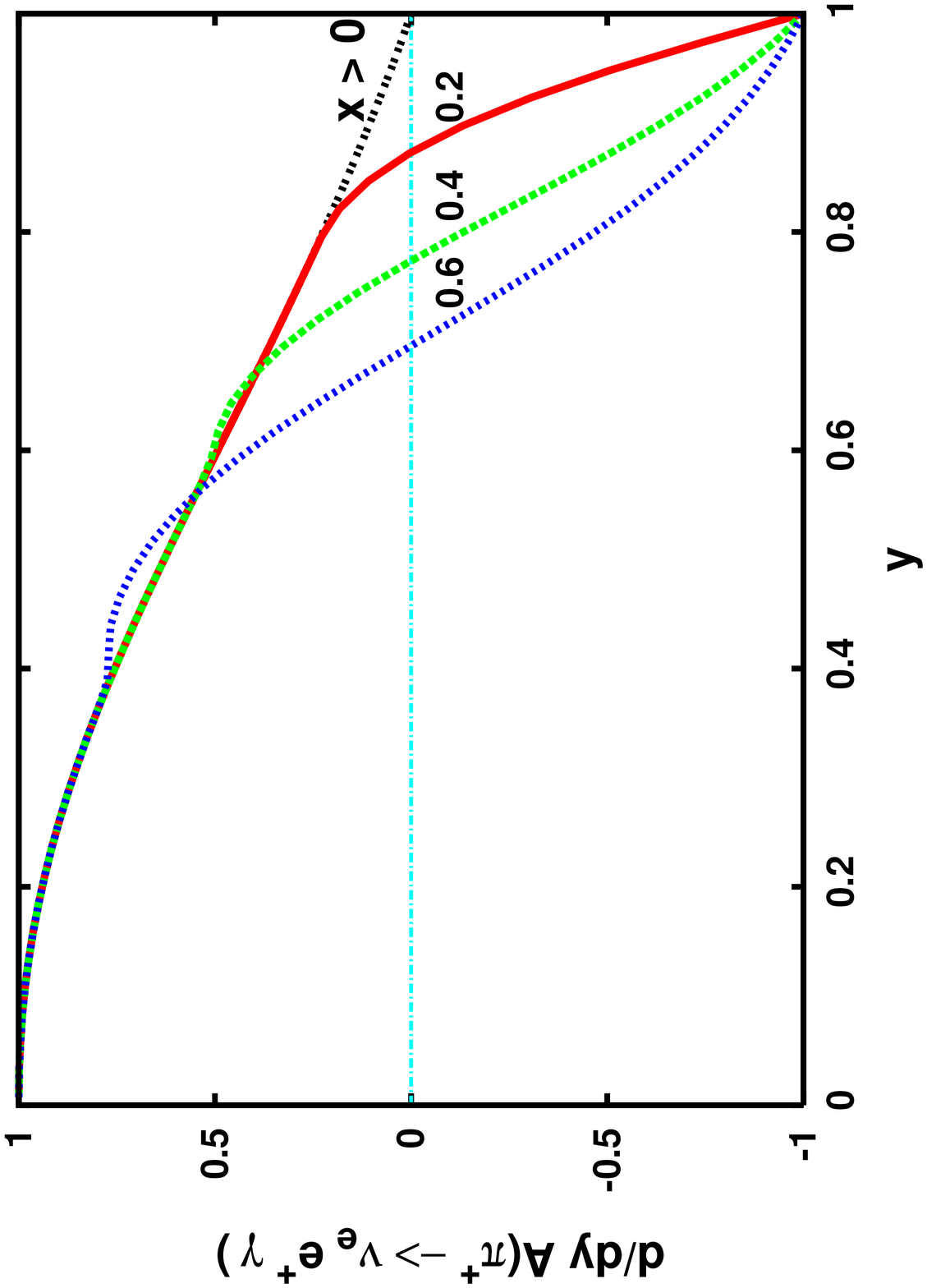}{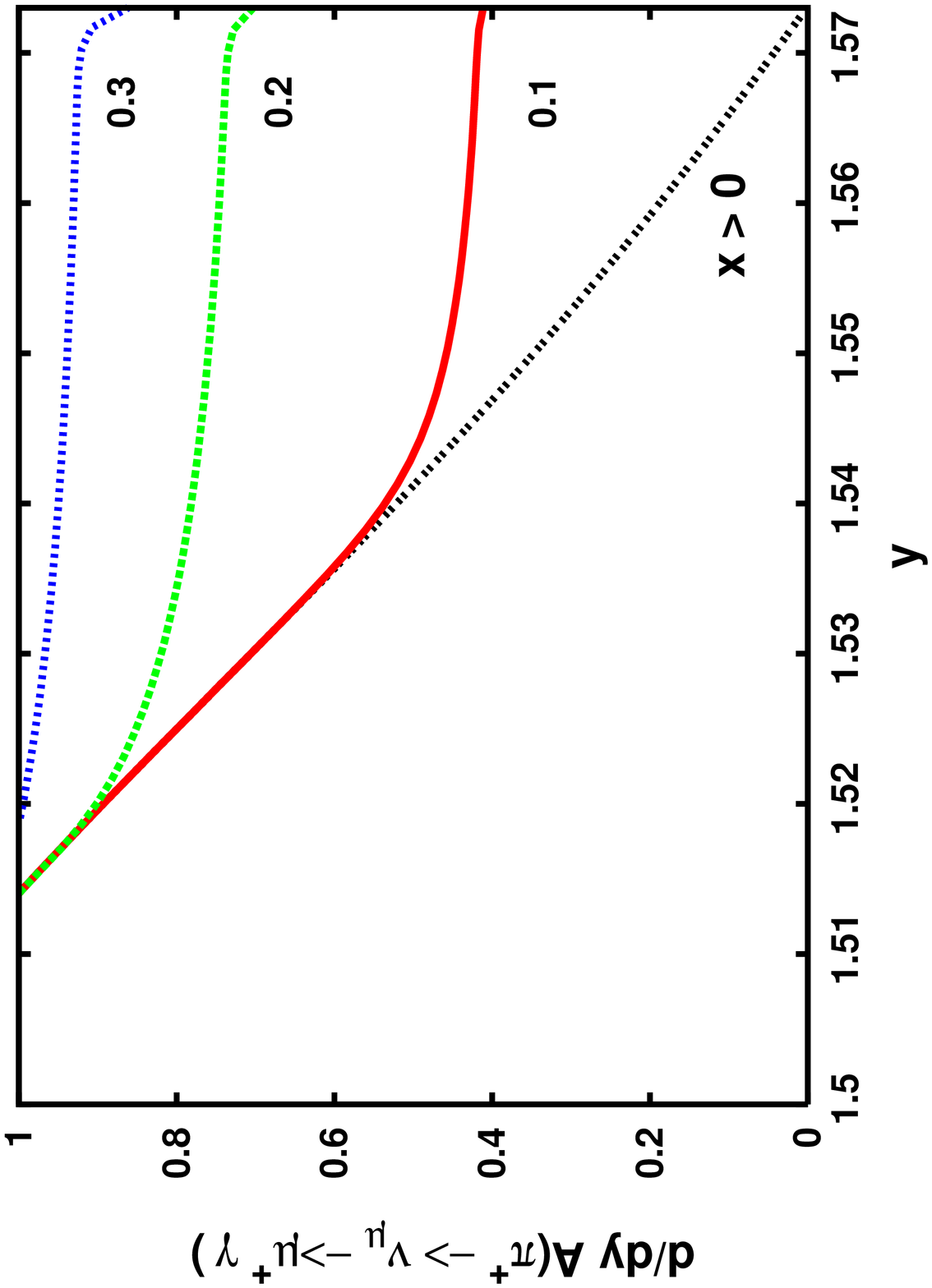}
\end{center}
\caption{\small 
The differential asymmetry $\frac{\rm d A_{\gamma}}{\rm dx}$ 
versus $x$ (top) and $\frac{\rm d A_{\gamma}}{\rm dy}$ versus $y$ (bottom), 
with kinematical cuts $\lambda >0,\,0.1,\,0.2,\,0.3$ (top-left), 
$\lambda > 0,\,0.8$ (top-right) and $y>0,\,0.2,\,0.4,\,0.6$ (bottom-left), 
$y> 0,\,0.1,\,0.2,\,0.3$ (bottom-right) respectively.}
\label{asymPion}
\end{figure}

Finally, in Fig.4, we present our results for the 
$x$- and $y$-distributions 
of the photon polarization asymmetry, as defined in Eq.(\ref{asym}).
In particular, in the top and bottom plots we report the results for the 
$dA_{\gamma}/dx$ and $dA_{\gamma}/dy$ respectively for several kinematical 
cuts, while the left and right plots correspond 
to the electron and muon channel decays respectively. 
A general property of
these results is that the $x$- and  $y$-distributions of asymmetry
vanish at $x=0$ and $y=1+r_l$ respectively.
This is a consequence of the fact that when the photon energy is 
approaching to zero, the polarized photon
densities of Dalitz plot tend to the same limit, due to the spin-decoupling 
property of soft photons, as discussed in section 2.
A remarkable aspect of these results is that, in the electron decay channel, 
the $d A_{\gamma}/dx$ distribution
becomes negative for some particular choices of cuts.
Analogously, the same effect can be achieved on the $y$-distribution
by increasing cuts on the photon energy. On the contrary, in the muon channel, 
the corresponding asymmetry  is always positive, 
as can be seen in the plots to the right in Fig.\ref{asymPion}.
In conclusion, we would like to emphasize that 
the position of the zeros of photon polarization 
asymmetry is particularly sensitive 
to the effects of the SD terms. This property could suggest 
a new experimental way for obtaining more precise measurements 
of form factors.

\subsection{Radiative $K^+$ decays}
In analogy with the radiative pion decays, we analyze here the corresponding 
ones in the kaon  sector, in particular
$K^+\to e^+ \nu_e \, \gamma$ and $K^+\to \mu^+ \nu_{\mu} \, \gamma$.
The expressions of amplitudes in terms of decay constants, masses
and form factors remain formally the same as in the pion decay.
However, the kaon electromagnetic form factors, as 
well as the decay constant $f_K$, and the ratios $r_e$, $r_{\mu}$ 
between leptons and kaon mass, are quite different from the pion case.
As we will see in the following,
these differences will sizeably affect the shape of distributions 
and asymmetries with respect to the corresponding pion decay. 

The most recent measurements of 
$V$ and $A$ form factors have been performed 
by the E787 collaboration \cite{kaon_exp} 
through radiative $K^+$ decay $K^+\to \mu^+\nu_{\mu}\, \gamma$.
In particular, the absolute value of $V+A$ has been determined finding
$|V+A|=0.165\pm 0.007\pm 0.011$, while a limit on $ -0.24<V-A<0.04$
has been set at 90\% C.L. These results have been obtained
by assuming constant form factors. The $|V+A|$ measurements are
consistent with previous results on $K\to e^+ \nu_e \gamma$, but they disagree
by almost 2 standard deviations 
with respect to predictions from leading order in ChPT \cite{Ksemilept}.
We recall here that the evaluation of form factors starts at one loop
in ChPT expansion, that is at ${\cal O}(p^4)$. 
At this order the chiral prediction, as for the pion case, 
gives constant form factors. 
The momentum dependence of the form 
factors starts then at the next-to-leading order, that is 
${\cal O}(p^6)$, and it is expected to be larger than in pion case,
due to sizeable effects of resonances exchange \cite{amet}.
In particular, by considering only a particular class of diagrams where
vector and axial-vector resonances are exchanged,
the form factors can be parametrized as
\bea
V(W^2)=\frac{V}{1-W^2/m_{K^{\star}}^2},~~~~~~
A(W^2)=\frac{A}{1-W^2/m_{K_1}^2},~~~~~~
\eea
where $W^2=m_K^2(1-x)$, and the masses $m_{K^{\star}}^2$ and 
$m_{K_1}^2$ correspond to vector and axial-vector resonances.
Then, in order 
to minimize the effects of resonance exchange, large $x$-regions should
be considered since $W^2\to 0$ when $x\to 1$, 
while low $x$-regions may be used to explore the 
$W^2$ dependence of $V$ and $A$.
The ${\cal O}(p^6)$ contributions, based on $SU(3)\times SU(3)$ symmetry
in ChPT, has been recently calculated in Ref.\cite{geng}.
Significant deviations of order of 10-20 \% 
have been found on $V$ and $A$ with respect to the leading order calculation. 
Moreover, while the vectorial form factor
is quite sensitive to photon energies, the axial one shows only
a modest effect \cite{geng}.

As for the pion decay, in order to simplify our analysis, we
will not take into account the momentum dependence in $V$ and $A$.
Then, consistently, we will take the $V,A$ predictions 
at the leading order in ChPT, re-absorbing the 
missing NLO contributions in the theoretical uncertainty.
In particular, our results are obtained by using the following values
\cite{Ksemilept}
\bea
V+A=-0.137,~~~~~~~~~V-A=-0.052\, .
\eea

\begin{figure}[tpb]
\begin{center}
\dofourfigs{3.1in}{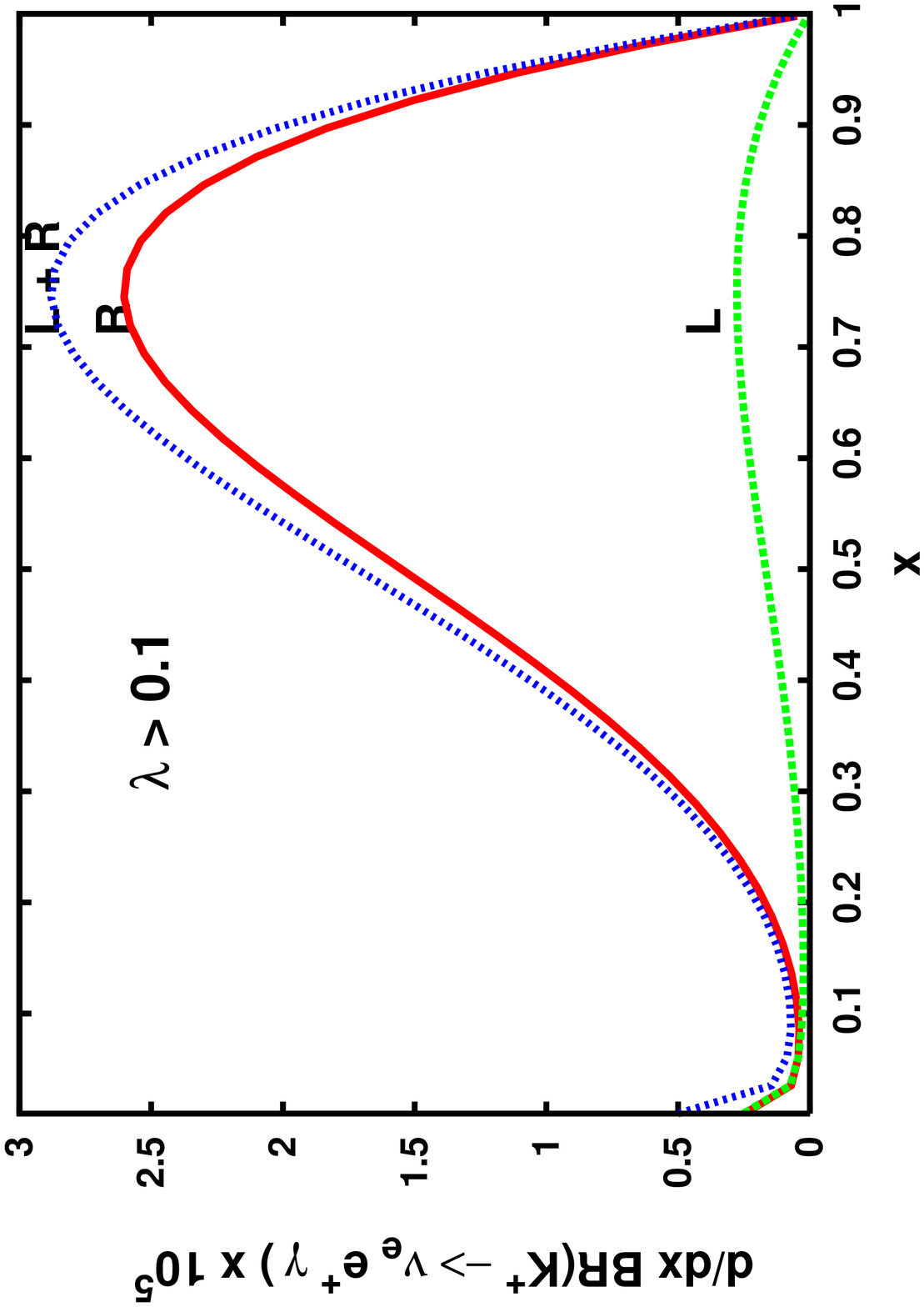}{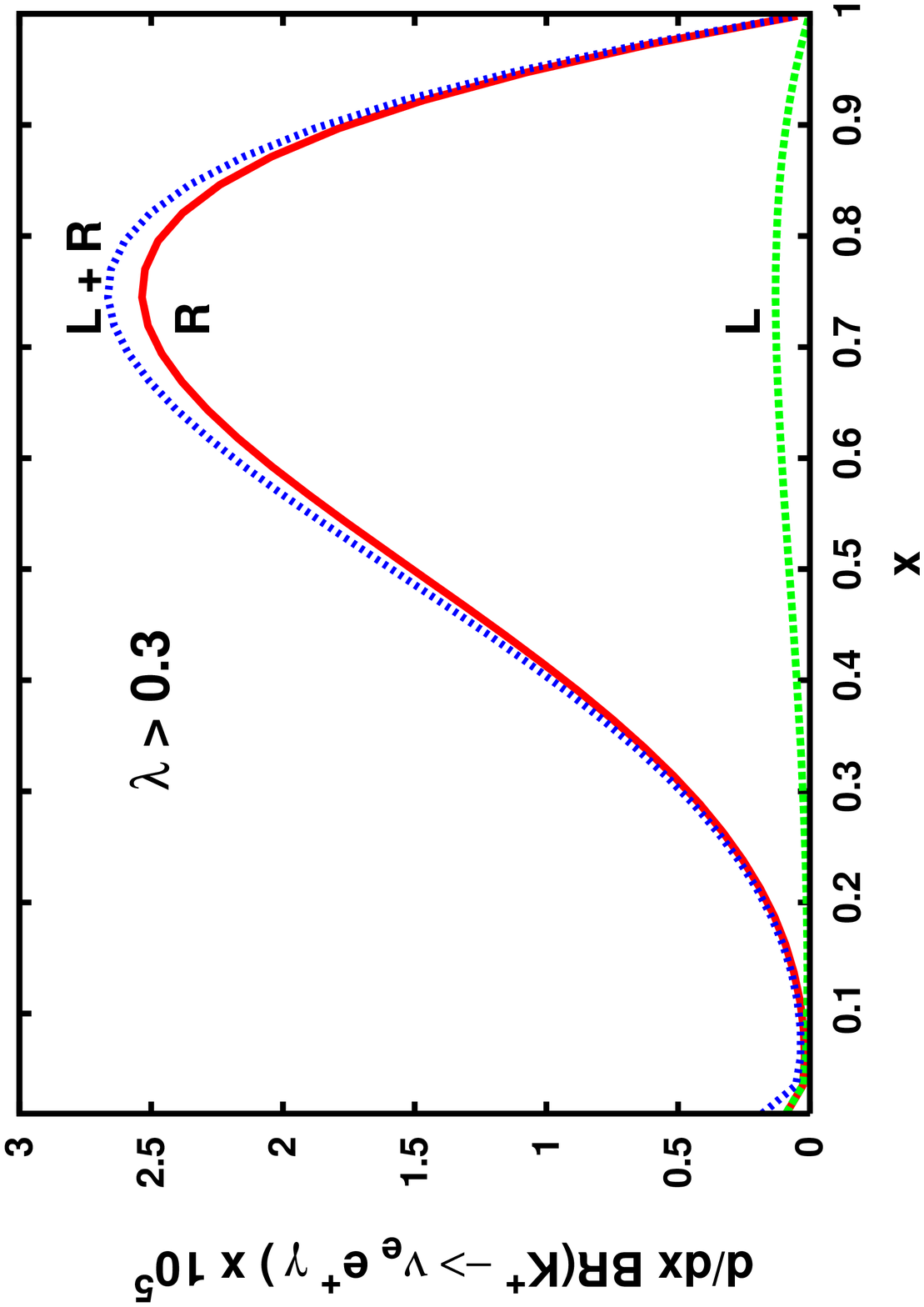}{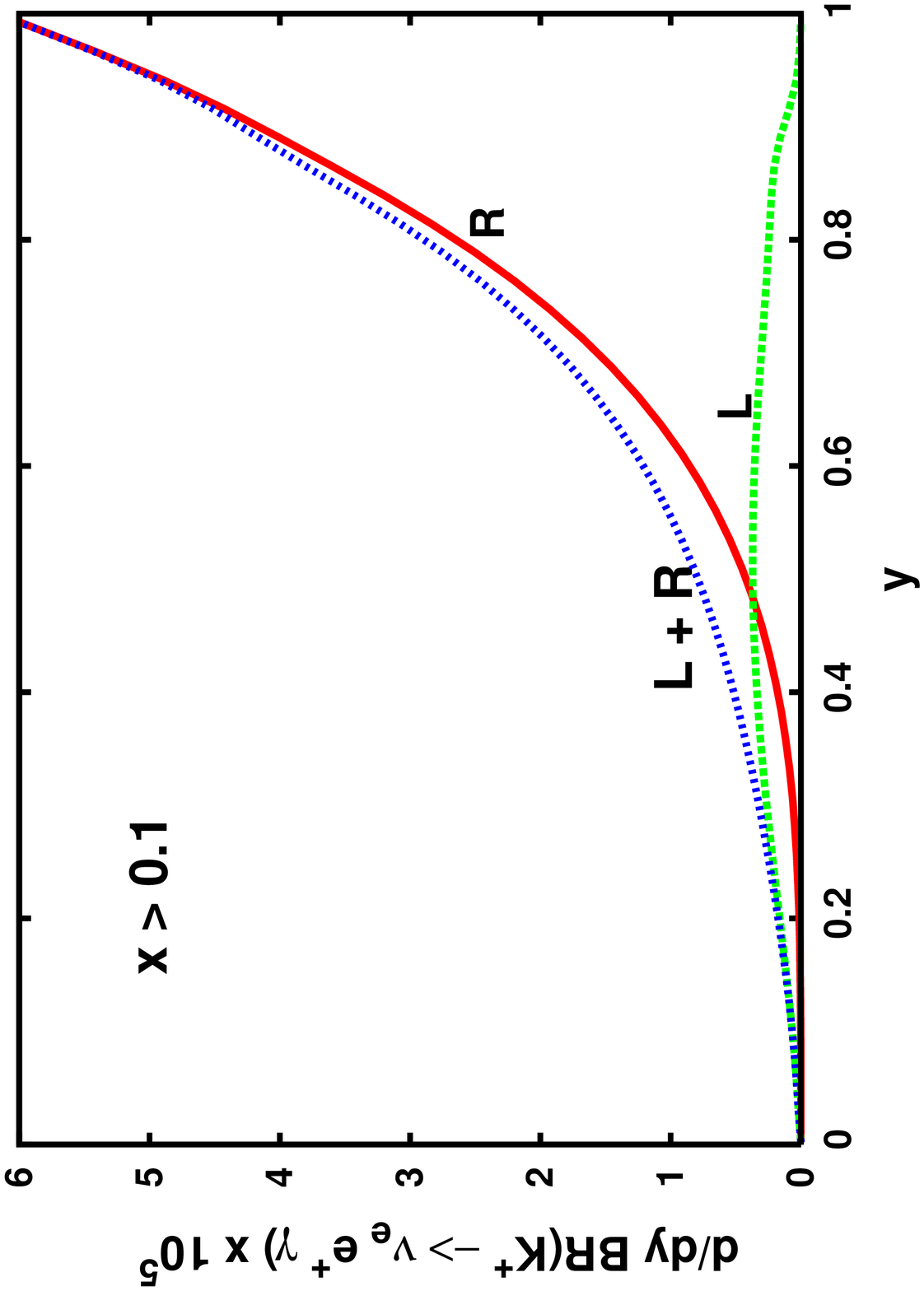}{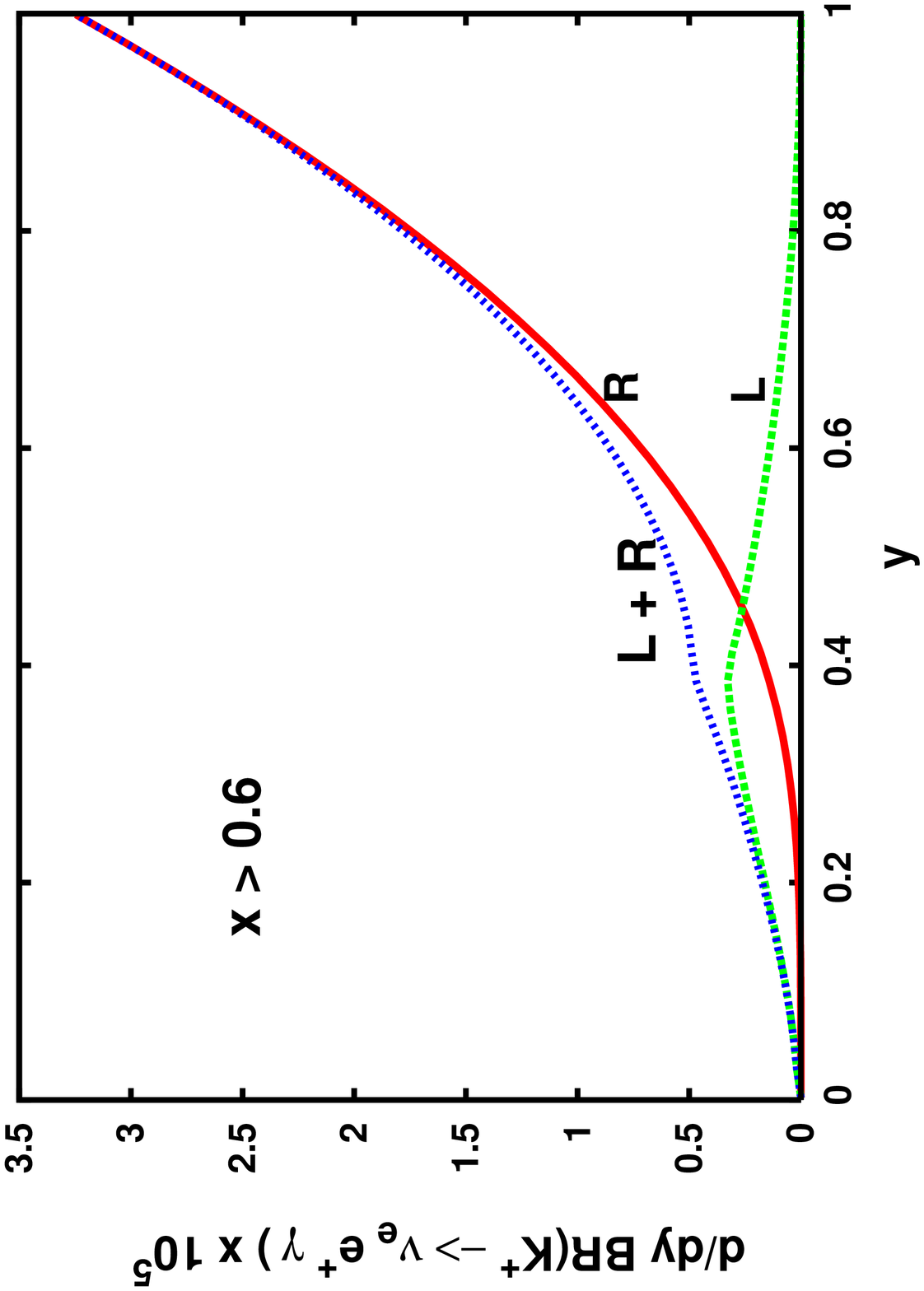}
\end{center}
\caption{\small As in Fig.~\ref{dBRx_Pe}, but for 
kaon decay in $K^+ \to \nu_{e} e^+ \gamma$. Curves correspond to 
kinematical cuts as reported in the figures.}
\label{dBRx_Ke}
\end{figure}

In Fig.\ref{dBRx_Ke} we show the $x-$ (top) and $y-$ (bottom)  distributions  
for the $K^+\to e^+ \nu_e\, \gamma$ decay. 
The general trend emerging from
these results is that in kaon decay, contrary to the pion case,
the right-handed photon production dominates over the lef-handed one, 
already for moderate cuts $\lambda >0.1$, as can be checked by comparing
results between Figs.\ref{dBRx_Pe}
and \ref{dBRx_Ke} with the same cuts on $x$ and $\lambda$.
This result can be roughly understood as follows. 
The IB contribution in the radiative $K^+$ decay is 
more ``chiral'' suppressed with respect to the corresponding $\pi^+$
due to the fact that $m_K\simeq 4m_{\pi}$.
Then, for the same values of $x$ and $\lambda$, 
the IB effects in pion decay will be larger than in the corresponding kaon one.
For instance, while the IB contributions in pion decay are 
still sizeable after cuts $\lambda>0.1$ and $x>2$ have been imposed, 
in $K^+$ decay these same cuts dramatically reduce 
the IB effects in favor of SD contributions. As already explained in 
section 4.1, when the photon is produced from the SD terms it is
mainly right-handed polarized.
In conclusion, the $d {\rm BR}/dx$ distributions in the 
top-plots of Fig.\ref{dBRx_Ke}, for $x>0.1$,
shows the same behavior of
the corresponding one in pion decay in the region 
$0.6<x<0.9$ and $\lambda >0.3$, where the right-handed photon contributions
are enhanced.
Moreover, as we can see from the top-plots in Fig.\ref{dBRx_Ke},
these curves have a maximum (for $0.1<x<1$ and $0.1<\lambda<0.3$)
around $x\simeq 0.75$.
In the bottom-plots of  Fig.\ref{dBRx_Ke} we report the analogous results 
for the positron energy distributions. As we can see, 
when the $y>0.5$ and $x>0.1$ cuts are imposed,
the right-handed photon gives the dominant effect.
\begin{figure}[tpb]
\begin{center}
\dofourfigs{3.1in}{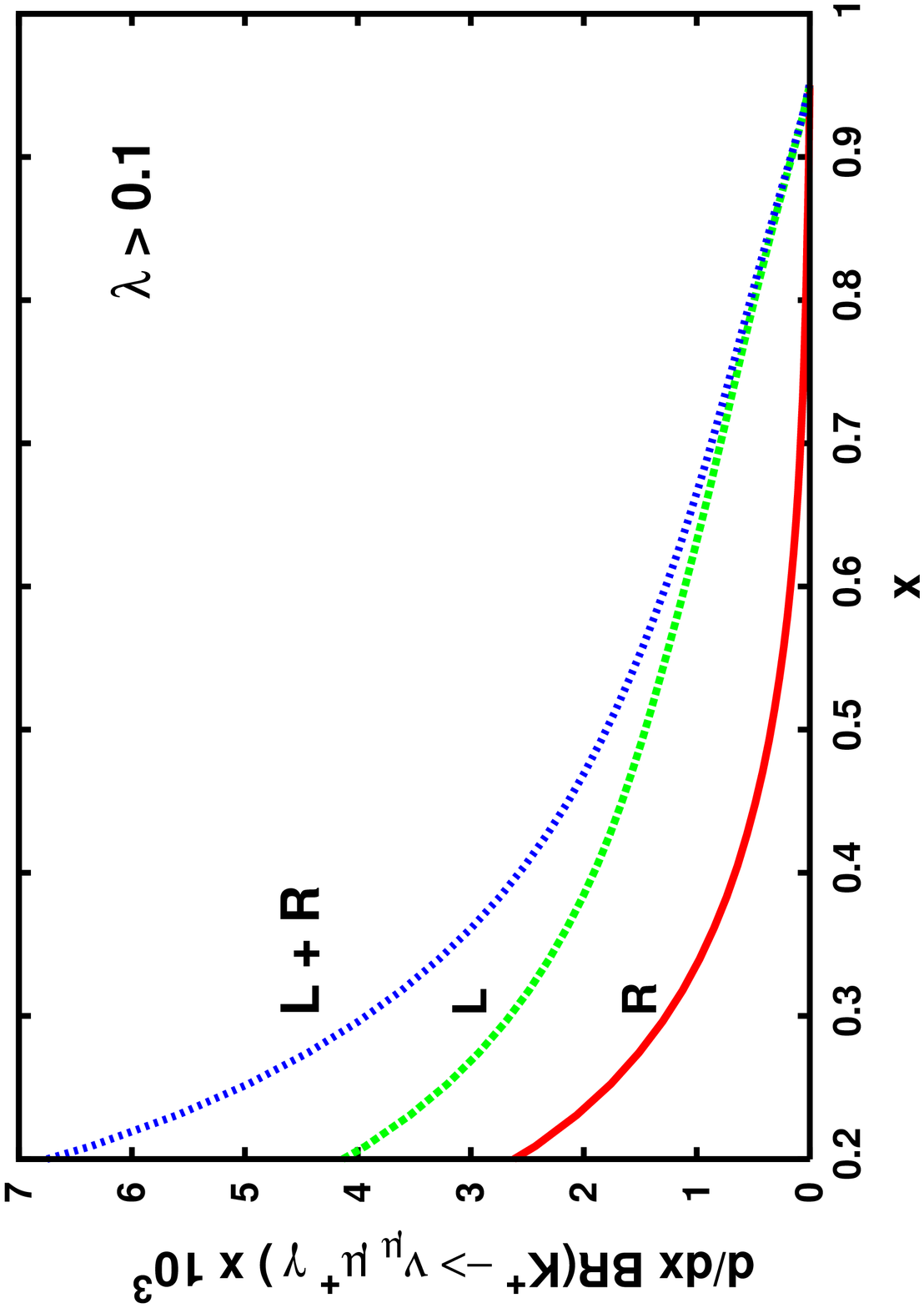}{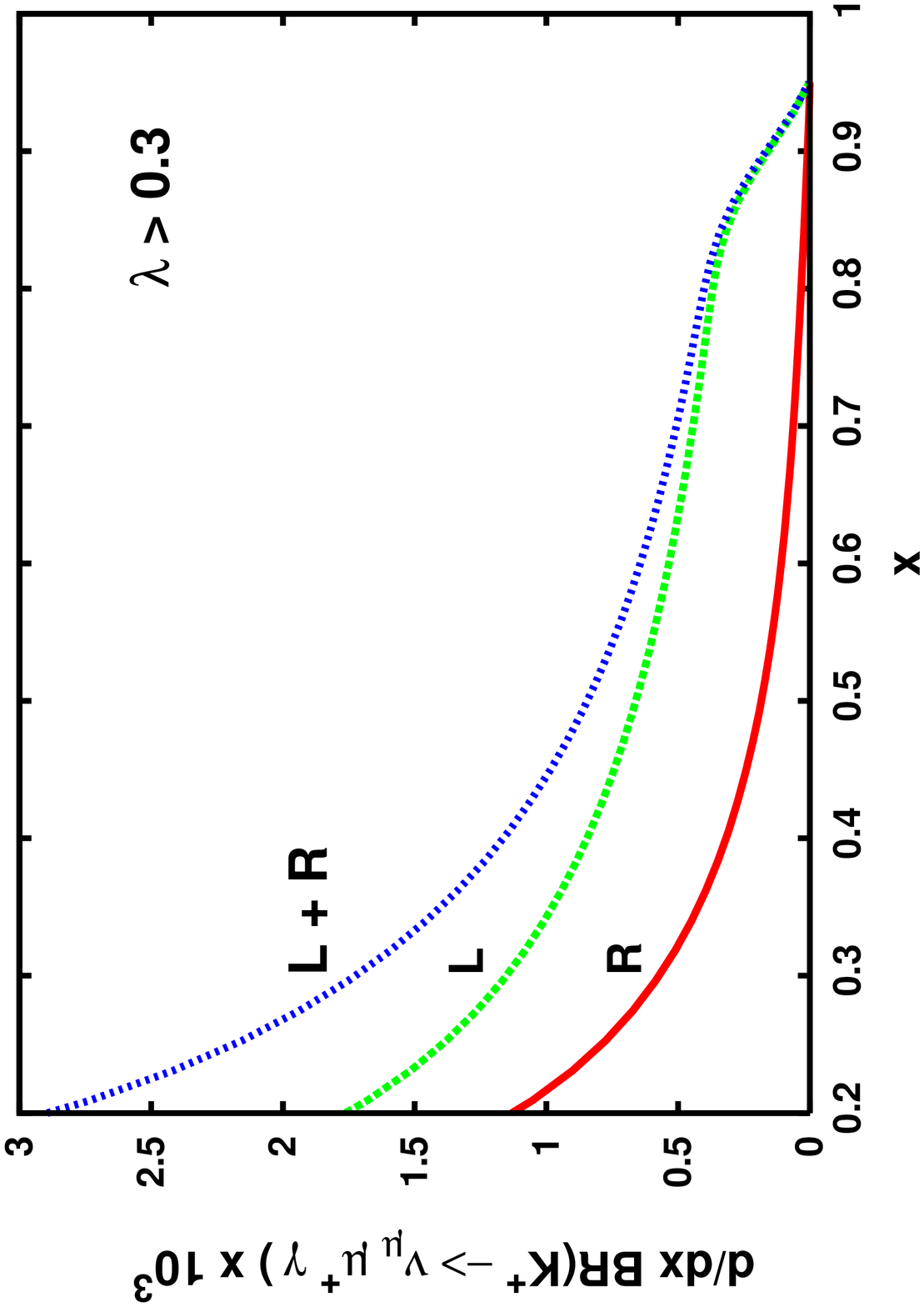}{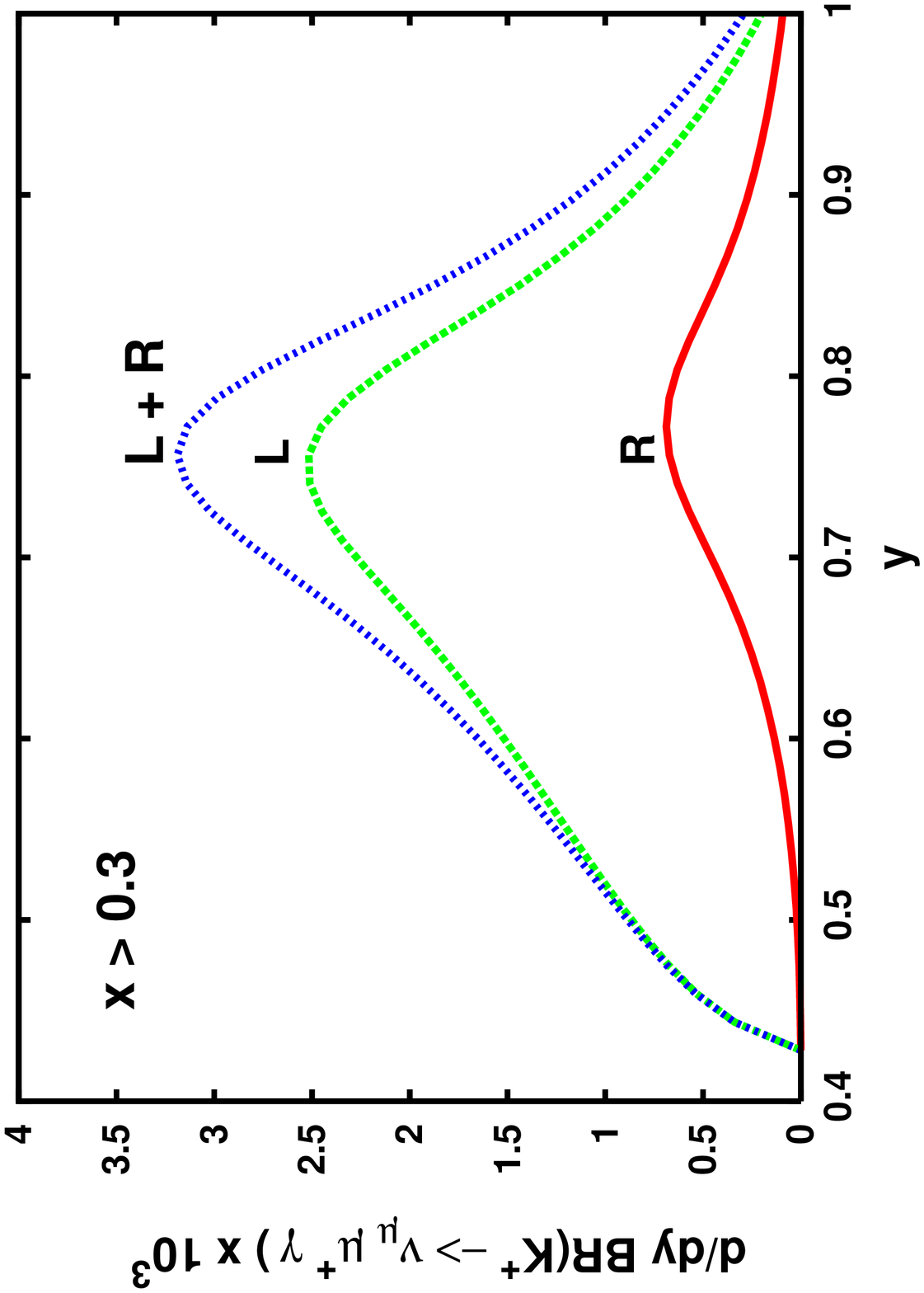}{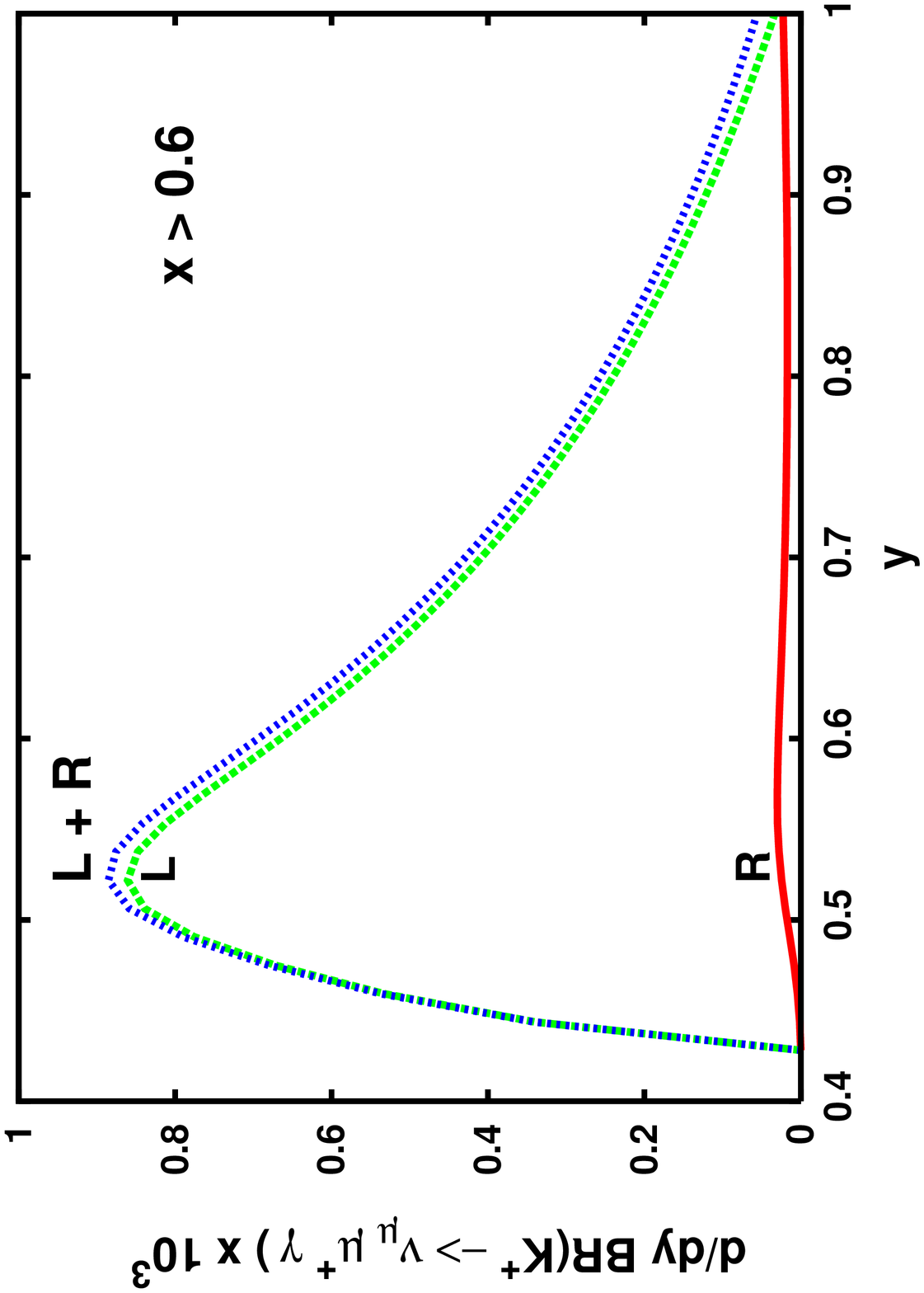}
\end{center}
\caption{\small As in Fig.~\ref{dBRx_Ke}, but for 
kaon decay in $K^+ \to \nu_{\mu} \mu^+ \gamma$.
Curves correspond to 
kinematical cuts as indicated in the figures.}
\label{dBRx_Kmu}
\end{figure}

In Fig.\ref{dBRx_Kmu} results for BR distributions in $x$ and $y$ 
are shown for the $K^+ \to \mu^+\nu_{\mu} \gamma$ decay.
The hierarchy between L and R curves,
and their shapes, are similar to the corresponding ones of
$\pi^+\to \mu^+\nu_{\mu} \gamma$ ( see Fig.\ref{dBRx_Ke}).
Notice that the available ranges of $x$ and $y$ are larger for the kaon decay
in the muon channel with respect to the corresponding pion decay, 
due to more available phase space of the former.
The same considerations regarding the shapes of $x$- and $y$-distributions 
of pion decay should hold here as well. 
As can be seen from these results, also in 
$K^+\to \mu^+\nu_{\mu}\gamma$ decay the left-handed photon polarization
gives the dominant effect for $x>0.2$ and $\lambda >0.3$ or analogously for
$x>0.3$ and $y>0.4$, as shown in the bottom plots for the $y$ distribution.
Numerical results for the total contribution L+R, are consistent 
with the corresponding ones in Ref.\cite{EG}.

\begin{figure}[tpb]
\begin{center}
\dofourfigs{3.1in}{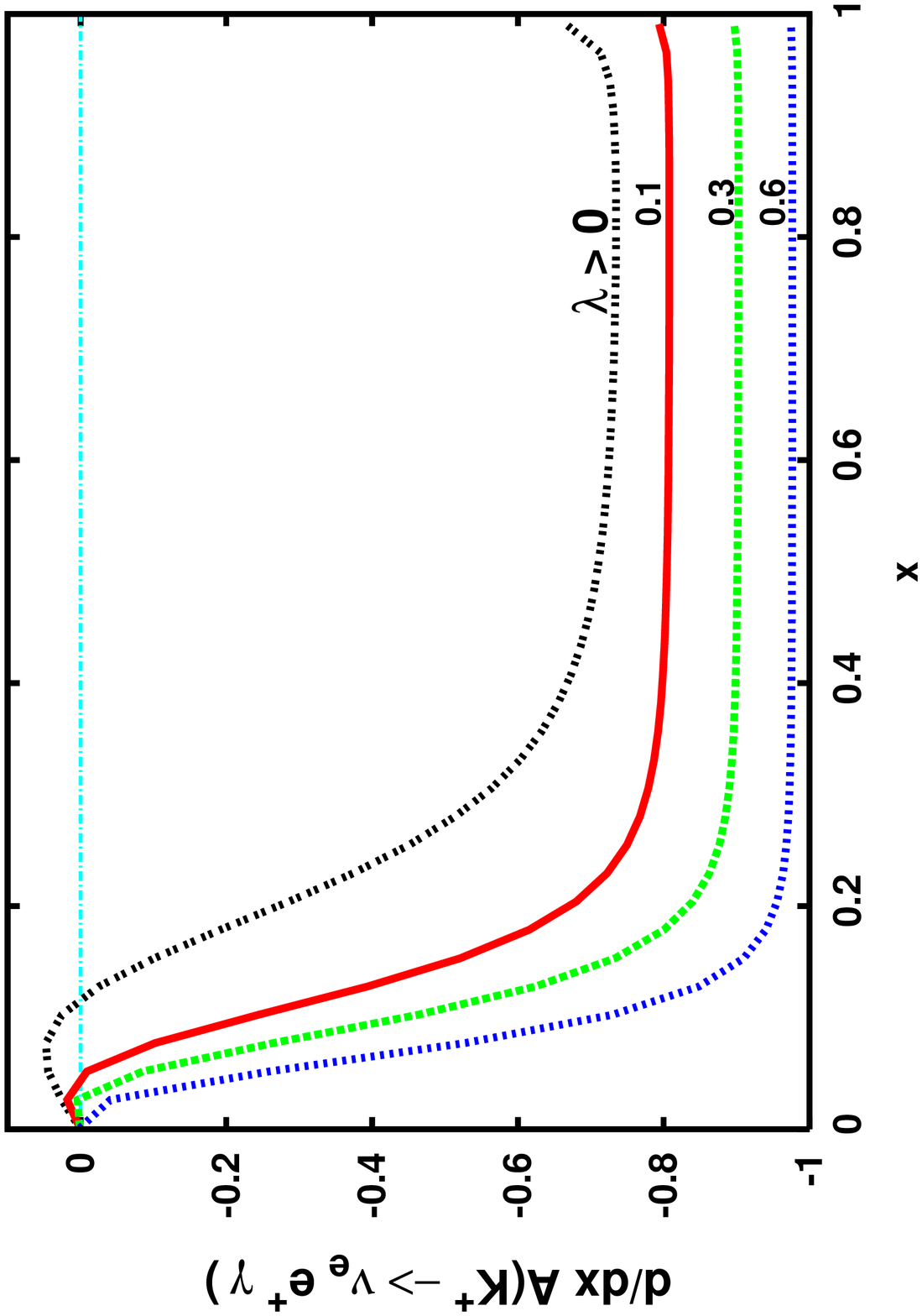}{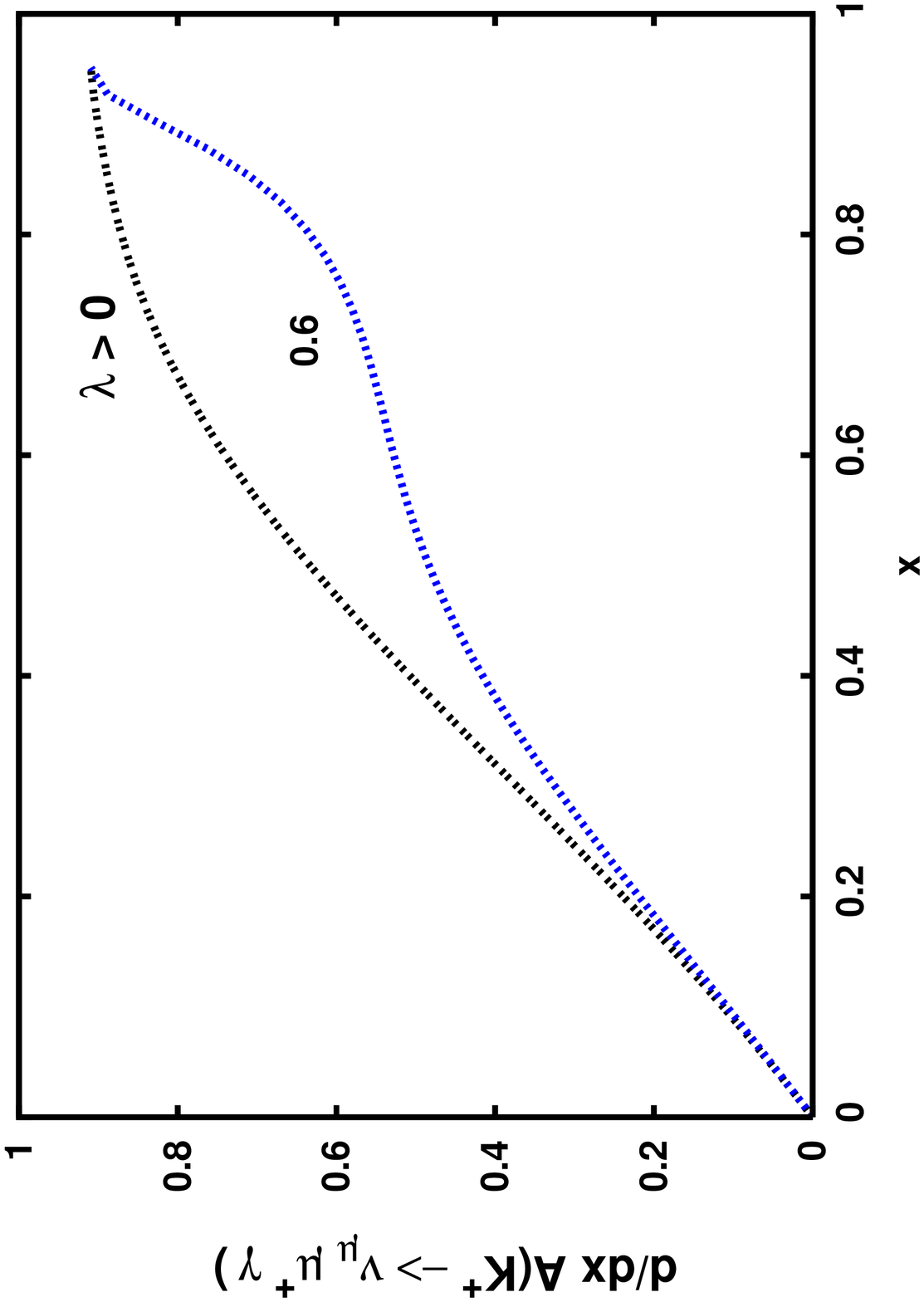}{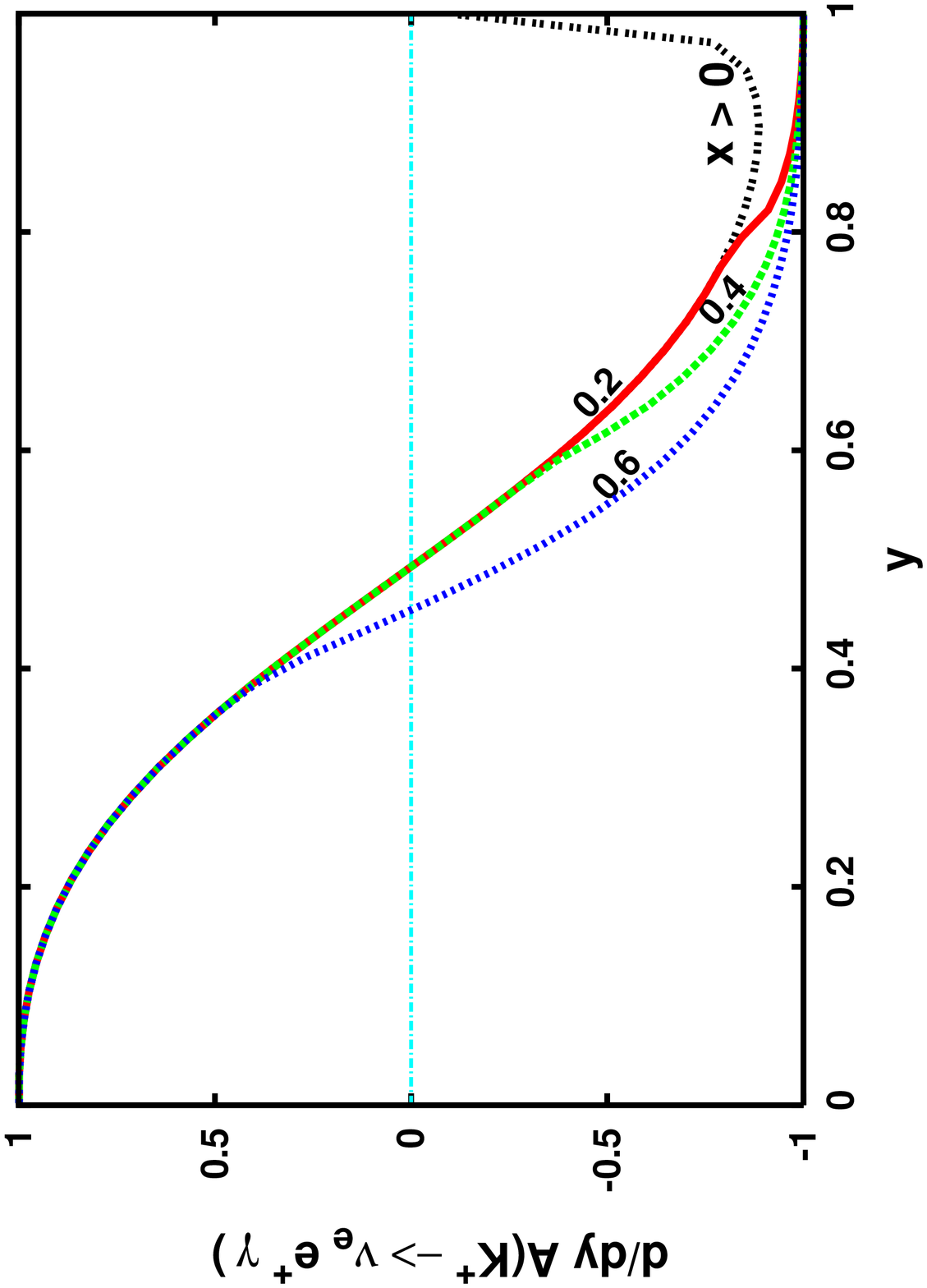}{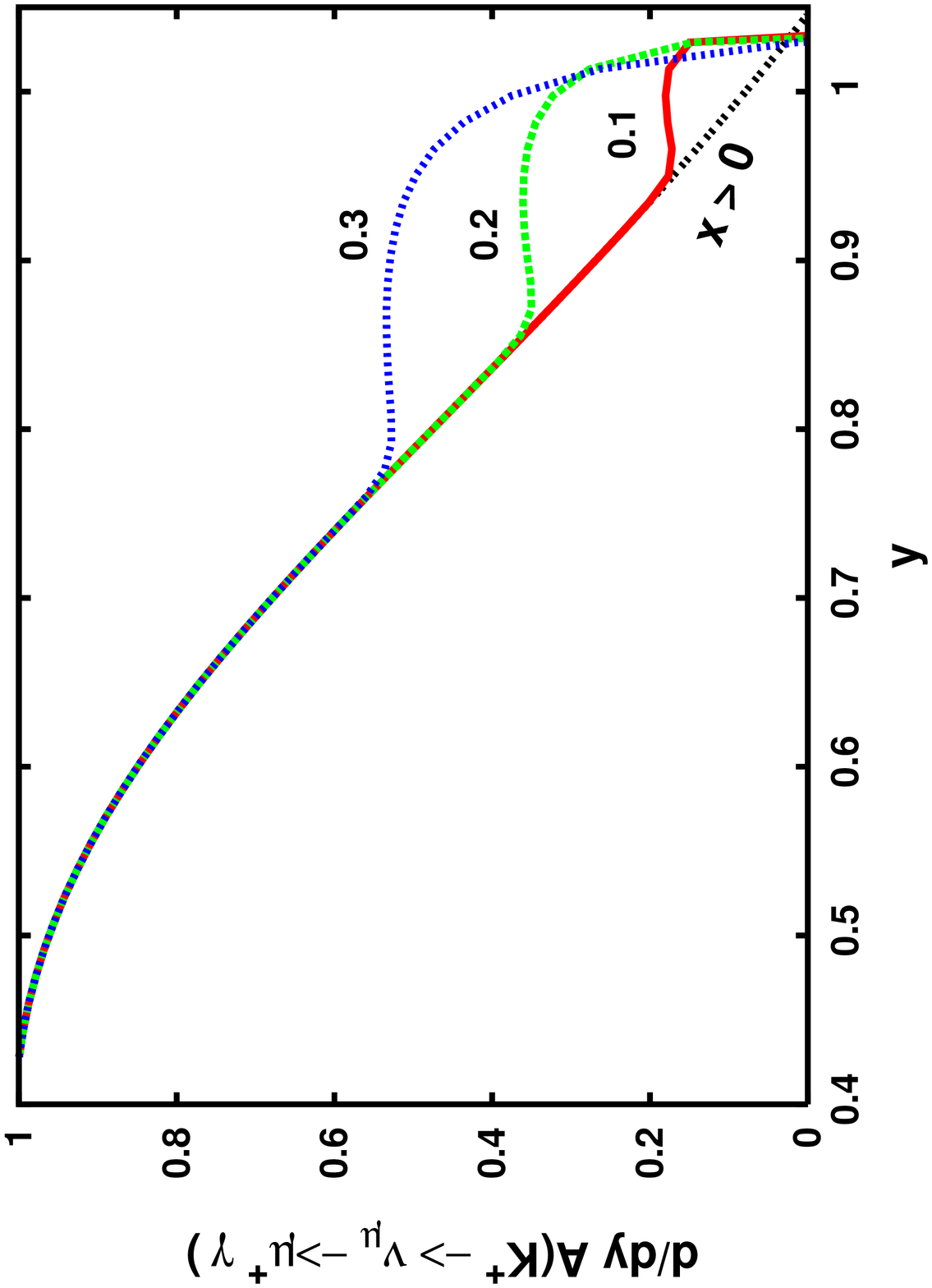}
\end{center}
\caption{\small Asymmetries as in Fig.~\ref{asymPion}, but for 
kaon decay in $K^+ \to \nu_{\mu} \mu^+ \gamma$. Numbers on the 
curves correspond to kinematical cuts.}
\label{asymkaon}
\end{figure}

In Fig.\ref{asymkaon} we show our results for the distribution of asymmetries
in the kaon decays. As we can see from left-top plots in Fig.\ref{asymkaon},
the $d{\rm A}_{\gamma}/dx$ distribution for $K^+\to e^+\nu_e\gamma$
is always negative in all range $x>0.1$, and 
tends to a constant value for $x > 0.3-0.4$, depending on the 
applied cuts on $\lambda$.
In particular, when $\lambda >0.6$, the $d{\rm A}_{\gamma}/dx$ already 
approaches its minimum value for $x>0.2$. 
By relaxing the constraints on $\lambda$,
we can see that the $d{\rm A}_{\gamma}/dx$ distribution 
could have a zero at $x\simeq 0.2$, and  analogously 
$d{\rm A}_{\gamma}/dy$ at $y\simeq 0.5$. 
On the right-plots we present the corresponding results for
the $K^+\to \mu^+\nu_{\mu}\gamma$ decay, and for some 
representative choices of cuts. As we can see, the
$d{\rm A}_{\gamma}/dx$ is more sensitive to cuts on $\lambda$ than 
the corresponding one in $\pi^+\to \mu^+\nu_{\mu}\gamma$. 
Analogous considerations hold 
for the $d{\rm A}_{\gamma}/dy$ distribution as well. In conclusion,
the photon polarization asymmetry for radiative meson decays
in muon channel, is always positive, vanishing only at the end point
$x=0$ or analogously
$y=1$, due to the spin decoupling property of the soft photon.

\subsection{Radiative $\mu^-$ decay}
In this section we will discuss the numerical results 
for the radiative muon decay $\mu^-\to \nu_{\mu} e^- \bar{\nu}_e  \gamma$.
As shown in section 3, this process is described by 
the leptonic Fermi interaction, where the photon is attached to 
external legs of muon and electron, see Fig.\ref{Muon}. 
Being a pure leptonic process, its decay rate 
can be calculated with high accuracy in perturbation theory.
In particular,
the 1-loop QED corrections have been evaluated in Ref.\cite{fks}
for the inclusive radiative muon decay, which 
corresponds to an accuracy of order ${\cal O}(\alpha^2)$ in the 
branching ratio.
However, studies of polarized radiative muon decays have been recently
published \cite{ss,fgkm}.  Also 
1-loop radiative corrections have been
included in the evaluation of the decay rate \cite{fgkm}.
However, in  these studies only the
polarization of fermions has been considered.
\begin{figure}[tpb]
\begin{center}
\dofigA{3.1in}{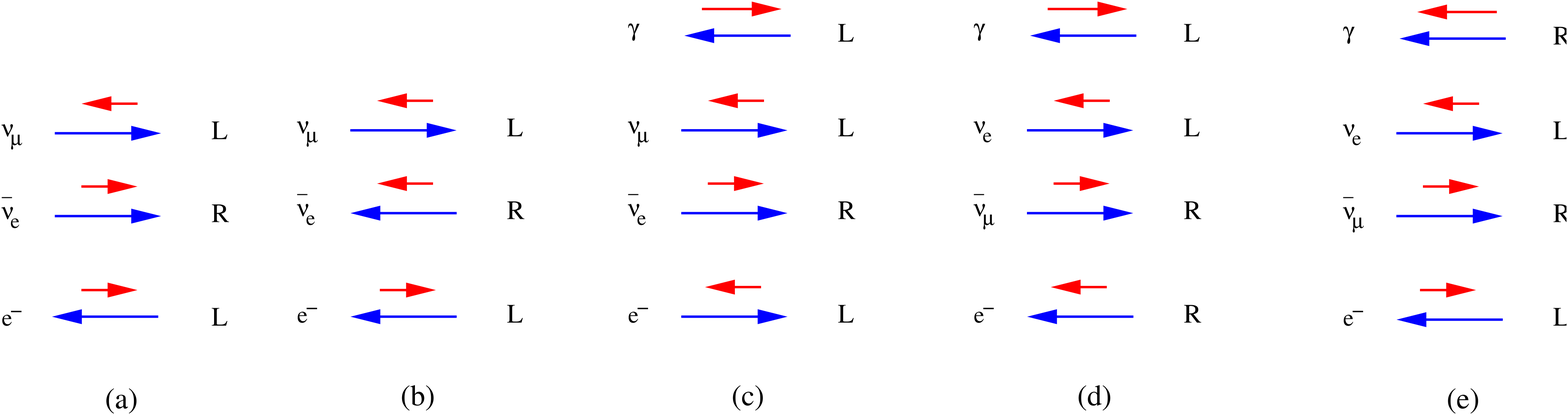}
\end{center}
\caption{\small Some helicity (in red) configurations of $\gamma$, $\nu_{\mu}$
$\bar{\nu}_e$ and $e^-$  for $\mu^-\to \nu_{\nu} \bar{\nu}_e e^- \gamma$ 
decay in $\mu^-$ rest frame, 
figures (c)-(e), when all momenta
(in blue) are aligned on the same axis. 
Figures (a) and (b) 
correspond to the non-radiative decay $\mu^-\to \nu_{\nu} \bar{\nu}_e$.
Direction of electron momentum in (a)-(b), as well as
photon momentum in (c)-(e) diagrams, is fixed by convention.}
\label{spin_muon}
\end{figure}

\begin{figure}[tpb]
\begin{center}
\dofourfigs{3.1in}{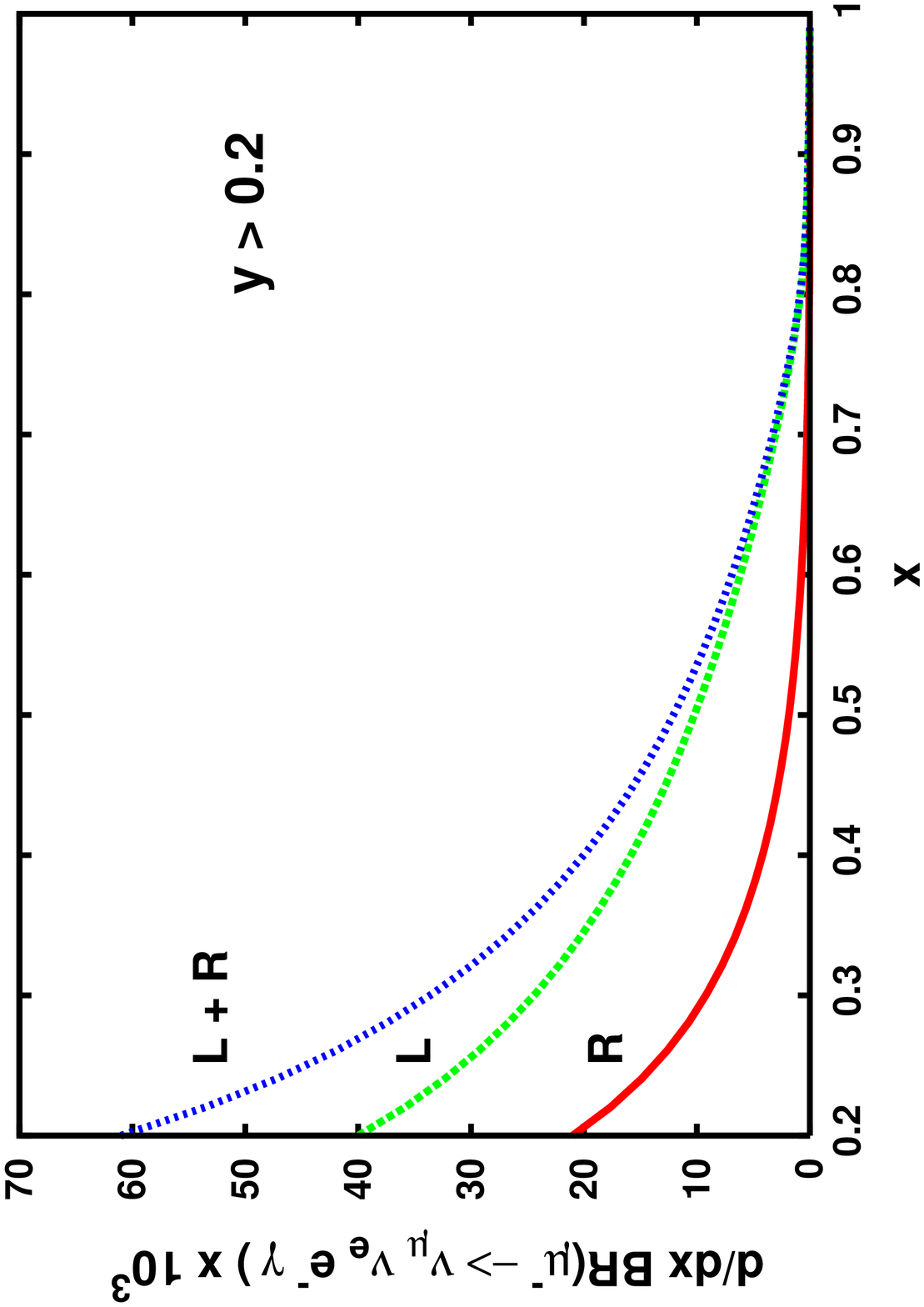}{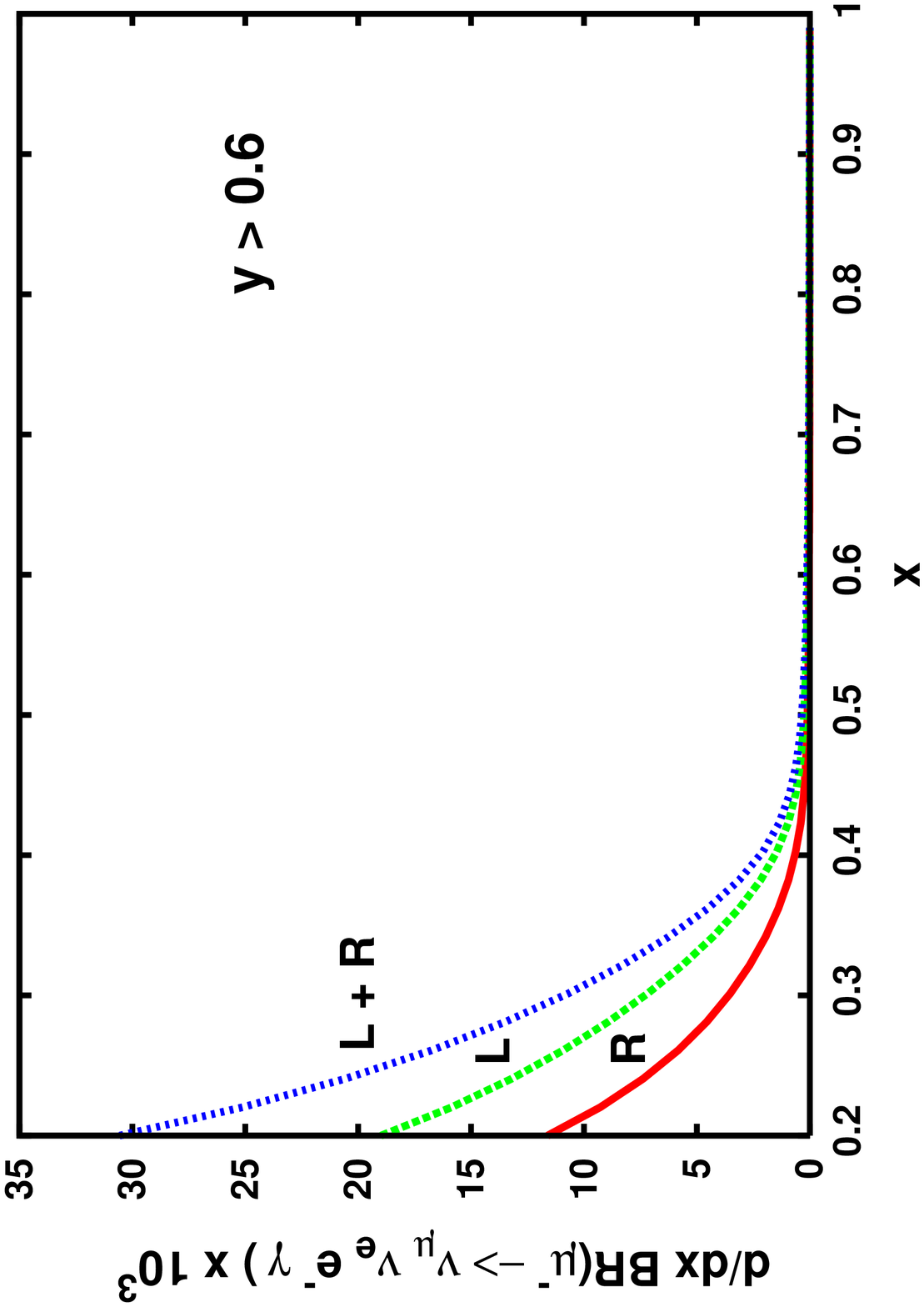}{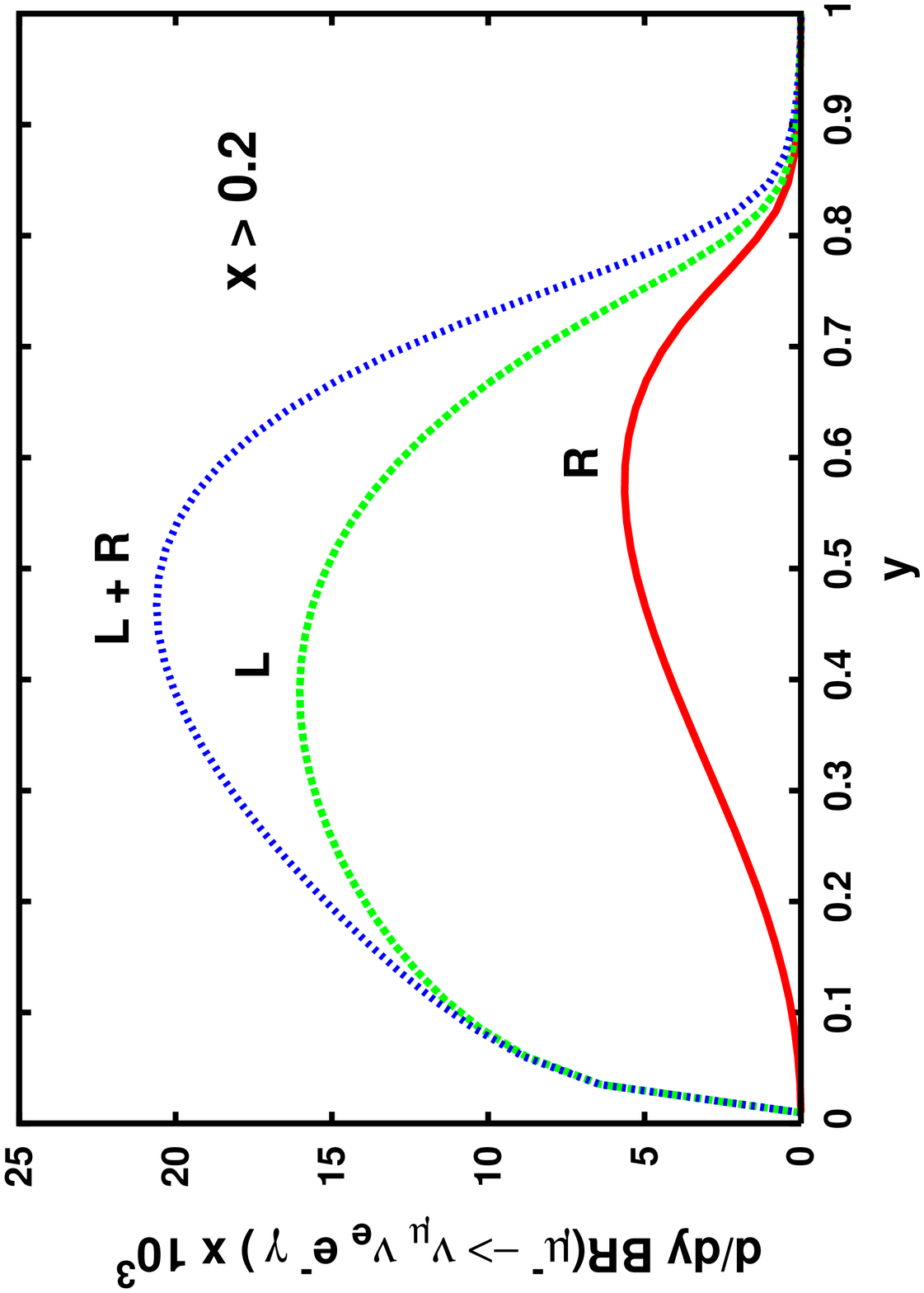}{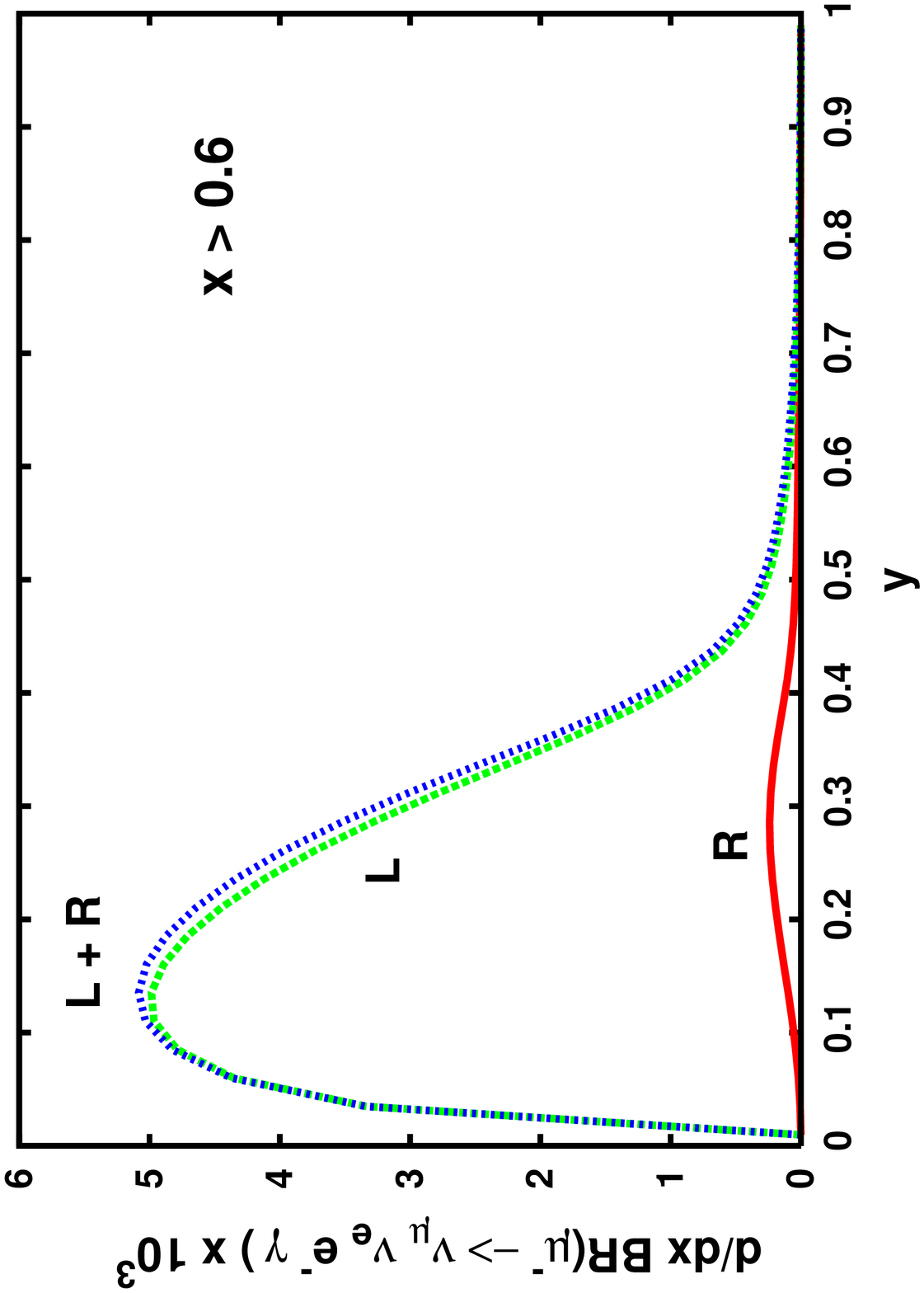}
\end{center}
\caption{\small
The photon energy ($E_{\gamma}$) spectrum 
$\frac{\rm d BR_{\gamma}}{\rm dx}$ versus $x=2\,E_{\gamma}/m_{\mu}$ (top) 
and electron energy ($E_{\gamma}$) spectrum 
$\frac{\rm d BR_{\gamma}}{\rm dy}$ versus $y=2\,E_e/m_{\mu}$
(bottom)
for muon decay $\mu^+ \to \bar{\nu}_{\mu} \nu_{e} e^+$
with left-handed (L) and right-handed (R) photon polarizations.
and for kinematical cuts  $y > 0.2$ (top-left), $y >0.6$ (top-right) 
and  $x > 0.2$ (bottom-left), $x >0.6$ (bottom-right) respectively.
}
\label{dBRx_muon}
\end{figure}

\begin{figure}[tpb]
\begin{center}
\dofigs{3.1in}{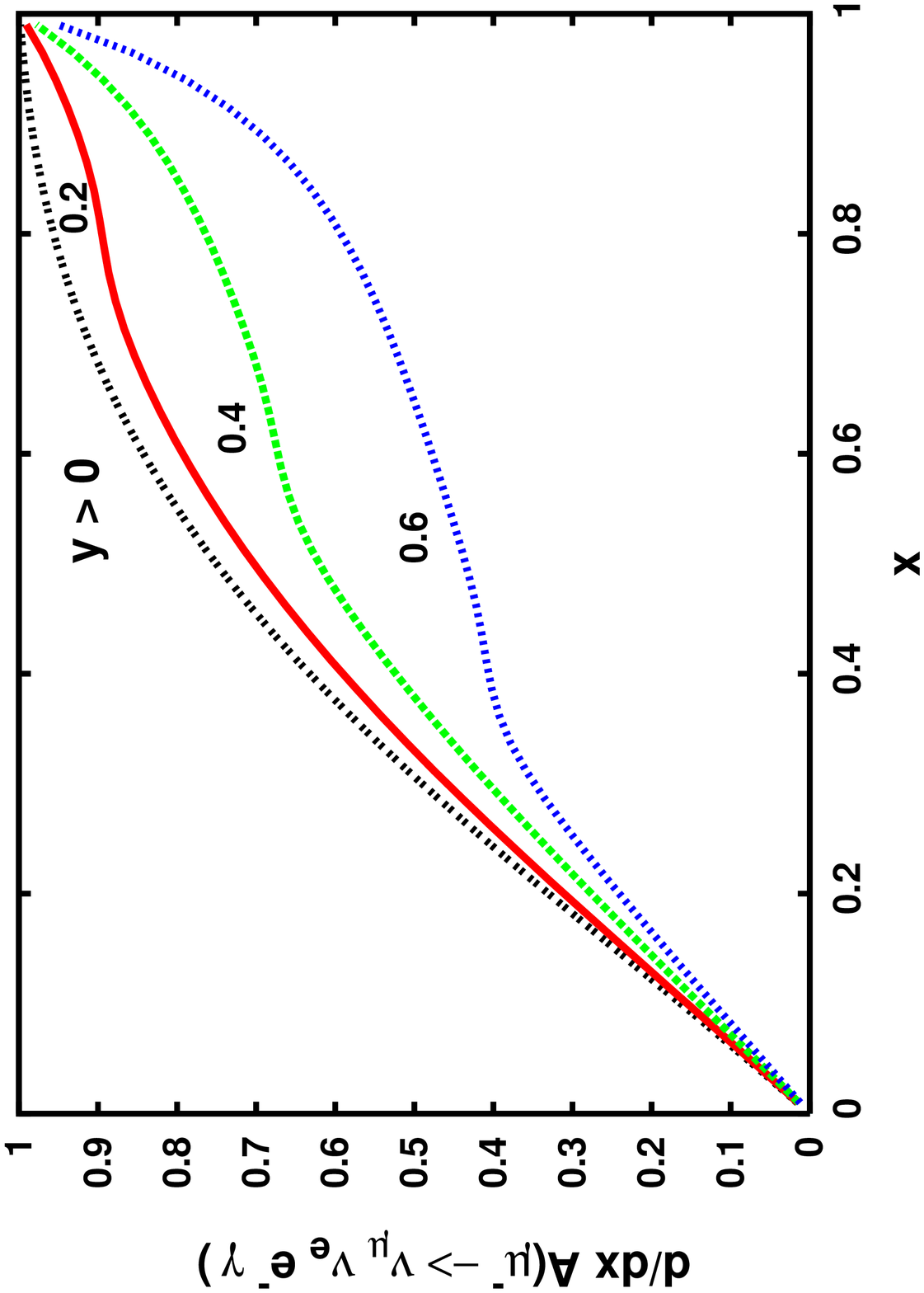}{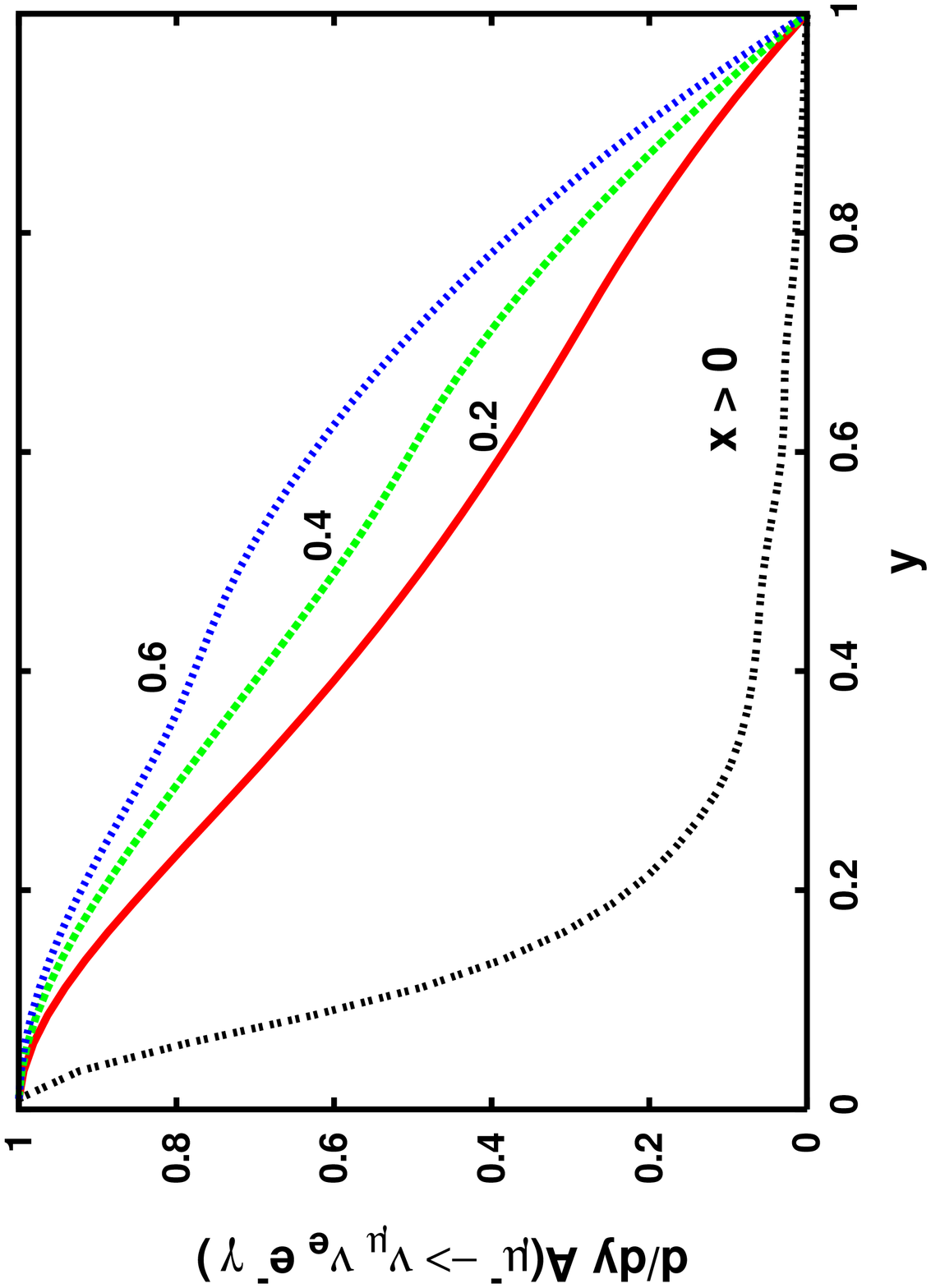}
\end{center}
\caption{\small As in Fig.\ref{dBRx_muon}, but
for the asymmetry $\frac{\rm d A_{\gamma}}{\rm dx}$ versus $x$ (left) and
$\frac{\rm d A_{\gamma}}{\rm dy}$ versus $y$ (right) 
and for kinematical cuts $y$ and $x~> \{0,\,0.2\,,0.4\,,0.6\}$ 
for left and right plots respectively.}
\label{asymMuon}
\end{figure}

Our results for the polarized photon distributions of branching ratios 
are shown in Fig.\ref{dBRx_muon}. These results, as well as 
the corresponding ones in Figs.\ref{asymMuon} and \ref{muePol},
have been obtained by integrating over the full the phase
space and by taking into account the full $r$ dependence.
From Fig.\ref{dBRx_muon} we can see that 
the main contribution to the radiative decay is provided by the 
left-handed photon polarization, while the right-handed one
is quite suppressed and decreases by increasing the photon energy. 
This behavior, again, can be explained by using angular momentum 
conservation and parity violation.
Notice that, due to the V-A nature of weak interactions,
the electron is mainly produced left-handed polarized in the muon decay, 
and chirality flip effects, needed to produce a right-handed 
electron, are always sub-leading, being proportional to
the electron mass. Moreover, due to the fact that 
we are integrating over the final phase space of neutrinos, 
the analysis is strongly simplified.
Indeed, after integration, the effect of neutrinos is re-absorbed in the
tensor $N^{\alpha \beta}$ appearing in Eqs.(\ref{Ntensor1}), (\ref{Ntensor2}).
Notice that $N^{\alpha \beta}$ is just a projector for the four-momentum 
$Q^{\alpha}\equiv -(p_e+k+p)^{\alpha}$ carried by the neutrinos pair,
and it can be seen 
as the sum over polarization states of a massive particle of spin 1.
In other words, regarding the spin content,
the neutrinos pair behave as a spin-1 particle of mass $Q^2$, having
three polarization states.
In the case of non-radiative muon decay,
the allowed spin configurations in the muon rest frame are 
shown in Fig.\ref{spin_muon}a,b, where all momenta are aligned on the same
axis $X$ and by convention the electron momentum is chosen along the  
negative direction. As we can see, if the electron is left-handed
($J_X=1/2$), the spin projection 
of neutrino anti-neutrino pair 
along the direction of their total momentum can be $J_X=-1$ or $J_X=0$, 
but not $J_X=+1$, being the muon a spin 1/2 particle. 

Let us now consider the radiative decay, with photon 
emission from the electron line.
It is known that, when hard photons are emitted parallel and forward
to the electron momentum, they can flip the electron
helicity, without paying any chiral mass suppression 
\cite{tv,fgkm,ss,dz,ln,fs}. The helicity-flip mechanism is illustrated below
\begin{center}
\epsfbox{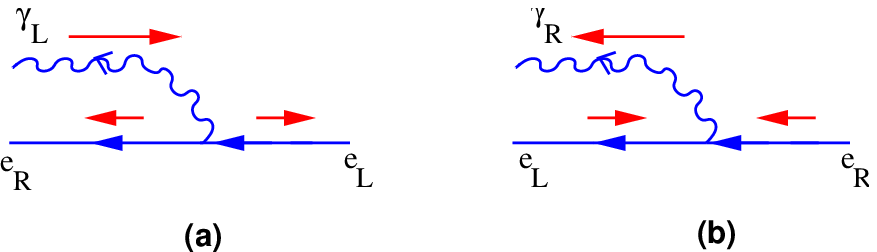}
\end{center}
for incoming left-handed $e_L$ (a)
and right-handed $e_R $ (b) electron by collinear photon bremsstrahlung. 
All momenta, indicated by blue arrows, 
are aligned on the same axis. Red arrows stand for the corresponding helicities
and an electron mass insertion is understood. 
We remind here that the chiral suppression
of the term $m_e^2$ appearing in 
the square modulus of the numerator due to the chirality flip, 
is compensated by the singular behavior in $\simeq 1/m_e^2$ appearing 
in the square modulus of the propagator for collinear photon emission.
The corresponding spin configurations in this case
are shown in Fig.\ref{spin_muon}d-e for the case of aligned momenta.
However, as for the meson case, the contributions coming
from the helicity-flip transitions are always smaller with respect to
the helicity conserving ones.
The largest contributions should come
from the photons emitted by the muon line. In this case 
the neutrinos and electron momenta are favored to be
busted backward with respect to the photon momentum, in order to compensate
for the missing momentum.
The corresponding spin configuration, in the particular 
limit in which all momenta are aligned on the same axis, 
is shown in Fig.\ref{spin_muon}c.
Since the favored spin of the $\nu+\bar{\nu}$ system is in this case $J_X=0$, 
the photon must be necessarily left-handed polarized.
This peculiar configuration should explain why the left-handed photon 
contribution is always dominant with respect to the right-handed one,
leading to an increasing relative gap as the photon energy increases.
This seems to be the case since, 
as the photon energy approaches the soft region $x\to 0$, 
the gap should decrease due to the 
spin decoupling property of soft photons.
Results concerning the branching ratio distributions for the production of
right-handed electron are shown in the next section.

In Fig.\ref{asymMuon} we show in the right (left) plots
the $d{\rm A}_{\gamma}/dy$ ($d{\rm A}_{\gamma}/dx$) 
distributions asymmetry for the radiative 
muon decay. 
We present our results for 
some representative choices of cuts, in particular 
$y>\{0,~0.2,~0.4,~0.6\}$ 
and $x>\{0,~0.2,~0.4,~0.6\}$ for the $d{\rm A}_{\gamma}/dx$ and 
$d{\rm A}_{\gamma}/dy$ respectively. 
As we can see from these results, the asymmetry in the muon case is
always positive, as a consequence of the dominant left-handed photon 
contribution as discussed above. The shapes of asymmetries 
are quite similar to the corresponding ones in 
$\pi^+\to \mu^+ \nu_{\mu} \gamma$ and $K^+\to \mu^+ \nu_{\mu} \gamma$, 
where the IB effects are larger than the SD terms. In particular, 
here the $d{\rm A}_{\gamma}/dx$ is monotonically
increasing with $x$, while analogously the $d{\rm A}_{\gamma}/dy$ is
monotonically decreasing. 

\section{Energy spectra of the polarized positron/electron}
In this section we discuss the results for the distributions of BRs 
in the positron energy for the decays $\pi^+\to e^+\nu_e \gamma$ and
$K^+\to e^+\nu_e \gamma$, and analogously in the electron energy for 
the radiative muon decay, for both lepton and photons polarizations.
As seen before, due to angular momentum conservation,
the IB contributions favors a left-handed positron 
in the soft photon energy region. 
However, in the case of an hard photon produced collinear with $e^+$, 
the IB favors a right-handed positron. 
This last effect is due to the helicity-flip 
mechanism discussed above.
However, if one imposes larger cuts on $x$ in order to reduce the IB effect,
then the relative contribution of the SD term grows up. In particular, 
when the positron is totally generated from the SD terms,
it is mainly right-handed polarized 
(see discussions in section 4.1), 
as can be seen from Eq.(\ref{SDr0}) in Appendix A,
since the photon is emitted from the hadronic vertex.
From these results one can
easily check that in the region of $y\to 1$, the $LR$ term  in Eq.(\ref{SDr0}) 
tends to zero and survives only the $RR$ one, corresponding to the production
of both positron and photon right-handed polarized, as required by angular
momentum conservation. In particular, for a generic meson $M=K,\pi$, we have
\bea
\lim_{y\to 1} \, 
\frac{d {\rm BR}^{(R,R)}}{dy}\,\simeq \,
{\rm BR}(M^+\to l^+\nu)\, \frac{\alpha}{2\pi} \frac{(V+A)^2}{48\, r_l}
\frac{m^2_{\pi}}{f^2_{\pi}}\left(1-4x_{\rm cut}^3+3x_{\rm cut}^4\right)\, ,
\label{BR_RR}
\eea
where $x_{\rm cut}$ is the cut on $x$,
${\rm BR}(\pi^+\to e^+\nu)=1.23\times 10^{-4}$ and 
${\rm BR}(K^+\to e^+\nu)=1.55\times 10^{-5}$ \cite{pdg}.
Notice that for a right-handed electron, the 
contribution from the other polarizations
vanishes in the massless lepton limit $r_l\to 0$ and for $y\to 1$.

\begin{figure}[tpb]
\begin{center}
\dofigs{3.1in}{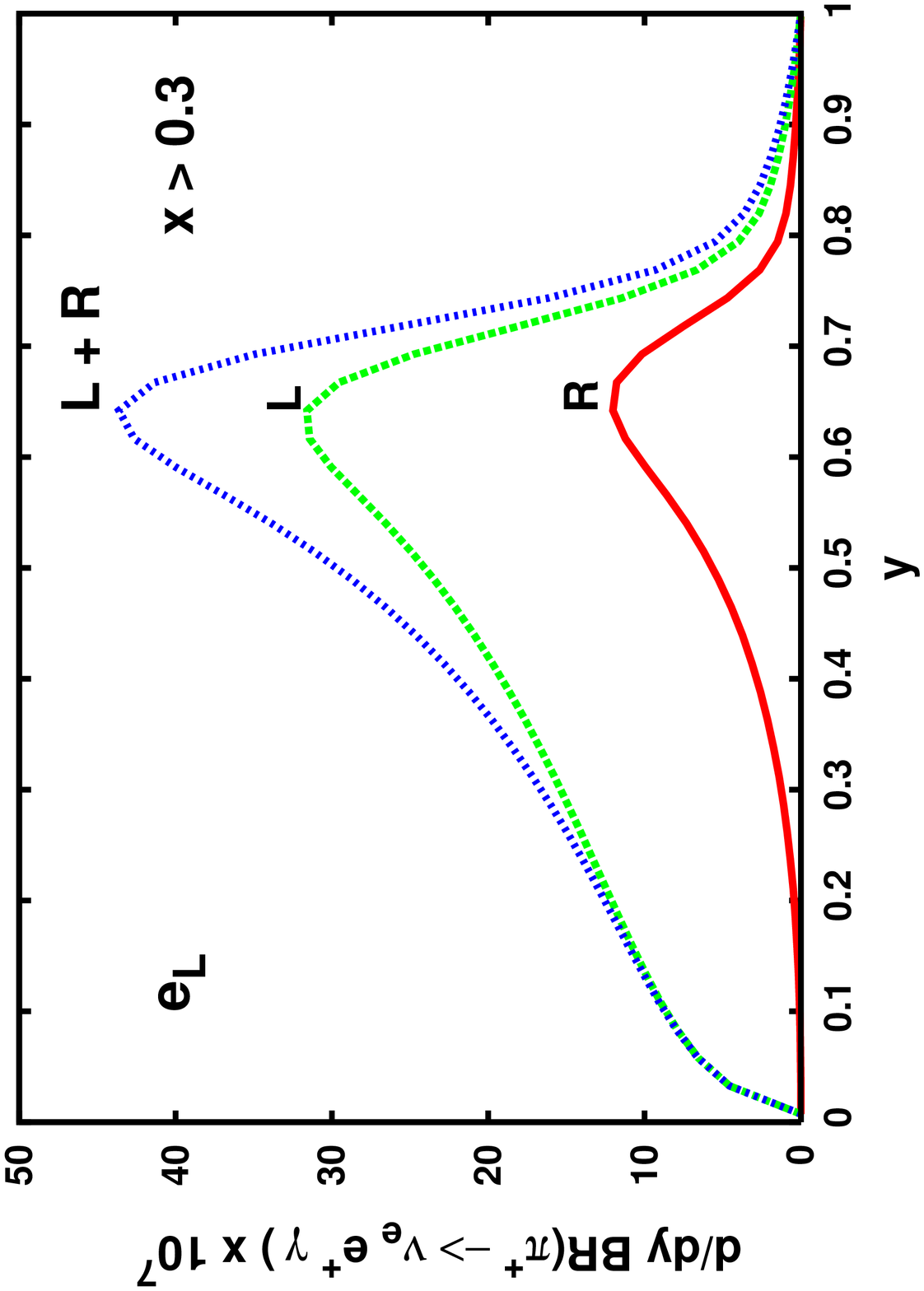}{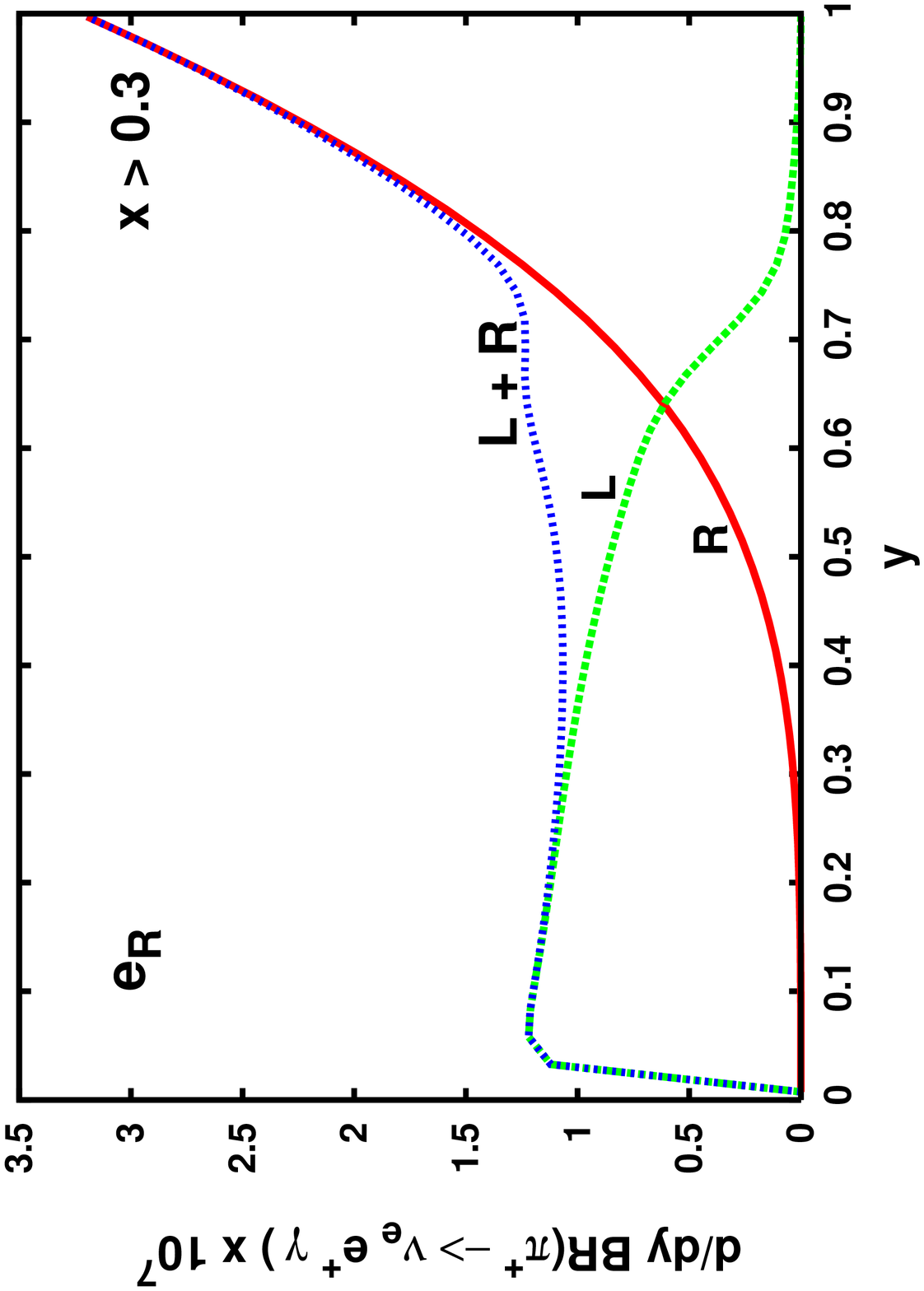}
\end{center}
\caption{\small The electron energy spectrum 
$\frac{\rm d BR}{\rm dy}$ versus $y$ for 
$\pi^+ \to \nu_e e^+ \gamma$, for left-handed (left plot) and
right-handed electron polarizations (right plot), 
with photon energy cut $x> 0.3$. As in previous figures, 
the labels $L$ and $R$ labels attached to 
the curves indicate pure left-handed and right-handed photon polarizations 
contributions respectively, while $L+R$ correspond to the sum. }
\label{PePol}
\end{figure}

In Fig.\ref{PePol} we show the plots 
corresponding to the $d{\rm BR}/dy$ of pion decay, where on the left and
right plots we report the case of left-handed ($e^+_L$) and  
right-handed ($e^+_R$) positron polarizations respectively.
These results are obtained for $x>0.3$.
As we can see, the contribution of $e^+_L$ is always dominant with respect to 
$e^+_R$ one. This is because the IB effect is still large
for $x>0.3$, and so a left-handed positron is favored.
Moreover, a peculiar aspect 
of these results is that for the $e^+_L$ production, 
the left-handed photon polarization
dominates in all the positron energy range. On the other hand,
in $e^+_R$ distribution, the right-handed photon gives the
leading effect for $y>0.7$, being totally induced by the SD terms.
The maximum of $d {\rm BR}/dy$ for $e^+_R$ production, achieved at 
$y\simeq 1$, can be easily checked by using the approximated
expression in Eq.(\ref{BR_RR}).

\begin{figure}[tpb]
\begin{center}
\dofigs{3.1in}{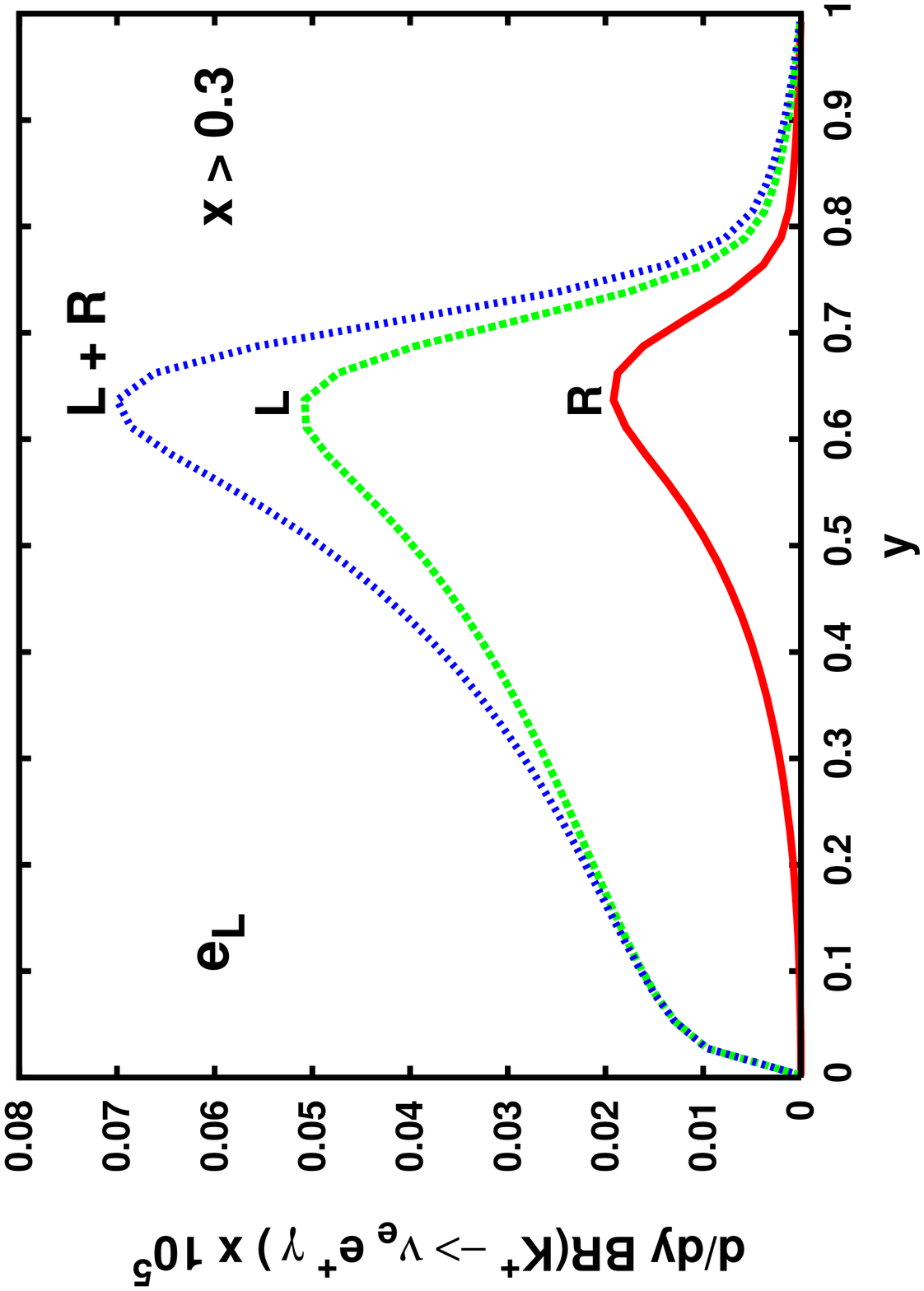}{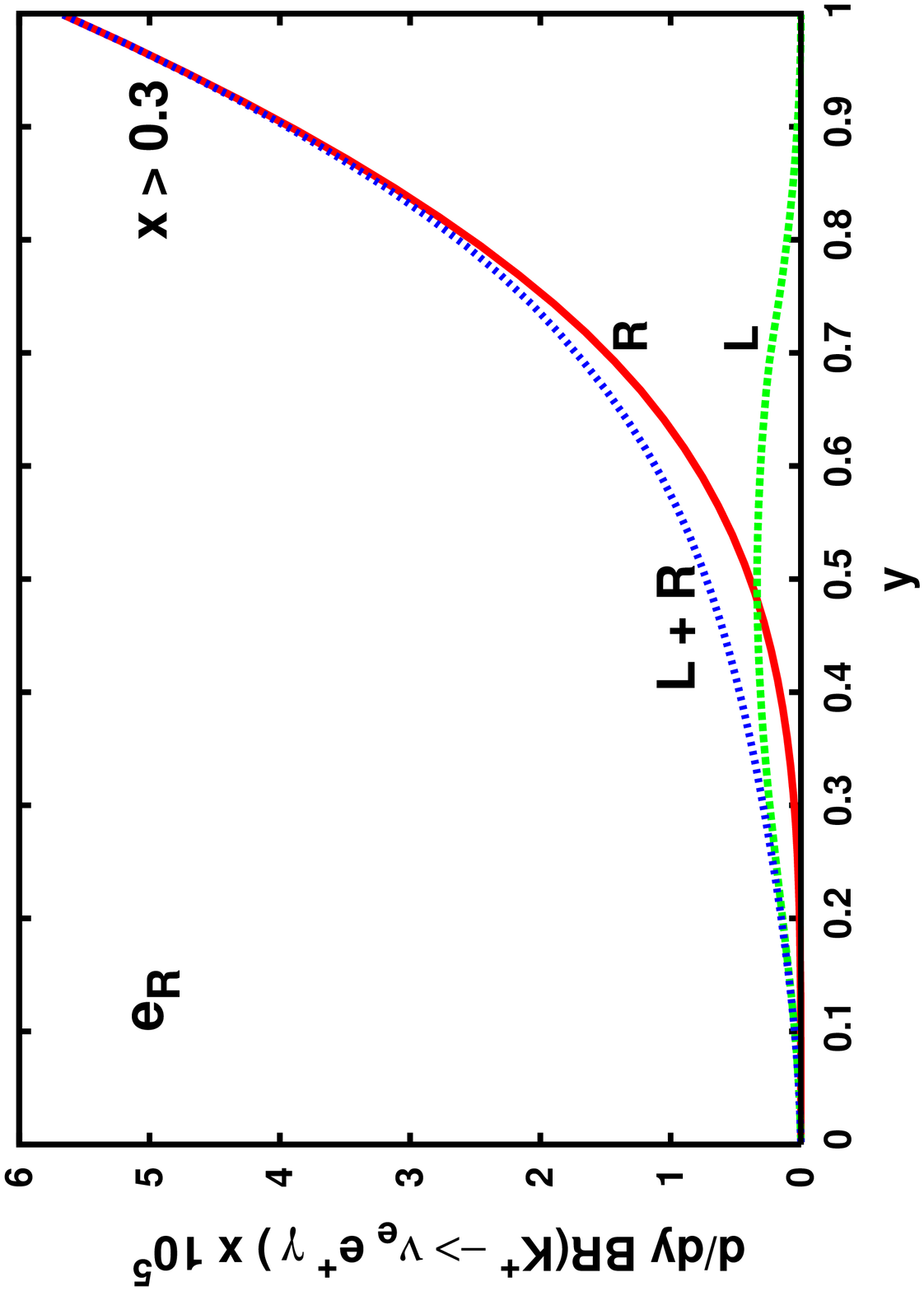}
\end{center}
\caption{\small As in Fig.\ref{PePol}, but for $K^+ \to \nu_e e^+ \gamma$.}
\label{KePol}
\end{figure}

Analogous results for the kaon decay are shown in Fig.\ref{KePol}.
Here, the situation is reversed with respect to the pion case.
The $e^+_R$ contribution gives the leading effect in the total BR 
already for $x>0.3$
and the right-handed photon polarization dominates for $y>0.5$.
On the other hand, in the $e^+_L$ distributions (see left plot in 
Fig.\ref{KePol}), the left-handed photon 
contribution provides the dominant effect.
As already mentioned in section 4.2, 
these differences with pion case 
are mainly a consequence of the fact that the IB amplitude
is more chiral suppressed in $K\to e\nu_e \gamma$ than 
in $\pi\to e\nu_e \gamma$.

\begin{figure}[tpb]
\begin{center}
\dofigs{3.1in}{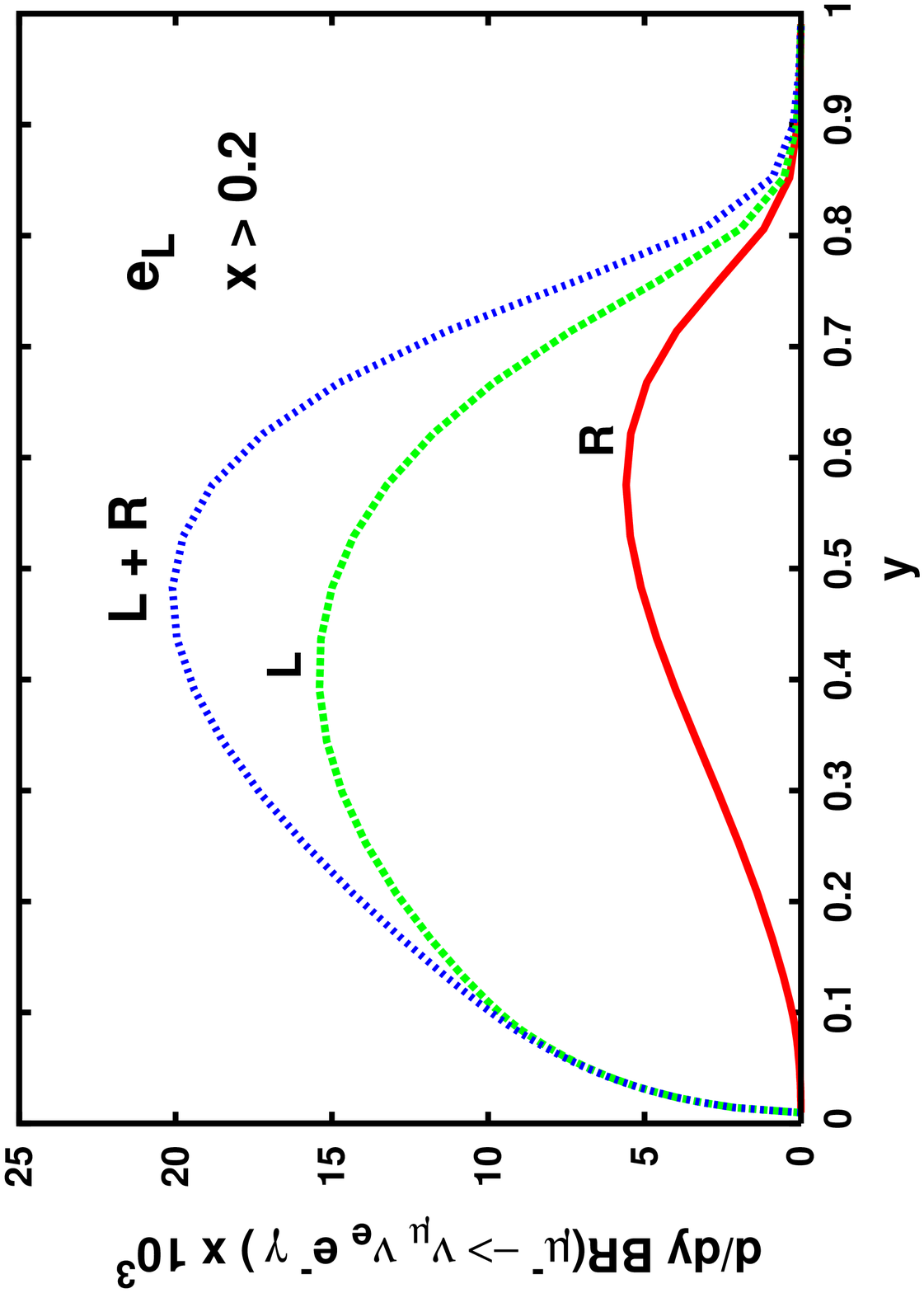}{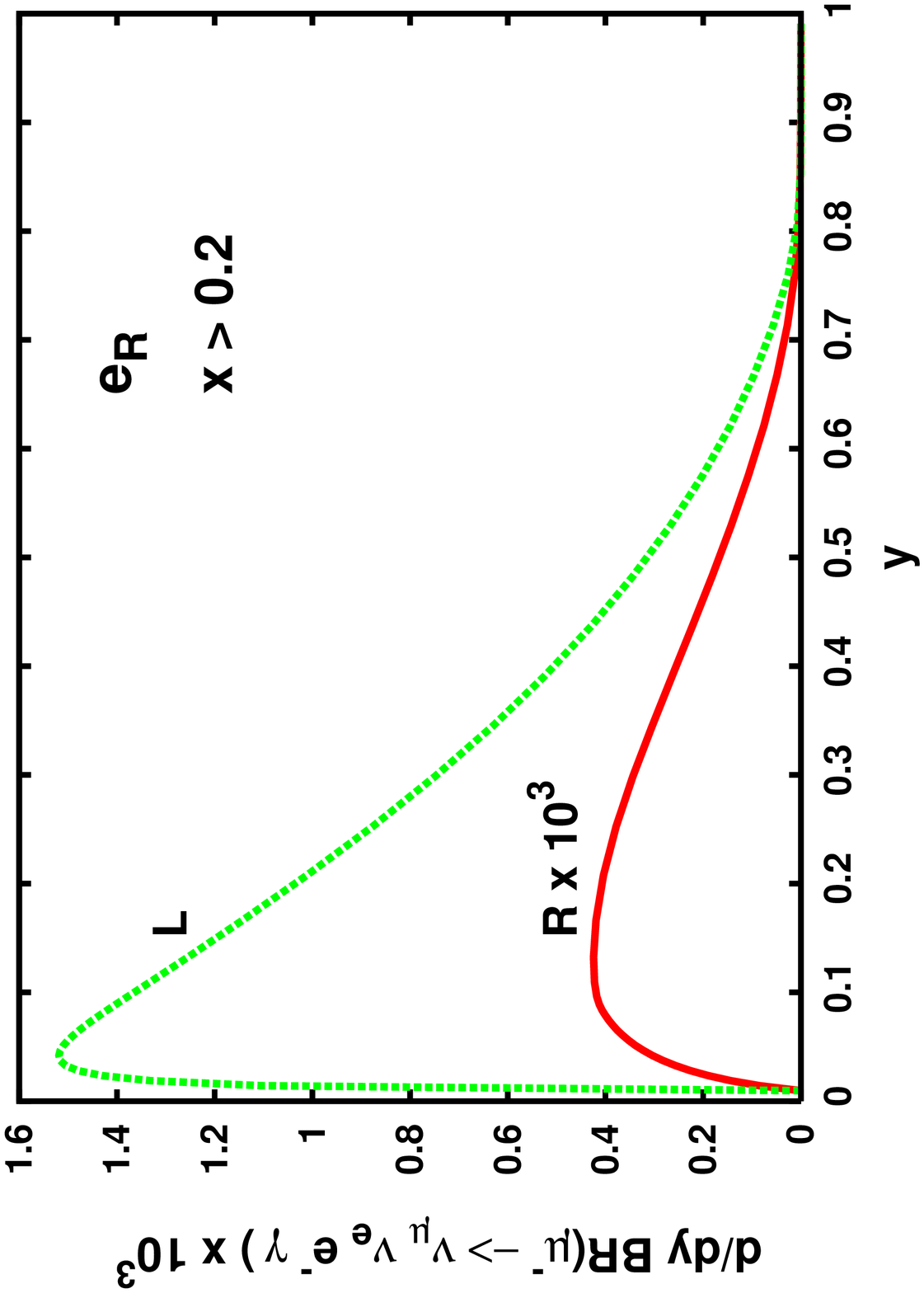}
\end{center}
\caption{\small 
The electron energy spectrum $\frac{\rm d BR}{\rm dy}$ versus $y$ for  
$\mu^+ \to \bar{\nu}_{\mu} \nu_{e} e^+$, for left-handed (left plot) and
right-handed electron polarizations (right plot), 
with photon energy cut $x> 0.2$. The curve in the right plot corresponding
to right-handed photon (R), has been multiplied by a 
rescaling factor of $10^3$.}
\label{muePol}
\end{figure}

Finally, in Fig.\ref{muePol} we present the 
electron energy distributions for the muon case. In this case we imposed
a cut $x>0.2$ on the photon energy.
We see that in 
both $e_R$ and $e_L$ distributions, the left-handed photon contribution
always provides the dominant effect, 
being this configuration favored by angular momentum
conservation. Moreover, in the $e_R$ case, the right-handed 
photon contribution is very tiny. As we explained in the previous section, 
these results are a consequence of the fact that in the radiative muon 
decay the electron is naturally produced left-handed due to the V-A theory.
On the other hand, the contribution 
of $e_R$ is mainly generated from the helicity flip mechanism of hard photons
emitted from the electron line and it is a sub-leading effect.

\section{Tensorial couplings}

Here we analyze the dependence of the photon polarization asymmetry,
in the radiative pion and kaon decays, induced by tensorial couplings.
The aim of this study is motivated by 
the recent measurements of the radiative 
pion decay $\pi^+\to \nu_e e^+ \gamma$ \cite{pibeta}, where
a significant discrepancy in the branching ratio, 
with respect to the SM predictions \cite{pibeta}, 
has been observed. This anomaly might be interpreted as
the effect of a centi-weak tensorial interaction beyond the V-A theory
\cite{Pobl1}.

An analogous discrepancy 
was noticed long time ago at the ISTRA experiment \cite{istra}, 
where radiative pions decays were studied in flight.
In that experiment, the $\pi^+\to \nu_e e^+ \gamma$ was investigated 
over a large phase space region, 
in particular $0.3< x<1$ and $0.2<\lambda< 1$.
The measured branching ratio $B_R^{\rm exp}=(1.6\pm 0.23)\times 10^{-7}$ 
\cite{istra} was found significantly smaller than the expected one
$B_R^{\rm th}=(2.41\pm 0.07)\times 10^{-7}$, 
based on the CVC hypothesis and V-A theory of SM. 
The fact that the measured number of events is less
than expected,  
cannot be explained by a missing unknown background.
This result was interpreted \cite{Pobl1} as a possible indication 
of a tensorial quark-lepton interaction with
coupling of order $10^{-2}$ in unity of $G_F$.
In particular, the suggested new contribution to 
the effective Hamiltonian for $\Delta S=0$ transitions is \cite{Pobl1}
\bea
H_{eff}^{\rm \Delta S=0}=\frac{f_T G_F}{2\sqrt{2}}\, V_{ud}
\left[ \bar{u}\sigma_{\mu\nu}(1-\gamma_5)d\right]
\left[ \bar{e}\sigma_{\mu\nu}(1-\gamma_5)\nu_e\right] + {\rm h.c.}
\label{Htensorial}
\eea
where $\sigma_{\mu\nu}=1/2[\gamma_{\mu},\gamma_{\nu}]$ and $f_T$
a dimensionless coupling.
Notice that tensorial interactions are 
not subject to the strong constraints coming
from the non radiative decay $\pi\to \nu e$ (as, for instance, for the scalar
interactions) simply because, the Lorentz covariance
forces the hadronic matrix element 
$\langle 0|\left[ \bar{u}\sigma_{\mu\nu}(1-\gamma_5)d\right]|\pi\rangle$
to vanish. On the other hand, $H_{eff}^{\rm \Delta S=0}$
can contribute to the 
amplitude ($M_T$) of the radiative decay $\pi^+\to \nu_e e^+ \gamma$
as
\bea
M_T=i\frac{e G_F}{\sqrt{2}}\,V_{ud}\, F_T
\epsilon^{\mu \star}\, q^{\nu}
\left[ \bar{e}\sigma_{\mu\nu}(1-\gamma_5)\nu_e\right]\, .
\label{ampl_tens}
\eea
The constant $F_T$ can be related to $f_T$ in 
Eq.(\ref{Htensorial}) by using low energy theorems and PCAC hypothesis 
\cite{Pobl1}, \cite{belyaev}
\bea
F_T^0=\frac{2}{3}\frac{\chi \langle \mu\rangle}{f_\pi}\, f_T^0
\eea
where $\langle \mu\rangle=\langle 0|\bar{q}q|0\rangle $ is 
the vacuum expectation 
for the quark condensate and $\chi$ is defined by
\cite{ioffe}
\bea
\langle 0|\bar{q}\sigma_{\mu\nu}q|\epsilon(k)\rangle \, =\, e_q\chi\, 
\langle 0|\bar{q}q|0\rangle\, F_{\mu\nu}\, ,
\eea
with $e_q$ the quark $(q)$ electric charge,
$F_{\mu\nu}=i(k_{\mu} \epsilon_{\nu}-k_{\nu} \epsilon_{\mu})$, and
$\epsilon_{\mu}(k)$ the photon polarization vector of momentum $k$.
Then, the destructive interference
between the SM and tensorial amplitudes 
accounts for the correct number of  ``missing'' events observed 
at ISTRA
if $F_T=(5.6\pm 1.7)\times 10^{-3}$ \cite{Pobl1}, 
corresponding to a tensorial coupling
$f_T\, \simeq\, (1.4\pm 0.4)\times 10^{-2}$ \cite{belyaev}, where 
$\langle \mu\rangle \, =\, -(0.24 \, {\rm GeV})^3$ and 
$\chi \, =\, -(5.7 \pm 0.6)\, {\rm GeV}^{-2}$ values have been used
\cite{ioffe}. 
This result is consistent with the limit 
$f_T^0<0.095$ (at 68\% confidence level)  coming from beta decay \cite{Pobl1}.

Clearly, if confirmed, this effect would be a clear signal of new physics.
Indeed, in the SM,
the tensorial coupling $f^{SM}_T$ is very small, being 
generated at two-loop level and chiral suppressed.
In Ref.\cite{belyaev} the supersymmetric (SUSY) origin of $f_T$ 
has been analyzed.
The leading SUSY contribution to $f_T$, 
given by charginos and squarks exchanges in penguin and box diagrams,
can be larger than SM one since it is induced at one-loop. However, present
bounds on SUSY particle spectra do not allow $f_T$ to be larger than
$f_T \simeq 10^{-4}$, 
too small for the required value suggested in \cite{Pobl1}.
Moreover, there have been also criticisms about the consistency of
such large tensorial couplings.
In Ref.\cite{voloshin} it was pointed out that, due to QED corrections,
an $f_T$ of order of ${\cal O}(10^{-2})$ might run in troubles.
Indeed, the operator in Eq.(\ref{Htensorial}) 
can mix under QED radiative corrections with
a scalar operator, whose contribution is
strongly constrained by $\pi^+\to e^+ \nu_{e}$ \cite{voloshin}.
In particular, an upper bound on $f_T < 10^{-4}$ can be set 
by imposing the strong constraints on scalar interactions coming 
from $\pi^+\to e^+ \nu_{e}$,  
which is two order of magnitude smaller than the required one \cite{Pobl1}.
However, more accurate analyses showed that 
it is possible to relax or even avoid the upper bound claimed in 
\cite{voloshin}. For instance, the simultaneous (fine-tuned) contributions 
of both tensorial and scalar interactions, as suggested 
by lepto-quark models \cite{herc}, might relax the upper bound in 
\cite{voloshin} and thus the interpretation given in \cite{Pobl1} 
cannot be regarded yet as ruled out.
There is also an 
alternative solution, proposed in Ref.\cite{chiz1}, where a modified 
tensorial interaction can formally avoid the mixing 
with scalar operator, while solving the ISTRA discrepancy.

In Ref.\cite{EG}, 
it was pointed out that an analogous effect might show up
in the kaon sector. In particular, if the origin of 
$f_T$ is flavour independent, then a tensorial interaction of
the same order is also expected in the $\Delta S=1$ transitions, leaving
to a large anomaly in radiative kaon decays, easily 
detected at present and future kaon factories \cite{EG}.

Recently, the PIBETA collaboration at Paul Scherrer Institute facility, 
has performed an accurate analysis of 
the $\pi^+\to \nu_e e^+ \gamma$ decay \cite{pibeta}
using a stopped pion beam.
More than 40,000 $\pi^+\to \nu_e e^+ \gamma$
events have been collected, allowing for a very precise
measurement of the branching ratio. In this experiment, a 
more significant discrepancy (about $8\sigma$ \cite{chiz2}) 
between data and SM predictions has been reported in the
kinematical region of high-energy photon/low-energy positron.
A significant number of expected events are missing.
As for the ISTRA anomaly, agreement with data can be improved  by 
adding a negative tensor term $F_T\simeq -0.002$, a bit smaller 
(in magnitude) than the corresponding one in \cite{Pobl1}.
More detailed analysis about the PIBETA experiment can be found in 
\cite{pocanic} and references therein.

Now we analyze the impact of a tensorial coupling 
on the photon polarization asymmetry $d{\rm A}_{\gamma}/dx$
in radiative pion and kaon decays.
In order to simplify the analysis, we will assume an universal tensorial 
interaction in both $\Delta S=0,1$ processes, parametrizing
all the effects in a phenomenological coupling $F_T$ as follows
\bea
M_T=i\frac{e G_F}{\sqrt{2}} F_T\, V_{uq}
\epsilon^{\mu \star}_L\, q^{\nu}
\left[ \bar{e}\sigma_{\mu\nu}(1-\gamma_5)\nu_e\right]\, ,
\label{t_amplitude}
\eea
where $q=d$ and $q=s$ for pion and kaon decays respectively.
As discussed in section 1, the photon emitted by 
the tensorial amplitude $M_T$ in Eq.(\ref{t_amplitude}) is purely left-handed.
This property can also be checked by noticing that the interference 
between $M_T$ and $M_{\SD}$ terms is proportional to the $V-A$ combination
\cite{Pobl1}. Below we provide the additional tensorial contributions 
to the (photon) 
polarized Dalitz plot density. In particular, the following term
$\rho^{(-1)}_T(x,\lambda)$ should be added to
$\rho^{(-1)}(x,\lambda)$ in Eq.(17) 
\bea
\rho^{(-1)}_T(x,\lambda)=2\,A_{\SD} F_T\left(
F_T f_{\scriptscriptstyle TT}(x,\lambda)+ 
2\sqrt{r_l}\frac{f_M}{m_M}\, f_{\scriptscriptstyle IBT}(x,\lambda)+
\sqrt{r_l}\,(V-A) f_{\scriptscriptstyle SDT}(x,\lambda)
\right)
\label{rho_tens}
\eea
where \cite{EG}
\bea
f_{\scriptscriptstyle TT}(x,\lambda)&=&\lambda\, x^3(1-\lambda),~~~
f_{\scriptscriptstyle IBT}
(x,\lambda)=x\left(1+r_l-\lambda-\frac{r_l}{\lambda}\right),~~~
\nonumber\\
f_{\scriptscriptstyle SDT}
(x,\lambda)&=&x^3 \left(1-\lambda\right)\, .
\label{f_tens}
\eea
\begin{figure}[tpb]
\begin{center}
\dofigs{3.1in}{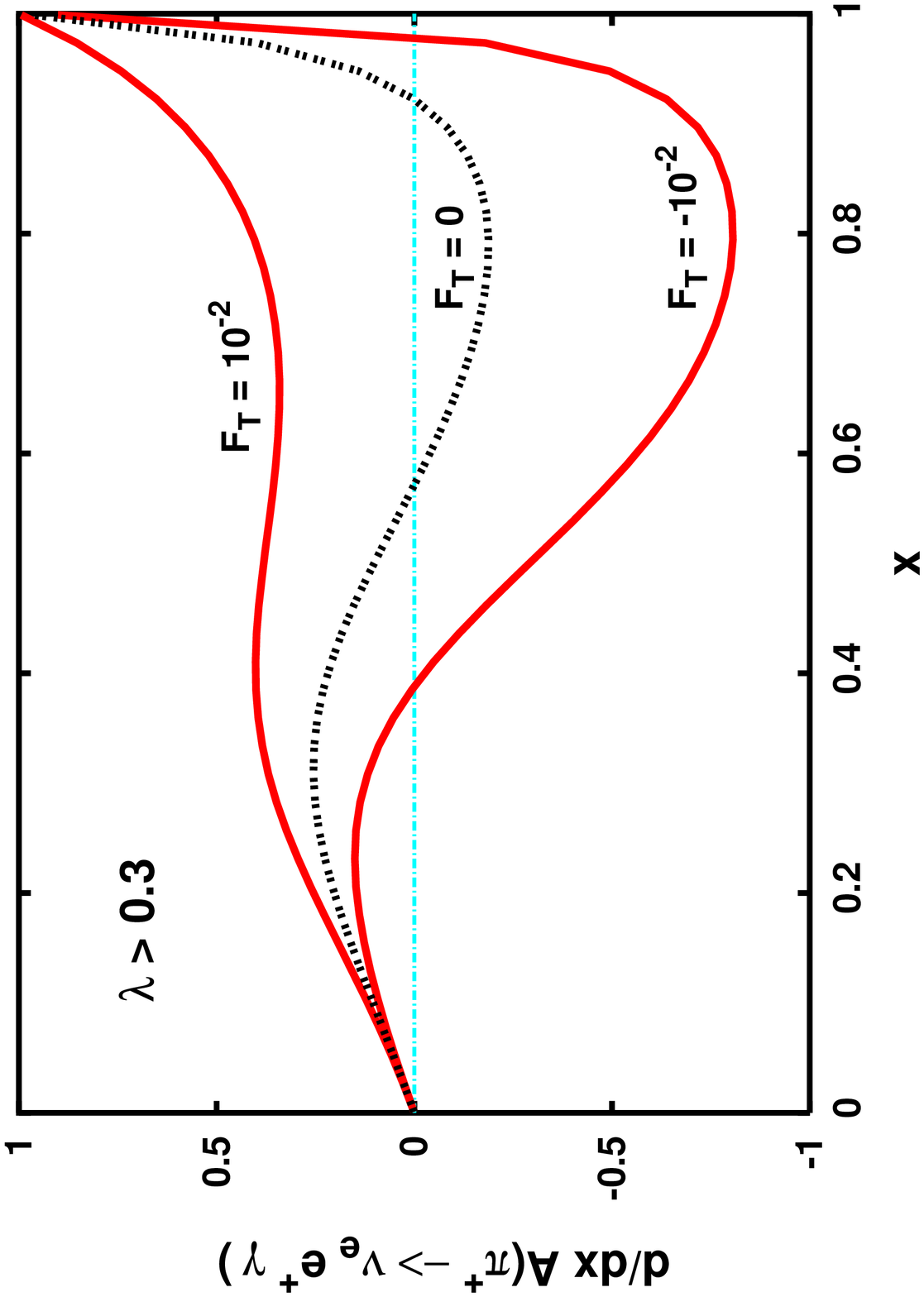}{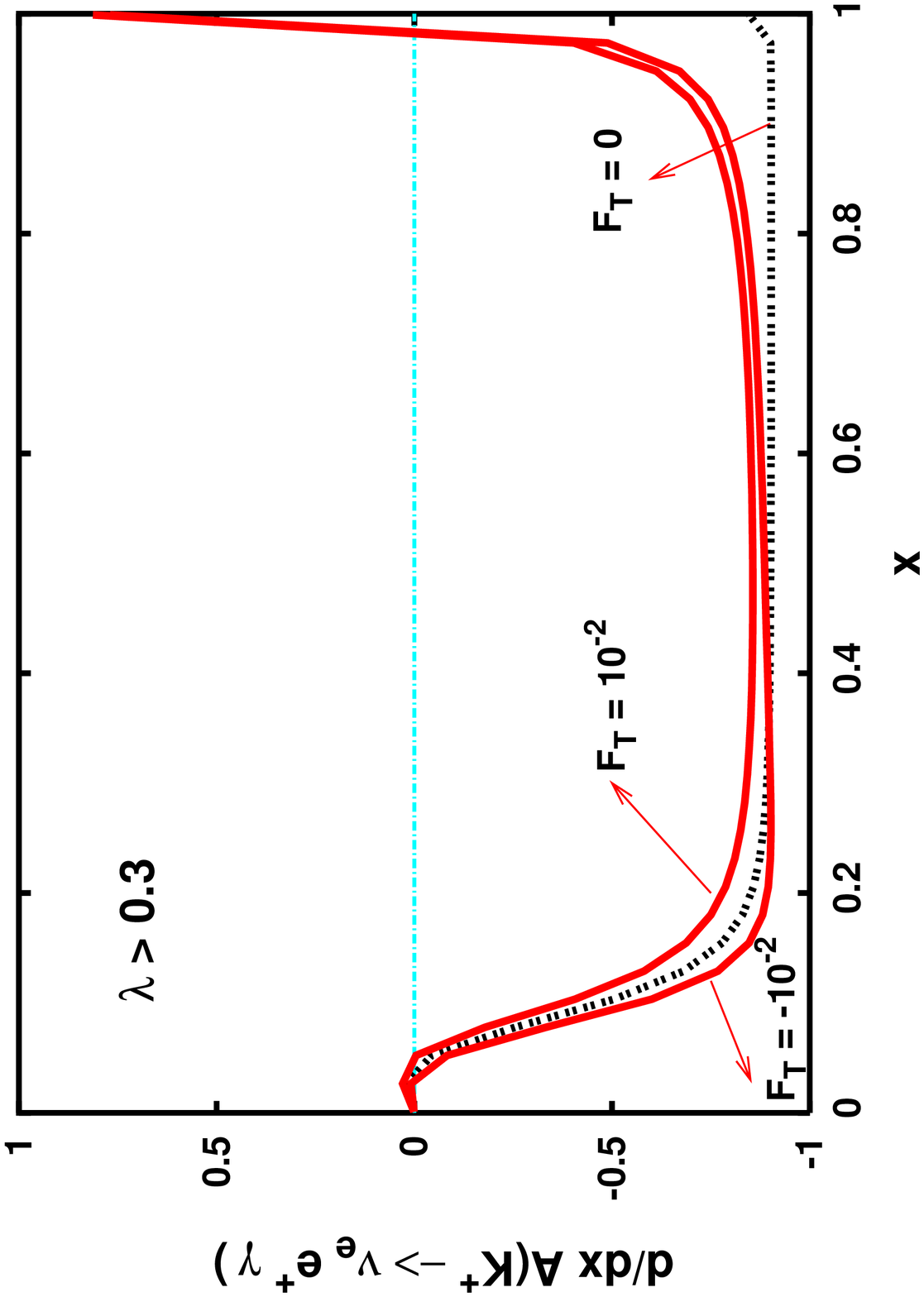}
\end{center}
\caption{\small Photon polarization asymmetry $d{\rm A}_{\gamma}/dx$ versus
$x$ and  with $\lambda>0.3$, for two values of tensorial coupling 
$F_T=\pm 10^{-2}$.
Left and right plots correspond to $\pi^+ \to \nu_e e^+ \gamma$ and 
$K^+ \to \nu_e e^+ \gamma$ decays respectively. The dark dashed curves 
stand for the standard model case ($F_T=0$).}
\label{tensor}
\end{figure}

In Fig.\ref{tensor} we show the $d{\rm A}_{\gamma}/dx$ asymmetry versus $x$,
for two representative values of $F_T=\pm 10^{-2}$ and 
cut $\lambda>0.3$, for the $\pi^+ \to \nu_e e^+ \gamma$ 
(left plot) and  $K^+ \to \nu_e e^+ \gamma$ (right plot).
 As we can 
seen from these results the shape of photon 
asymmetry is quite sensitive to a 
tensorial coupling in the range of $|F_T|\simeq 10^{-2}$. In particular, 
in the pion case, this sensitivity is more pronounced, and the position of
zeros of the asymmetry strongly depend on $F_T$. 
On the other hand, in the kaon decay,
large deviations should appear only in the region of large $x$, where
the tensorial effect is enhanced.
This difference in the two decays can be explained
due to the fact that the tensorial interference, which is the dominant
effect when $F_T<10^{-2}$, is always chiral suppressed,
being proportional to $m_e/m_M$. Thus, in the radiative kaon decay, 
this effect is more suppressed than in the pion case, 
due to the larger meson mass.
We have explicitly checked that, in the corresponding pion and kaon 
decays in muon channel,
the sensitivity of the asymmetry to $F_T$ is very modest and
we do not show the corresponding results.
In conclusion, we suggest that the possibility to measure the photon 
polarization in pion, or even in kaon decays, could be very useful to
clarify the controversial question of tensorial 

\section{Cancellation of mass singularities}
In this section we discuss the mechanism of mass singularities cancellation
and the way it takes place in meson and muon polarized radiative decays.  
As will be seen a new peculiar cancellation pattern shows up in the particular 
case of the polarized amplitudes differently from the well known cancellation 
taking place in the inclusive unpolarized amplitudes.

In a theory with massless particles a crucial test  of  the consistency of 
the computation is represented by the absence of mass singularities in any 
obtained physical quantity. 
Mass singularities are of two types:  infrared and  collinear.
Infrared divergences originate from massless particles with a 
vanishing momentum in the small energy soft limit. Physical states as, 
for example,  a single charged particle, are degenerate with states made 
by the same particle accompanied by soft photons. This corresponds to the 
impossibility of distinguishing a charged particle from the one accompanied 
by given number of soft photons due to the finite resolution of any 
experimental apparatus. An infrared divergence appears in 
QED when the energy $E_{\gamma}$ of the photon goes to zero 
as a factor of the form:
\bea
I=\int_{0}^1\frac{d\epsilon}{\epsilon}    
\eea
where $\epsilon=\frac{E_{\gamma}}{E}$ is the fraction of the energy of 
the photon with respect to the total available energy ${E}$ for the process.
The Bloch Nordsiek theorem \cite{bn} assures the cancellation of  infrared 
divergences in any inclusive cross section.
Collinear divergences, instead, come from massless particles 
having a vanishing value of the relative emission angle. 
In QED, specifically, when one or more photons, in the limit of zero 
fermion mass, are in a collinear configuration i.e. with emission 
angle $\theta \simeq 0$.   Physical states containing a massless charged 
particle are degenerate with states containing the same particle and a 
number of collinear photons. Any experimental apparatus, having a finite 
angular resolution, cannot distinguish between them. 
The angular separation of two massless particles with momenta $p$ and $k$ 
is such that  they move parallel to each other with a combined invariant 
mass for $\theta \rightarrow 0$:
\bea
q^2=(p+k)^2=2\,{p_0}\,{E_{\gamma}}(1-\cos\theta)\rightarrow\;0
\eea
even though neither $p$ nor $E_{\gamma}$ are soft. Here $\theta$ is the
emission angle of a photon with respect to the fermion. The inclusive 
procedure of integrating over the photon emission angles by keeping the 
fermion mass finite does not give rise to any collinear singularity. 
The divergence appears in the limit $\theta \rightarrow 0$ as the presence 
of a logarithm of the form $\log(\frac{E}{m})\simeq \log(\theta)$.

For collinear singularities, as well as for  infrared ones, the case 
for inclusive unpolarized  processes is well known and it is governed 
by the Kinoshita-Lee-Nauenberg (KLN) theorem \cite{kln}.
For the collinear singularities the KLN 
theorem guarantees that collinear divergences cancel out if 
one performs a sum of the amplitude over all the sets of 
degenerate states order by order in the perturbative expansion. 
For the amplitude of a single photon emission a combination of 
collinear and infrared singularities gives, for instance, 
contributions of the type:
\bea
R= \frac{\alpha}{\pi}\;\int_0^1 \; \frac{d\epsilon}{\epsilon} \;
\int_0^1\; \frac{d\theta}{1-\cos\theta} \, .
\eea

In order to discuss the above aspects on the cancellation of 
lepton mass singularity in the polarized pion decay, 
we first recall the mechanism taking place in the unpolarized case
\cite{bk,ms}.
In general, the decay rate is made free from mass singularities in the 
ordinary way: the  cancellation of divergences occurs in the total 
inclusive decay rate at  order ${\cal O}(\alpha)$, namely in the pion case
\bea
\Gamma^{\rm (incl)}=\Gamma(\pi\to \nu e)+
\Gamma(\pi \to \nu e \gamma)\, ,
\eea
when the full  ${\cal O}(\alpha)$ order contributions are included,
i.e. those relative to real and virtual photon emission \cite{bk,ms}.
However, for a pointlike (structureless) pion, 
due to the chirality flip of final charged lepton,
the pion decay amplitude is always proportional to $m_l$
and vanishes in the $m_l\to 0$ limit. In other words, in the 
limit $m_l\to 0$ the decay rate is made finite from mass singularities
in a trivial way. For example, as we will see later on,
a term proportional to ${\rm Log}(m_l)$ will remain 
in the inclusive width due to the mass renormalization of
the charged lepton in the virtual contributions to $\Gamma(\pi\to \nu e)$.
However, since it will be multiplied by $m_l^2$, it will give no troubles
since $\Gamma_0\to 0$ tends to zero at the same time.
However, as pointed out by Kinoshita in \cite{bk}, the leading
${\log}(m_l)$ terms in the IB contribution to $\Gamma(\pi\to \nu e\gamma)$
cancel out exactly when one adds the virtual contributions. In other words,
the mass singularities in the ${\log}(m_l)$ 
terms should cancel independently from the fact that the effective 
coupling in the pion decay is proportional to the charged lepton mass or not.
The cancellation mechanism of these Log terms
shows a non trivial aspect of the KLN theorem in the pion decay.
For this reason, in the following discussion we will consider
the following ratios 
$\Gamma(\pi\to \nu e \gamma)/\Gamma_0$ and $\Gamma(\pi\to \nu e )/\Gamma_0$
which survives the limit $m_l\to 0$.

Let us now consider the case of radiative polarized decays within the 
soft and collinear region for the radiated photon. We will investigate 
in this section the mechanism which will assure the finiteness of the 
lepton distribution against the appearance of infrared and collinear 
singularities on the above ratios of widths.
Let us start by the inclusive distributions in terms 
of the final lepton energy $y$ as listed in Eq.(\ref{dBRy}). 
The Inner Bremsstrahlung contribution contained in Eq.(\ref{IBr0}) 
in the $r_l\rightarrow0$ limit is composed by the four expressions 
corresponding to the various polarization states of the final photon 
and lepton respectively. The last, $RR$ polarized term is identically zero. 
The remaining three are related to the left-handed ( first and third ) 
and right-handed ( second ) lepton respectively. 

The logarithms $L_1,L_2$ do correspond to collinear contributions. 
By integrating the double-inclusive distribution of Eq.(\ref{dBRxy}) 
one gets in the expression of Eq.(\ref{FR}) for the $IB$ case 
that the expressions for $F^{(\lambda_{\gamma},\lambda_l)}_{i}(y)$ 
depend on the logarithms $L_1$ and $L_2$ respectively 
\bea
L_1=\log{\left(\frac{y+{\rm A}_l-2\,r_l}{y-{\rm A}_l-2\,r_l}\right)},~~~~~~~~
L_2=\log{\left(\frac{y+{\rm A}_l-2}{y-{\rm A}_l-2}\right)}
\nonumber
\eea
These terms do give rise to two kinds of ``collinear'' logarithms: 
\bea
L_1&=&\log\frac{E_l}{m_l}  ,~~~~~~~~ 
L_2=\log\frac{E_l+\sqrt{E_l^2+m_l^2}-m_M}{E-\sqrt{E_l^2+m_l^2}-m_M}\, .
\nonumber
\eea
The first logarithm represents the case of the photon being parallel to 
the lepton, the second collinear  logarithm for $m_l\rightarrow 0$ and 
$m_M\rightarrow 0$ corresponds to the case where the photon is 
parallel to the decaying meson \cite{tv}. Clearly, it is only $L_1$ which is
affected by the true collinear divergence in the limit $m_l\to 0$.

With respect to the unpolarized inclusive case some differences are 
worth to be noticed here:
\begin{itemize}
\item
Different polarization amplitudes do represent independent 
observables in the decay. Therefore if we consider the two 
cases of a right-handed and left-handed lepton they have to 
be also separately finite. 
\item
At zero order in the pion decay the angular momentum conservation 
imposes to the lepton to be left-handed. By radiating a photon a 
total zero angular momentum is assigned to the final state even 
if a right-handed lepton emits a left-handed polarized photon.  
This contribution is represented by the second term in Eq.(\ref{IBr0}).
\item
For $y\rightarrow 1$ only the second term in Eq.(\ref{IBr0}), 
corresponding to the $LR$ polarization, contribution is finite, 
i.e. it is zero: 
\bea
\lim_{{r_l\to 0}\;\;{y\to1}}\frac{1}{\Gamma_0}\frac{d\Gamma_{\IB}^{(\rm L,R)}}{dy}=
\frac{\alpha}{2\pi}\Big(1-y\Big)=0\, .
\nonumber
\eea
\end{itemize}
This fact shows that the $LR$ term, corresponding 
to the anomalous term \cite{anomaly}, it is finite by itself 
without the need of any cancellation mechanism in the 
infrared $y\rightarrow 1$ and collinear limit $r_l \rightarrow 0$. 
A detailed discussion of the undergoing dynamics can be found in 
Ref.\cite{tv}

Analogous conclusions, regarding the finiteness of 
the right-handed lepton contribution in the  
$r_l \rightarrow 0\, , y\rightarrow 1$ limits to the structure dependent terms
$|SD|^2$ and the $IB \times SD$, 
see Eqs.(\ref{SDr0}) and (\ref{INTr0}) respectively,
hold there as well.

Let us now consider the contributions of the type $LL$ and $RL$ 
giving rise to a left handed lepton.
Manifestly the first and third term of  Eq.(\ref{IBr0}) are 
divergent in the $r_l\rightarrow 0 \;,~ y \rightarrow 1$ limits. 
The expression obtained by adding first and third contributions 
in Eq.(\ref{IBr0}) is:
\bea
\frac{1}{\Gamma_0}\Big[\frac{d\Gamma_{\IB}^{(\rm L,L)}}{dy}+
 \frac{d\Gamma_{\IB}^{(\rm R,L)}}{dy}\Big]
=
\frac{\alpha}{2\pi}\,\frac{1}{y-1}\Big[ \Big(1+y-{\hat L}_1 -{\hat L}_2\Big)
+
\Big(y\left(y+1\right)-
{\hat L}_1\,y^2
+ {\hat L}_2\left(1-2\,y\right) \nonumber
\Big)\Big]
\eea
which is divergent both in the collinear and in the infrared limit. 
The coefficients of the collinear logarithms remain different 
from zero as $r_l\rightarrow 0$ and $y\rightarrow 1$, leaving to a
divergent expression.
As for the unpolarized case for the left-handed lepton contributions 
one needs to consider the additional virtual contributions in order 
to cancel infrared singularities \cite{bk}.

The case of the left-handed lepton includes also the diagram of 
the virtual photon i.e. the one with a photon  line connecting  
meson and charged lepton. This diagram does not add any angular 
momentum to the zeroth order term since a virtual particle does 
not add angular momentum to the final state. The amplitude containing 
the virtual photon gives rise, therefore, to a  lepton neutrino final state
having the same helicities as the ones of the tree level amplitude. 
The combination of real and virtual contributions should, in the 
left-handed lepton channel, add among each other to give a finite result.  
This mechanism is the same taking place for the cancellation of singularities 
for the inclusive, unpolarized amplitudes as we will discuss in more details
below.

According to \cite{bk} the total width for the unpolarized 
IB contribution to $\Gamma(\pi \to \nu e \gamma)$ is given by
\bea
\frac{\Gamma_{IB}(x_0)}{\Gamma_0}&=&\frac{\alpha}{\pi}
\Big\{b(r_l)\left(\log{\frac{x_0}{2}}-\log(1-r_l)-\frac{1}{4}
\log(r_l)+\frac{3}{4}\right)
\nonumber\\
&-&\frac{r_l\left(10-7r_l\right)}{4\left(1-r_l\right)^2}\log(r_l)
+\frac{2\left(1+r_l\right)}{1-r_l}L(1-r_l)+\frac{15-21r_l}{7(1-r_l)}\Big\}\, ,
\label{IBtot}
\eea
where $x_0$ is the minimum photon energy which 
regularizes the infrared divergence in the photon mass, 
the function $b(x)=\frac{1+x}{(1-x)}\log(x)+2$, and 
$L(x)=\int_0^x\log(1-t) dt/t$.
For a generalization of the result in Eq.(\ref{IBtot}) to the 
inclusion of the leading logarithmic terms
to all orders in perturbation theory see Ref.\cite{k,kf,nt}.

For the virtual 1-loop contribution one has to consider the radiative 
corrections to the operator $gm_l^{0}Q$, where 
$Q=\bar{\psi}_{l}(1-\gamma_5)\psi_{\nu}\varphi_{\pi}$, with
$\varphi_{\pi}$ is the pion field and $m_l^{0}$ is the `bare' mass of the
charged lepton. These corrections 
split in two separate contributions: $\Gamma^{(1)}$ given by the 
correction to the operator $Q$
and  $\Gamma^{(2)}$ arising when one try to express 
the bare mass $m_l^{0}$ in terms of the renormalized lepton mass $m_l$, namely 
$m_l^0=m_l-\delta m_l$ with 
$\delta m_l=\frac{3\alpha}{2\pi}m_l\left(\log(\Lambda/m_l)+\frac{1}{4}\right)$
\cite{ms}.
For $\Gamma^{(1)}$, one has \cite{bk}
\bea
\frac{\Gamma^{(1)}}{\Gamma_0}=\frac{\alpha}{\pi}\,\left\{
-b(r_l)\left(\log(\frac{x_0}{2})-\frac{1}{4}\log(r_l)+\frac{3}{4}\right)
+\frac{r_l}{2(1-r_l)}\log(r)+\frac{1}{2}\right\}
+\frac{3\alpha}{2\pi}\log(\frac{\Lambda}{m_{\pi}})
\label{Gamma1}
\eea
Notice that the last term, containing the ultraviolet cut-off $\Lambda$
needed to regularize the UV divergency,  can in principle
be absorbed in a re-definition of $f_{\pi}$ at order $\alpha$, 
see Ref.\cite{ms} for more details.

As can be seen by comparing the results in
Eqs.(\ref{IBtot}) and (\ref{Gamma1}), the
$\log{r_l}$ terms surviving the limit $r_l\to 0$ cancel out in the sum
of virtual and real emission contributions as a consequence of the KNL theorem.
Finally, for the total contribution to the unpolarized inclusive decay rate 
at order $\alpha$, including the contribution of $\Gamma^{(2)}$,  
one gets 
\cite{bk,ms}:
\bea
\frac{\Gamma^{(\rm incl)}}{\Gamma_0}=
1+\frac{\alpha}{\pi}\, \Big\{
\frac{3}{2}\log(r_l)+\frac{13}{8}-\frac{\pi^2}{3}\, \Big\}\, ,
\label{Gammainc}
\eea
where we retained only the leading terms in $m_e\to 0$ limit.
As previously mentioned, the appearance of
the $\log{r_l}$ term in (\ref{Gammainc}) is 
due to the renormalization of the charged lepton mass which 
does not follow the same pattern of collinear mass singularities 
discussed above
\cite{bk}.
For simplicity, we omitted in (\ref{Gammainc}) 
the term containing a  $\log(\Lambda/m_{\pi})$,
since it can be absorbed into a re-definition of $f_{\pi}$ 
at order $\alpha$ inside $\Gamma_0$.

As stated above,  in the right-handed case, on the contrary,
the mass singularities cancellation occurs with a different mechanism.  
Infrared and collinear limits in the ratio $\Gamma_{IB}^{(LR)}/\Gamma_0$
give separately a finite result. 
In particular, the coefficient of the collinear logarithms for 
the right handed lepton case is the lepton mass, instead of the 
usual correction factor coming from the soft and the virtual photon 
contributions.

The particular cancellation mechanism occurring in the right-handed 
radiative decay is originated by the combined constraints of the angular 
momentum conservation in the pion vertex and the one of the helicity 
flip in the photon-lepton vertex \cite{tv}.

Let us now consider the case of the muon decay. As shown in 
Eq.(\ref{integrated}) as for the meson case also in the muon 
decay lepton distribution we see that the $LR$-photon-lepton 
polarized distribution is free from collinear and infrared singularities 
and goes to zero in the infrared limit $y\rightarrow 1$.
The remaining $RL$ and $LL$ distributions, apart from the identically 
zero $RR$ term, do give a finite contributions in the $y\rightarrow 1$ 
limit, provided that the same $x_0$ cut-off is also set free to go to 
the soft kinematical limit i.e. $x_0 \rightarrow 0$. In the muon case 
the pattern of  singularities cancellation repeats itself as for the 
meson case.

\section{Conclusions}

We have computed polarized distributions in radiative 
meson and muon decays, by taking
into account final lepton and photon helicity degrees of freedom.
The definition of photon polarization asymmetry
has been introduced,
allowing a new approach to investigate interaction dynamics via a finite
and universal quantity directly associated to parity violation.
Analytical and numerical results 
for the polarized distributions and branching ratios, as well as 
for the photon polarization asymmetry, have been explicitly derived.

The main results of the photon polarization analysis in meson decays,
inclusive in the spin degrees of freedom of the final lepton, 
can be summarized as follows.
In the pion case, the production of  
{\it hard} photons in association with {\it soft} positrons,
are mainly favored to be left-handed polarized.
However, when the positron energy increases,
the relative gap between left-and right-handed photons 
decreases, due to the increasing contributions 
of hadronic structure dependent terms.
Remarkably, in the kaon decay,
when energy cuts $E_{\gamma} \gsim 25$ MeV and $E_{e^+} \gsim 120$ MeV are
imposed, both photon and positron are mainly right-handed polarized
and a large and negative photon polarization asymmetry is expected.
Regarding the corresponding meson decays in muon channel, 
the left-handed photon production always gives the leading effect.
The same behavior is observed in the radiative muon decay $\mu^-\to \nu_{\mu} e^- 
\bar{\nu}_e\, \gamma$.
All these results can be easily explained in terms
of angular momentum conservation and parity violation.

We have also systematically analyzed the mechanisms of cancellation of infrared and 
collinear divergences in polarized meson and muon decays. 
It has been shown that the finiteness of the polarized amplitudes  takes place
in a different way for  left-  with respect to right-handed final leptons
when inclusive results in the photon polarization 
degrees of freedom are taken into account.

Finally, we propose a possible test using photon polarization 
in order to solve the
controversial issue of large tensorial couplings in lepton-quark interactions, 
as suggested by the recent observed anomaly at the PIBETA experiment.
In particular, it is argued that the measurement of 
the photon polarization asymmetry
may constitute a sensible test to resolve such 
controversial issue in radiative pion decay, providing
a sensitive probe to hadronic form factors as well as to
new physics effects in meson radiative decays.

We believe that all these new results could open a more extended
perspective into the physics of the semileptonic weak decays.
In particular, the less inclusive approach to the polarized processes, 
by explicitly taking into account lepton and photon 
polarization degrees of freedom, could allow, when the experimental 
conditions make it compatible, a new quantitative approach and a 
more detailed inspection of meson and muon decays.
Moreover, we are confident that Standard Model physics as 
well as signals of physics beyond the Standard Model could be 
put under scrutiny and more closely investigated by using 
tests involving polarized quantities.

\section*{Acknowledgments}
We acknowledge a useful exchange of mail messages with L.M. Sehgal.
We wish to thank CERN TH-PH Department for the warm hospitality extended to
us during the preparation of this work.
E.G. would like to thank the Academy of Finland for partial financial support
(Project numbers 104368 and 54023).

\newpage
\section*{Appendix A}

Here we provide the most general results for  
the polarized meson radiative decay in the meson rest frame,
as a function of $x,y$ variables defined in Eq.(\ref{variables}), and 
for both charged lepton $l=e,\mu$ 
and photon helicities $\lambda_l$ and $\lambda_{\gamma}$
respectively. For later convenience, we will 
use the same notation adopted in section 2, where
the symbols $L$ and $R$ are associated to particle helicities 
$\lambda=-1$ and $\lambda=1$ respectively.
In particular, for the differential radiative decay rate 
normalized to its non radiative one, we obtain
\bea
\frac{1}{\Gamma_0}\frac{d^2 \Gamma^{(\lambda_{\gamma},\lambda_l)}}{dx\, dy}=
\frac{\alpha}{2\pi}\frac{1}{(1-r_l)^2}\,
\rho^{(\lambda_{\gamma},\lambda_l)}(x,y)
\label{dBRxy}
\eea
where 
\bea
\rho^{(\rm L,\lambda_l)}(x,y)&=&
f_{\IB}^{(\rm L,\lambda_l)}(x,y)
+\frac{m_M^2}{f_M^2}\frac{(V-A)^2}{4\,r_l} f_{\SD}^{(\rm L,\lambda_l)}(x,y)
+\frac{m_M}{f_M}(V-A)\, f_{\INT}^{(\rm L,\lambda_l)}(x,y)~~~ 
\nonumber\\
\rho^{(\rm R,\lambda_l)}(x,y)&=&
f_{\IB}^{(\rm R,\lambda_l)}(x,y)
+\frac{m_M^2}{f_M^2}\frac{(V+A)^2}{4\,r_l} f_{\SD}^{(\rm R,\lambda_l)}(x,y)
+\frac{m_M}{f_M}(V+A)\, f_{\INT}^{(\rm R,\lambda_l)}(x,y)~~~  
\nonumber
\eea
where the functions 
$f_{IB}^{(\lambda_{\gamma},\lambda_l)}(x,y)$,
$f_{SD}^{(\lambda_{\gamma},\lambda_l)}(x,y)$, and
$f_{INT}^{(\lambda_{\gamma},\lambda_l)}(x,y)$ are given by\footnote{
The symbol ${\rm A}_l=\sqrt{y^2-4r_l^2}$ appearing
below should not be confused with the axial form factor $A$.}

\bea
f_{\IB}^{(L,\lambda_l)}(x,y)&=&
\frac{ 1 - y +r_l }{2\,{\rm A}_l\,{x^2}\,{z^2}}
\,\Big\{
{\rm A}_l\,\left(\, x + y -1 + 
          r_l\,\left(\,-3x - y + x\,y \right)
+ {r_l^2}
 \right)
\nonumber\\
&-& 
 \lambda_l
\left[
       y\,\left(x + y -1\right)  + 
       r_l\,\left( 4 - 6\,x - 6\,y + 3\,x\,y + {y^2} - 
          x\,{y^2} \right)    
\right.
\nonumber\\
&+& \left.
{r_l^2}\,\left( 4 + 2\,x - y \right)
         \right]\Big\}
\nonumber\\
\nonumber\\
f_{\IB}^{(R,\lambda_l)}(x,y)&=&
\frac{1-y+r_l}{{2\,{\rm A}_l\,{x^2}\,z^2}}\,
     \Big\{{\rm A}_l\,\left( x -1 +r_l\right)  
       -\lambda_l\Big[\left( x -1 \right) \,y - 
       r_l\,\left(2\,x + y -4\right) \Big]\Big\}
\nonumber\\
&\times& \Big\{\left(x -1 \right) \,
        \left( x + y -1\right) +r_l \Big\} 
\nonumber\\
\nonumber\\
f_{\SD}^{(L,\lambda_l)}(x,y)&=&
\frac{1-y+r_l}{2{\rm A}_l} \,
     \Big\{ {\rm A}_l\,\left( 
          \left( x - 1\right) \,\left( y - 1 \right) +r_l
          \right) 
\nonumber\\
 &-&\lambda_l\Big[ 
       y\,\left( x + y -1 - x\,y \right)
+ r_l\,\left( 2\,x - y \right) 
\Big]
\Big\}
\nonumber\\
\nonumber\\
f_{\SD}^{(R,\lambda_l)}(x,y)&=& \frac{(x-1)(x+y-1)+r_l}{2{\rm A}_l}\, 
\Big\{
{\rm A}_l\left(1-x-y+r_l\right)
\nonumber\\
&-&\lambda_l\Big[
y\left(x+y-1\right)-r_l\left(2x+y\right)\Big]\Big\}
\nonumber\\
\nonumber\\
f_{\INT}^{(L,\lambda_l)}(x,y)&=&
\frac{1 - y +r_l }{2\,{\rm A}_l\,xz}\,
\Big\{{\rm A}_l\,\left( 1 - 2\,x - y + x\,y +r_l \right)
\nonumber\\
&-&\lambda_l\Big[2 - 2\,x   - 
       3\,y + 2\,x\,y + {y^2} - x\,{y^2} + r_l\,\left( 2 + 2\,x - y \right)
         \Big]
\Big\}
\nonumber\\
\nonumber\\
f_{\INT}^{(R,\lambda_l)}(x,y)&=&
\frac{1 - y +r_l }{2\,{\rm A}_l\,xz}\,
     \Big\{{\rm A}_l + \lambda_l\Big[2\,x + y -2\Big] \Big\} \,
     \Big\{\left(1-x\right) \,
        \left( x + y -1 \right)  -r_l \Big\}
\eea
where $\lambda_l=1$ and $\lambda_l=-1$ correspond to right- (R) and 
left-handed (L) fermion polarization respectively, and symbols
$\Gamma_0=\Gamma(M^+\to \nu_l l^+)$, $z=x+y-1-r_l$,
and ${\rm A}_l=\sqrt{y^2-4r_l}$.

By integrating equations above in the photon energy $x$ range
\bea
1-\frac{1}{2}\left(y+{\rm A}_l\right)\, \leq \, &x&\, \leq \, 
1-\frac{1}{2}\left(y-{\rm A}_l\right)
\nonumber\\
2\sqrt{r_l} \,\leq &y&\, \leq\, 1+r_l\, ,
\eea
we obtain 
\bea
\frac{1}{\Gamma_0}\frac{{d\Gamma}^{(\lambda_{\gamma},\lambda_l)}}{dy}&=&
\frac{\alpha}{2\pi}\frac{1}{(1-r_l)^2}
F^{(\lambda_{\gamma},\lambda_l)}(y)
\label{dBRy}
\eea
where 
\bea
F^{(L,\lambda_l)}(y)&=&
\frac{F^{(L,\lambda_l)}_{\IB}(y)}{2{\rm A}_l\left(1-y+r_l\right)}
+\frac{m_M^2}{f_M^2}\frac{(V-A)^2}{4\,r_l}\,F^{(L,\lambda_l)}_{\SD}(y)
+\frac{m_M}{f_M}\, (V-A)\,\frac{F^{(L,\lambda_l)}_{\INT}(y)}{2\,{\rm A}_l}
\nonumber\\
F^{(R,\lambda_l)}(y)&=&
\frac{F^{(R,\lambda_l)}_{\IB}(y)}{2{\rm A}_l\left(1-y+r_l\right)}
+\frac{m_M^2}{f_M^2}\frac{(V+A)^2}{4\,r_l}\,F^{(R,\lambda_l)}_{\SD}(y)
+\frac{m_M}{f_M}\, (V+A)\,\frac{F^{(R,\lambda_l)}_{\INT}(y)}{2\,{\rm A}_l}
\nonumber
\eea
where the functions $F^{(\lambda_{\gamma},\lambda_l)}_{i}(y)$ are given by
\bea
F^{\rm (L,\lambda_l)}_{\IB}(y)
&=& {\rm A}_l\,\Big\{\left(L_2+L_2\right)\,\left( 1 + r\,(1 - y) \right)  +
 {\rm A}_l\, \left( y - 3 + r \right)\Big\} 
\nonumber\\
&+&\lambda_l\,\Big\{
\left(L_1+L_2\right)\,\left( 2\,{r^2} - y + 
     r\,\left( 2 + y - {y^2} \right)  \right)  + 
 {\rm A}_l\, \Big( 2 
+ r\,\left(y -6\right)  - y + {y^2} \Big)
\Big\}
\nonumber\\
\nonumber\\
F^{\rm (R,\lambda_l)}_{\IB}(y)
&=&
{\rm A}_l\,\Big\{ 
    L_1\,\left( r + {r^2} - 3\,r\,y + {y^2} \right)
 -\left( L_2\,
       \left( 1 + r - 2\,y + r\,y \right)  \right)
\nonumber\\
&-& {\rm A}_l\,\left( 1 - 3\,r + y \right) \Big\}+\lambda_l\,
\Big\{
L_1\,\left( 2\,{r^2} + 2\,{r^3} + r\,y - 
     5\,{r^2}\,y + 3\,r\,{y^2} - {y^3} \right)  
\nonumber\\
&+& 
L_2\,\left( y - 2\,{y^2} - 
     r\,\left( 2 - 7\,y + {y^2} \right)  -2\,{r^2}  \right)  + {\rm A}_l\,
  \left(y + {y^2} -6\,r + 2\,{r^2} - r\,y \right)
\Big\}
\nonumber\\
\nonumber\\
F^{\rm (L,\lambda_l)}_{\SD}(y)
&=&
\frac{{\rm A}_l}{4}\,\Big\{
{\left(y -1 \right) }^2\,y - 
       r\,\left(y + {y^2} -2 \right)+ 2\,{r^2}  \Big\}
+\frac{\lambda_l}{4}\,\Big\{
\left(y -1 \right) \,
     \Big(  \left(y -1 \right) \,{y^2} 
\nonumber\\
&-& 
       r\,\left( 4\,y + {y^2} -4 \right) + 4\,{r^2}  \Big)
\Big\}
\nonumber\\
\nonumber\\
F^{\rm (R,\lambda_l)}_{\SD}(y)
&=&
\frac{{\rm A}_l}{24}\Big({y^3} - 
       2\,r\,y\,\left( 2 + y \right)  +
8\,{r^2} \Big) +\frac{\lambda_l}{24}\,\Big(y^2-4\,r\Big)^2
\nonumber\\
\nonumber\\
F^{\rm (L,\lambda_l)}_{\INT}(y)
&=& (1-y+r)\Big\{
{\rm A}_l\, \Big(L_2+L_1\,(y-1)\Big)
+\lambda_l\,(L_2 (y-2)+L_1\,((y-1) y -2\,r))
\Big\}
\nonumber\\
\nonumber\\
F^{\rm (R,\lambda_l)}_{\INT}(y)
&=&
-\left( 1 - y +r\right)\Big\{
{\rm A}_l\,\left( {\rm A}_l + L_2 + L_1\,r \right)\,+
\lambda_l\,
\Big( 2\,{\rm A}_l\,\left( r -1  \right)  + 
      L_1\,r\,\left( 2\,r - y \right)  
\nonumber\\
&+& 
      L_2\,\left(y -2\right)  \Big) \,
\Big\}
\label{FR}
\eea
where 
\bea
L_1&=&\log{\left(\frac{y+{\rm A}_l-2\,r_l}{y-{\rm A}_l-2\,r_l}\right)},~~~~~~~~
L_2=\log{\left(\frac{y+{\rm A}_l-2}{y-{\rm A}_l-2}\right)}
\eea
Now we expand the formulas above in the large electron energy
region $y\gg \sqrt{r}$. By retaining only 
the leading terms in $r_l$ expansion we obtain
\begin{itemize}
\item {\bf $\bf |\bf IB|^2$ contribution}
\bea
\lim_{r_l\to 0}\,
\frac{1}{\Gamma_0}\frac{d\Gamma_{\IB}^{(\rm L,L)}}{dy}&=&
\frac{\alpha}{2\pi}\,\frac{1}{y-1}\Big(1+y-{\hat L}_1 -{\hat L}_2\Big)
\nonumber \\
\lim_{r_l\to 0}\,
\frac{1}{\Gamma_0}\frac{d\Gamma_{\IB}^{(\rm L,R)}}{dy}&=&
\frac{\alpha}{2\pi}\Big(1-y\Big)
\nonumber\\
\lim_{r_l\to 0}\,
\frac{1}{\Gamma_0}\frac{d\Gamma_{\IB}^{(\rm R,L)}}{dy}&=&
\frac{\alpha}{2\pi}\,\frac{1}{y-1}\Big(y\left(y+1\right)-
{\hat L}_1\,y^2
+ {\hat L}_2\left(1-2\,y\right)
\Big)
\nonumber\\
\lim_{r_l\to 0}\,
\frac{1}{\Gamma_0}\frac{d\Gamma_{IB}^{(\rm R,R)}}{dy}&=&0
\label{IBr0}
\eea
\item {\bf $\bf |\bf SD|^2$ contribution}
\bea
\lim_{r_l\to 0}\,
\frac{1}{\Gamma_0}\frac{d\Gamma_{\SD}^{(\rm L,L)}}{dy}&=&0
\nonumber\\
\lim_{r_l\to 0}\,
\frac{1}{\Gamma_0}\frac{d\Gamma_{\SD}^{(\rm L,R)}}{dy}&=&
\frac{\alpha}{2\pi}\,\frac{m_M^2}{f_M^2}\,
\frac{(V-A)^2}{8\,r_l}\,y^2\,(1-y)^2
\nonumber\\
\lim_{r_l\to 0}\,
\frac{1}{\Gamma_0}\frac{d\Gamma_{\SD}^{(\rm R,L)}}{dy}&=&
0
\nonumber\\
\lim_{r_l\to 0}\,
\frac{1}{\Gamma_0}\frac{d\Gamma_{\SD}^{(\rm R,R)}}{dy}&=&
\frac{\alpha}{2\pi}\, \frac{m_M^2}{f_M^2}\,
\frac{(V+A)^2}{48\,r_l}\,y^4
\label{SDr0}
\eea
\item {\bf $\bf IB\times SD + c.c.$ contribution}
\bea
\lim_{r_l\to 0}\,
\frac{1}{\Gamma_0}\frac{d\Gamma_{\INT}^{(\rm L,L)}}{dy}&=&
\frac{\alpha}{2\pi}\,\frac{m_M}{f_M}\,
(V-A)\,\hat{L}_2\,\frac{1-y}{y}
\nonumber\\
\lim_{r_l\to 0}\,
\frac{1}{\Gamma_0}\frac{d\Gamma_{\INT}^{(\rm L,R)}}{dy}&=&
-\frac{\alpha}{2\pi}\,
\frac{m_M}{f_M}\,
(V-A)\,\Big(\hat{L}_2+\hat{L}_1\,y\Big)\,\frac{\left(1-y\right)^2}{y}
\nonumber\\
\lim_{r_l\to 0}\,
\frac{1}{\Gamma_0}\frac{d\Gamma_{\INT}^{(\rm R,L)}}{dy}&=&
\frac{\alpha}{2\pi}\,
\frac{m_M}{f_M}\,
(V+A)\,\Big(2\,\hat{L}_2+y\,(2+y)\Big)\,\frac{y-1}{2\,y}
\nonumber\\
\lim_{r_l\to 0}\,
\frac{1}{\Gamma_0}\frac{d\Gamma_{\INT}^{(\rm R,R)}}{dy}&=&
\frac{\alpha}{2\pi}\,\frac{m_M}{f_M}\,
(V+A)\,\Big(2\,\hat{L}_2\,(y-1)+y\,(y-2)\Big)\,\frac{y-1}{2\,y}
\label{INTr0}
\eea
\end{itemize}
where ${\hat L}_{1,2}=\lim_{r_l\to 0}\, \left(L_{1,2}\right)$ and so
\bea
{\hat L}_1&=&2\log{y}-\log{(1-y)}-\log{r_l}\, +\,{\cal O}(r)
\nonumber\\
{\hat L}_2&=&\log{(1-y)}\,+\,{\cal O}(r) \, .
\eea
Notice that the lepton mass inside $L_1$ is needed 
in order to regularize the collinear divergences. 
The above results in Eqs.(\ref{IBr0})-(\ref{INTr0})
are obtained in the approximation
$r/y^2\ll 1$ and are not valid near the region of minimum $y\simeq \sqrt{r}$.
Nevertheless, there is always a real 
infrared singularity in the photon energy spectrum, which is present
in the terms $\log(1-y)$ when $y\to 1$ 
even if the electron mass is taken into account,
corresponding to the known soft photon singularity.
This divergent term for $y\to 1$ is necessary in order to cancel the
infrared singularity appearing in the one-loop corrections
of non-radiative decay, as required by the KLN theorem.
Details of the cancellation mechanism for polarized decays 
are reported in section 7.
\newpage
\section*{Appendix B}
In this appendix we provide  the expressions for the 
basic functions $g_{0,1}$, 
$\bar{g}_{0,1}$,  $G_{10,1}$,  and $\bar{G}_{0,1}$ appearing in section 2
for the differential decay rates of radiative muon decay.
In the muon rest frame, the differential decay width is given in 
Eq.(\ref{BRmuon}) 
\bea
\frac{1}{\Gamma_0}
\frac{d \Gamma^{(\lambda_{\gamma}\,,\,\lambda_e)}}{dx\,dy\,dcos{\theta}}=
\frac{\alpha}{8\pi}\,\frac{1}{x\,z^2}\,
\Big[A_e\, (g_0+\lambda_{\gamma}\, \bar{g}_0)+\lambda_e\,
(g_1+\lambda_{\gamma}\, \bar{g}_1)\Big]
\eea
where $z=\frac{x}{2}(y-{\rm A}_e\, \cos{\theta})$, ${\rm A}_e=\sqrt{y^2-4r}$,
and the functions $g_{0,1}$ and $\bar{g}_{0,1}$ are
\bea
g_0&=&
  z\,\Big\{ -2\,{x^4} + {x^3}\,
      \left( 3 - 6\,y + 2\,z \right)  - 
     2\,{x^2}\,\left( -3\,y + 4\,{y^2} - z - 4\,y\,z + 
        {z^2} \right)  
\nonumber\\
&+& 2\,z\,
      \left( -3\,y + 2\,{y^2} + 3\,z - 4\,y\,z + 2\,{z^2}
        \right)  + x\,\Big( 6\,{y^2} - 4\,{y^3} - 6\,z + 
        2\,y\,z + 8\,{y^2}\,z 
\nonumber\\
&-& 5\,{z^2} - 6\,y\,{z^2} + 
        2\,{z^3} \Big)  \Big\}\,
+
  r\,\Big\{ 4\,{x^4} + {x^3}\,
      \left( -6 + 8\,y - 5\,z \right)  + 
     2\,{z^2}\,\left( 4 - 3\,y + 3\,z \right) 
\nonumber\\
&+&     2\,{x^2}\,\left( -3\,y + 2\,{y^2} + 3\,z - y\,z + 
        {z^2} \right)  + x\,z\,
      \left( -8\,y + 6\,{y^2} - 6\,z - 6\,y\,z + 3\,{z^2}
        \right)  \Big\}\,
\nonumber\\
&+&
2\,{r^2}\,{x^2}\,\Big( 4 - 3\,x - 3\,y + 3\,z \Big) 
\nonumber\\
\nonumber\\
\bar{g}_0&=&
   z\,\Big\{2\,{x^4} + {x^3}\,
      \left( -3 + 6\,y - 2\,z \right)  + 
     2\,{z^2}\,\left( 1 - 2\,y + 2\,z \right)  
\nonumber\\
 &+&    {x^2}\,\left( -6\,y + 4\,{y^2} - 2\,z - 2\,{z^2} \right)
+       x\,z\,\left( 6 - 6\,y + 4\,{y^2} + z - 6\,y\,z + 
        2\,{z^2} \right)  \Big\}
\nonumber\\
 &-& 
  r\,\Big\{ 4\,{x^4} + x\,\left( -2 + 6\,y - 3\,z \right) \,
      {z^2} - 6\,{z^3} 
- 2\,{x^2}\,z\,
      \left( 1 - 3\,y + z \right)  
\nonumber\\
&+& 
     {x^3}\,\left( -6 + 4\,y + z \right)  \Big\}
+2\,{r^2}\,{x^2}\,\left( x + 3\,z \right)
\nonumber\\
\nonumber\\
g_{1}&=&
  y\,z\,\Big\{ 2\,{x^4} + 
     {x^3}\,\left( -3+ 6\,y - 2\,z\right)  + 
     2\,{x^2}\,\left( -3\,y + 4\,{y^2} - z - 4\,y\,z + 
        {z^2} \right)  
\nonumber\\
&-& 2\,z\,
      \Big( -3\,y 
+ 2\,{y^2} + 3\,z - 4\,y\,z + 2\,{z^2}
        \Big)  
\nonumber\\
&+& x\,\left(  -6\,{y^2} + 4\,{y^3} + 6\,z - 
        2\,y\,z - 8\,{y^2}\,z + 5\,{z^2} + 6\,y\,{z^2} - 
        2\,{z^3} \right)  \Big\}  
\nonumber\\
&+& 
  r\,\Big\{ -4\,{x^5} + {x^4}\,
      \left( 6 - 8\,y + 4\,z \right)  + 
     2\,{z^2}\,\left( -12 + 8\,y + {y^2} - 8\,z - y\,z \right)
\nonumber\\
&+& {x^3}\,\left( 6\,y - 8\,{y^2} - 4\,z + 7\,y\,z - 
        4\,{z^2} \right)  + 
     x\,z\,\Big( 24\,y - 16\,{y^2} - 2\,{y^3} + 16\,z + 
        18\,y\,z 
\nonumber\\
&+& 2\,{y^2}\,z - y\,{z^2} \Big)  + 
     {x^2}\,\left( 6\,{y^2} - 4\,{y^3} - 22\,y\,z + 
        6\,{y^2}\,z + 6\,{z^2} - 10\,y\,{z^2} + 4\,{z^3} \right)
         \Big\}
\nonumber\\
 &+& 2\,{r^2}\,
   \Big\{ {x^4} + 4\,x\,y\,z - 4\,{z^2} + 
     {x^3}\,\left( 8 + y + 2\,z \right)  
\nonumber\\
&+&  {x^2}\,\Big( -12 + 8\,y + {y^2} 
- 8\,z - y\,z + {z^2}
         \Big)  \Big\} 
-8\,{r^3}\,{x^2}
\nonumber\\
\nonumber\\
\bar{g}_{1}&=& 
-  y\,z\,\Big\{ 2\,{x^4} + 
     {x^3}\,\left( -3 + 6\,y - 2\,z \right)  + 
     2\,{z^2}\,\left( 1 - 2\,y + 2\,z \right)  
\nonumber\\
&+& 
     {x^2}\,\left( -6\,y + 4\,{y^2} - 2\,z - 2\,{z^2} \right)
+ x\,z\,\left( 6 - 6\,y + 4\,{y^2} + z - 6\,y\,z + 
        2\,{z^2} \right)  \Big\}  
\nonumber\\
&-& 
  r\,\Big\{ -4\,{x^5} + 2\,\left( 4 + y \right) \,{z^3} + 
     {x^4}\,\left( 6 - 8\,y + 4\,z \right)  
\nonumber\\
&+& 
     x\,{z^2}\,\left( -8 - 2\,y - 2\,{y^2} + y\,z \right)  - 
     2\,{x^2}\,z\,\left( -5\,y + 5\,{y^2} + z - 5\,y\,z + 
        2\,{z^2} \right)  
\nonumber\\
&+&
     {x^3}\,\left( 6\,y - 4\,{y^2} - 4\,z - 3\,y\,z + 
        4\,{z^2} \right)  \Big\}
-
2\,{r^2}\,{x^2}\,\Big\{ 3\,{x^2} + 
     x\,\left( -4 + 3\,y - 2\,z \right)  
\nonumber\\
&+&
     \left( 4 + y - z \right) \,z \Big\} \, .
\label{g_0}
\eea
\\
After integrating the above distributions in $\cos{\theta}$, on the range 
$-1<\cos{\theta}<1$, we get
\bea
\frac{1}{\Gamma_0}
\frac{d \Gamma^{(\lambda_{\gamma}\,,\,\lambda_e)}_{\rm res}}{dx\,dy}=
\frac{\alpha}{24\pi}\,\frac{1}{{\rm A}_e\, x}\,
\Big[G_0+\lambda_{\gamma}\, \bar{G}_0+\lambda_e\,
(G_1+\lambda_{\gamma}\, \bar{G}_1)\Big]
\label{dGamxy}
\eea
where the functions $G_{0,1}$ and $\bar{G}_{0,1}$ are given by
\bea
G_{0}&=&
-6\,{\rm A}_e\,L\,\Big( 2\,x + 2\,y -3\Big) \,
   \left( {x^2} + 2\,x\,y + 2\,{y^2} \right)  + 
  {y^2}\,\Big\{ 24\,y\,\left( 2\,y -3\right)  
\nonumber\\
&+& 
     6\,x\,\left( 4 + y \right) \,\left( 4\,y -3\right)  + 
     {x^2}\,\left( 36 + \left( 33 - 10\,y \right) \,y \right)
         + 2\,{x^3}\,\left( 6 + 
        y\,\left( 2\,y -3\right)  \right)  \Big\}
\nonumber\\
&+& 
  r\,\Big\{ -6\,x\,\left( {\rm A}_e\,L\,\left( 5\,x -6\right)  + 
        8\,\left(x\,\left( 3 + x \right)  -6\right)  \right)
         - 12\,\Big( -24 + {\rm A}_e\,L\,\left( 4 + x \right)  
\nonumber\\
&+& 
        x\,\left( 26 + \left( 11 - 2\,x \right) \,x \right) 
         \Big) \,y + 4\,
      \left( 9\,{\rm A}_e\,L -24 +
        x\,\left( \left( 11 - 5\,x \right) \,x -42\right) 
         \right) \,{y^2} 
\nonumber\\
&+& 
     9\,\left(x -4\right) \,\left( 2 + x \right) \,{y^3}
      \Big\}  + 4\,{r^2}\,
   \Big\{ 24\,\left( 3\,y -4\right)  + 
     x\,\Big( 9\,{\rm A}_e\,L + x\,\left( 4\,x - 9\,y -4\right)  
\nonumber\\
&+& 
        18\,\left( 4 + y \right)  \Big)  \Big\} 
\nonumber\\
\nonumber\\
\bar{G}_0&=&
x\,\Big\{ 6\,{\rm A}_e\,L\,\left( x + 2\,y \right) \,
     \left( 2\,x + 2\,y -3\right)  
+ 
    {y^2}\,\Big( 12\,\left( 6 - x\,\left( 3 + x \right) 
           \right)  
\nonumber\\
&-& 
3\,\left( 18 + 
          x\,\left( 2\,x -1\right)  \right) \,y 
+
       2\,\left( 6 + x\,\left( 2\,x -5\right)  \right) \,
        {y^2} \Big)  
+ r\,
     \Big( 48\,\left( x\,\left( 3 + x \right)  -6 \right)  
\nonumber\\
&+& 
       12\,\left( 18 + x\,\left( 2\,x -1\right)  \right) \,
        y - 4\,\left( x -1\right) \,
        \left( 5\,x -6\right) \,{y^2} 
+ 
       9\,\left( x -2\right) \,{y^3} 
\nonumber\\
&-& 
       6\,{\rm A}_e\,L\,\left( x + 6\,y -2\right)  \Big) 
+ 
    4\,{r^2}\,\Big( 9\,{\rm A}_e\,L + 
       4\,\left( x -3 \right) \,\left( 2 + x \right)  - 
       9\,\left( x -2\right) \,y \Big)  
 \Big\}
\nonumber\\
\nonumber\\
G_1&=& {\rm A}_e\,\Big\{ 12\,\left( 3 + r - 2\,x \right) \,
      \left( {x^2} + r\,\left( {x^2} - 8\right)\right)  + 
     4\,\Big[ -3\,x\,\left( x - 1\right) \,
         \left( 6 + x \right)  
\nonumber\\
&+& 
        r\,\Big( 48 + x\,
            \left( 18 + x\,\left( 4\,x  -13 \right)  \right) 
            \Big)  \Big] \,y - 
     3\,\Big( -24 + r\,\left(x -4\right) \,
         \left( 2 + x \right)  
\nonumber\\
&+&
        x\,\left( 26 + \left( 11 - 2\,x \right) \,x \right) 
         \Big) \,{y^2} - 
     2\,\left( 24 + x\,\left( 12 + 
           x\,\left( 2\,x -5\right)  \right)  \right) \,{y^3}
      \Big\}  
\nonumber\\
&+&
6\,L\,\Big\{ 2\,{r^2}\,\left( -4 + x \right) \,
      \left( 2\,x - y \right)  + 
     y\,\left( 2\,x + 2\,y -3 \right) \,
      \left( {x^2} + 2\,x\,y + 2\,{y^2} \right)  
\nonumber\\
&+&
     r\,\Big( 4\,\left( x -1\right) \,{x^2} + 
        \left( 24 + x\,\left( 7\,x -22\right)  \right) \,y + 
        2\,\left( 3\,x -8\right) \,{y^2} - 2\,{y^3} \Big) 
      \Big\}
\nonumber\\
\nonumber\\
\bar{G}_1&=&x\,\Big\{ {\rm A}_e\,\Big[ 12\,\Big( 8\,r + 
          \left(r -3\right) \,\left( 1 + r \right) \,x
- 2\,\left( r -1\right) \,{x^2} \Big)  + 
       4\,\Big( 3\,\left( x -1\right) \,
           \left( 6 + x \right)  
\nonumber\\
&+&  r\,\left( x -4\right) \,\left( 3 + 4\,x \right) 
           \Big) \,y - 3\,
        \left( -18 + r\,\left( x -2\right)  + x - 
          2\,{x^2} \right) \,{y^2} 
\nonumber\\
&-&       2\,\left( 6 + x\,\left( 2\,x -5\right)  \right) \,
        {y^3} \Big]  
+ 6\,L\,
     \Big[ 2\,{r^2}\,\left( 2\,x - y -4\right)  - 
       y\,\left( x + 2\,y \right) \,
        \left(2\,x + 2\,y -3\right)  
\nonumber\\
&+& 
       r\,\left( -4\,{x^2} + 10\,\left( y -1\right) \,y + 
          x\,\left( 4 + 3\,y \right)  \right)  \Big]  \Big\}\, ,
\eea
where 
$L=\log \left( \frac{y+{\rm A}_e}{y-{\rm A}_e}\right)$,~ 
and $r=m_e^2/m_{\mu}^2$~ . The above results in Eq.(\ref{dGamxy}) 
completely agree with the 
corresponding ones in Ref.\cite{ss} after summing over the photon helicities.
\end{document}